\documentclass[prd,notitlepage,showpacs,nofootinbib,superscriptaddress]{revtex4-2}

\usepackage[utf8]{inputenc} 

\usepackage{amsmath}
\usepackage{amssymb}
\usepackage{float}
\usepackage{comment}
\usepackage{slashed}
\usepackage[normalem]{ulem}
\usepackage{bbold}
\usepackage{cancel}
\usepackage{graphicx}
\usepackage{multirow}
\usepackage{rotating}
\usepackage{caption}
\usepackage{subcaption}
\usepackage{tabulary}

\newcolumntype{K}[1]{>{\centering\arraybackslash}m{#1}}
\def\gsim{\raise0.3ex\hbox{$\;>$\kern-0.75em\raise-1.1ex\hbox{$\sim\;$}}}
\def\lsim{\raise0.3ex\hbox{$\;<$\kern-0.75em\raise-1.1ex\hbox{$\sim\;$}}}

\usepackage[usenames,dvipsnames]{color}

\newcommand {\ignore}[1]{}

\usepackage[colorinlistoftodos]{todonotes}
\usepackage[colorlinks=true,linkcolor=darkblue,citecolor=darkblue,urlcolor=darkblue, pdfborder={0 0 0}]{hyperref}
\usepackage[normalem]{ulem}

\usepackage{caption}
\usepackage{subcaption}
\definecolor{linkcolor}{rgb}{0,0,0.8}

\definecolor{darkgreen}{rgb}{0,0.5,0}
\definecolor{darkred}{rgb}{0.6,0,0}
\definecolor{brown}{rgb}{0.59, 0.29, 0.0}
\definecolor{mightnightblue}{RGB}{25,25,112}
\definecolor{darkblue}{rgb}{0,0,0.8}

\newcommand{\U}{\mathbf{U}}

\newcommand{\Y}{\mathbf{Y}}

\def\Y{\mathbf{Y}}

\def\Mnu{\mathbf{M}_\nu}

\def\Mnuh{\widehat{\mathbf{M}}_\nu}

\def\U{\mathbf{U}}

\newcommand{\dmatm}{\Delta m^2_{31}}
\newcommand{\dmsol}{\Delta m^2_{21}}
\newcommand{\M}{\mathbf{M}}

\def\Y{\mathbf{Y}}

\def\Mnu{\mathbf{M}_\nu}
\def\Mnuh{\widehat{\mathbf{M}}_\nu}

\def\U{\mathbf{U}}

\def\dmatm{\Delta m^2_{31}}
\def\dmsol{\Delta m^2_{21}}

\newcommand{\AddrCFTP}{%
	Departamento de F\'{\i}sica and CFTP, Instituto Superior T\'ecnico, Universidade de Lisboa, Av. Rovisco Pais 1, 1049-001 Lisboa, Portugal}

\newcommand{\AddrISEL}{ISEL - Instituto Superior de Engenharia de Lisboa, Instituto Polit\'ecnico de Lisboa, Rua Conselheiro Em\'{\i}dio Navarro, 1959-007 Lisboa, Portugal}

\usepackage[force]{feynmp-auto}

\begin{document}
	
	\title{ Minimal U(1) two-Higgs-doublet models for quark and lepton flavour}
	
	\author{\textbf{J.~R. Rocha}}\email{jose.r.rocha@tecnico.ulisboa.pt}
	\affiliation{\AddrCFTP}
	
	\author{\textbf{H.~B. C\^amara}}\email{henrique.b.camara@tecnico.ulisboa.pt}
	\affiliation{\AddrCFTP}
	
	\author{\textbf{R.~G. Felipe}}\email{ricardo.felipe@isel.pt}
	\affiliation{\AddrISEL} \affiliation{\AddrCFTP}
	
	\author{\textbf{F.~R. Joaquim}}\email{filipe.joaquim@tecnico.ulisboa.pt }
	\affiliation{\AddrCFTP}

	\begin{abstract}
		\vspace{0.2cm}
		\begin{center}
			{ \center \bf ABSTRACT}\\    
		\end{center}
		In the context of the 2HDM, and assuming that neutrinos acquire masses via the Weinberg operator, we perform a systematic analysis to determine the minimal quark and lepton flavour patterns, compatible with masses, mixing and CP violation data, realisable by Abelian symmetries. We determine four minimal models for quarks, where the number of independent parameters matches the number of observables. For the lepton sector, three minimal predictive models are identified. Namely, we find scenarios with a preference for the upper/lower octant of the $\theta_{23}$ atmospheric mixing angle, that exhibit lower bounds on the lightest neutrino masses currently probed by cosmology and testable at future neutrinoless double beta decay experiments, even for a normally-ordered neutrino masses. We investigate the phenomenology of each model taking into account all relevant theoretical, electroweak precision observables, scalar sector constraints, as well as stringent quark flavour processes such as $\overline{B} \rightarrow X_s \gamma$, $B_s \rightarrow \mu^- \mu^+$ and meson oscillations, and the charged lepton flavour-violating decays $e_\alpha^{-} \rightarrow e_\beta^{-} e_\gamma^{+} e_\delta^{-}$ and $e_\alpha \rightarrow e_\beta \gamma$. We show that, in some cases, Abelian flavour symmetries provide a natural framework to suppress flavour-changing neutral couplings and lead to scenarios featuring heavy neutral/charged scalar masses below the TeV scale within the reach of current experiments.
	\end{abstract}
	
	\maketitle
	\noindent
	
	\section{Introduction}
	\label{sec:intro}
	
	In the Standard Model~(SM), quark mixing is encoded in the unitary Cabibbo-Kobayashi-Maskawa~(CKM) matrix, appearing in weak charged-current~(CC) interactions~\cite{Cabibbo:1963yz,Kobayashi:1973fv}. This non-trivial flavour structure leads to the only known source of charge-parity~(CP) violation in Nature, first observed in the neutral Kaon system $K^0-\overline{K^0}$ in 1964~\cite{CP_violation_kaon}. Nowadays, the CKM matrix parameters have been determined with extreme precision providing a fundamental test of the SM. Nonetheless, the SM does not incorporate a guiding principle to explain the observed fermion masses and mixing patterns -- this is often referred to as the flavour puzzle. Besides the flavour puzzle in the quark sector, there is also an analogous one in the lepton sector. In fact, the observation of neutrino oscillations~\cite{Kajita:2016cak,McDonald:2016ixn} requires the existence of neutrino masses and lepton mixing, thus providing evidence for physics beyond the SM~(BSM). The growing experimental neutrino program has shed light on the properties of neutrinos, with oscillation experiments measuring the neutrino-mass squared differences and the mixing angles, i.e. the parameters of the lepton mixing matrix, also known as the Pontecorvo-Maki-Nakagawa-Sakata~(PMNS) matrix, as well as the CP-violating Dirac phase with increasing precision. Global fits to the data provide current values of the neutrino observables, some of which are known to an accuracy of less than one percent~\cite{deSalas:2020pgw,Esteban:2020cvm,Capozzi:2021fjo}. However, some fundamental questions about neutrinos remain unanswered. In particular, it is still unknown whether neutrinos are Majorana or Dirac particles, what their mass ordering and absolute mass scale are and whether or not there is leptonic CP violation. At the same time, the search for neutrinoless double beta decay~($0\nu\beta\beta$), which is sensitive to Majorana CP violation, will be crucial for exploring the particle nature of neutrinos (see~\cite{Bilenky:2014uka,DellOro:2016tmg,Dolinski:2019nrj} for recent contributions on this topic). From a theoretical perspective, and interpreting the SM as an effective theory, it is natural to consider that neutrinos are Majorana particles. In this context, naturally small Majorana masses are generated from the unique dimension-five Weinberg operator $\nu_L \nu_L \phi_0 \phi_0 / \Lambda$~\cite{Weinberg:1979sa}~\footnote{The ultraviolet completions of the SM that realise the Weinberg operator at tree-level lead to the well-known seesaw mechanisms~\cite{Minkowski:1977sc,Gell-Mann:1979vob,Yanagida:1979as,Schechter:1980gr,Glashow:1979nm,Mohapatra:1979ia}, with a plethora of possibilities for the radiative incarnations of the operator~\cite{Cai:2017jrq}.}, when the neutral Higgs doublet component acquires a non-zero vacuum expectation value~(VEV) after electroweak symmetry breaking (EWSB).
	
	A common approach to address the above mentioned flavour puzzles is to extend the SM with additional particle content and supplement the new framework with continuous and/or discrete symmetries that impose the required fermion mass and mixing structures to reproduce the observed experimental data. Several frameworks have been put forward to tackle this problem, and we wish to highlight in this work one of the simplest approaches that relies on implementation of texture zeros in the Yukawa coupling and mass matrices~\cite{Ludl:2014axa,Ludl:2015lta,GonzalezFelipe:2016tkv,Cebola:2015dwa}, realised by horizontal Abelian symmetries that can be discrete $\mathbb{Z}_N$ and/or continuous U(1)~\cite{Grimus:2004hf,Dighe:2009xj,Adhikary:2009kz,Dev:2011jc,GonzalezFelipe:2014zjk,Samanta:2015oqa,Kobayashi:2018zpq,Rahat:2018sgs,Nath:2018xih,Barreiros:2018ndn,Barreiros:2018bju,Correia:2019vbn,Camara:2020efq,Barreiros:2020gxu,Barreiros:2022aqu}. However, within the SM, Abelian flavour symmetries are not viable, as they would lead either to massless fermions and/or vanishing mixing angles~\cite{GonzalezFelipe:2014zjk,Correia:2019vbn,Camara:2020efq} since all fermions would couple to the same Higgs field. Consequently, one of the simplest extensions of the SM that allows for the implementation of such symmetries is the two-Higgs-doublet model (2HDM)~\cite{Branco:2011iw}. In fact, Abelian symmetries have been previously employed within the 2HDM to tackle separately the quark flavour puzzle~\cite{Ferreira:2010ir}, as well as the lepton one in a variety of neutrino mass generation schemes~\cite{GonzalezFelipe:2014zjk,Correia:2019vbn,Camara:2020efq}. In these 2HDM scenarios, dangerous tree-level flavour-changing neutral couplings (FCNCs) can appear. The latter are naturally controlled in the so-called Branco-Grimus-Lavoura (BGL) models~\cite{Branco:1996bq,Botella:2009pq,Botella:2012ab}, as a result of an exact symmetry of the Lagrangian, which is then spontaneously broken by the VEVs of the neutral Higgs doublet components. Therefore, Abelian flavour symmetries can be a natural framework to control the FCNCs, which are heavily constrained experimentally by quark flavour processes such as $\overline{B} \rightarrow X_s \gamma$, $B_s \rightarrow \mu^- \mu^+$ and neutral meson oscillations, among others. In the lepton sector the new heavy scalars will mediate charged lepton flavour violation~(cLFV), such as $\mu \rightarrow e \gamma$ and $\mu \rightarrow 3 e$ processes, which are the object of an ongoing experimental program with the aim to detect new physics~(NP) signals.
	
	In this work, we consider minimal Abelian flavour symmetries in the 2HDM. We take an agnostic approach to neutrino mass generation by extending the renormalisable Lagrangian by the effective Weinberg operator. The horizontal Abelian symmetries are minimal in the sense that the quark and lepton mass matrices are maximally-restricted containing the minimal number of independent parameters required to reproduce the observed fermion masses, mixing and CP violation data, while satisfying relevant phenomenological constraints. The paper is organised as follows. In Section~\ref{sec:framework}, we set up our framework by presenting the quark and lepton Yukawa sector, as well as the Weinberg operator for Majorana neutrino mass generation within the 2HDM. We identify the most restrictive flavour structures realised by Abelian flavour symmetries in Section~\ref{sec:texturesrealisable}, by performing a systematic search of all possible quark and lepton mass matrix texture-zero combinations compatible with data. We then analyse, in Section~\ref{sec:neutrino}, the lepton sector predictions for Dirac CP violation, the atmospheric mixing angle, the lightest neutrino mass and the effective Majorana neutrino mass parameter relevant for $0 \nu \beta \beta$ decay. In Section~\ref{sec:pheno}, we study the phenomenological implications of the realisable cases, taking into account theoretical, electroweak precision, scalar sector, quark and lepton flavour constraints. Finally, our concluding remarks are presented in Section~\ref{sec:concl}. Details on the scalar sector and scalar-fermion interactions in the mass-eigenstate basis can be found in the appendices.
	
	\section{Yukawa sector in the 2HDM with Majorana neutrinos}
	\label{sec:framework}
	
	In the 2HDM, the SM is extended with a second Higgs doublet, parameterised as
	\begin{equation}
	\Phi_a = \begin{pmatrix} \phi^+_a \\ \phi^0_a \end{pmatrix}
	\; , \; 
	a = 1,2 \; ,
	\end{equation}    
	with $\phi^+_a$ and $\phi^0_a$ being the charged and neutral components of the doublets, respectively. The Yukawa Lagrangian is
	\begin{align}
	-\mathcal{L}_\mathrm{Yuk.} & = \overline{q_L}\left(\Y^d_1 \Phi_1 + \Y^d_2 \Phi_2\right)d_R 
	+ \overline{q_L} \left(\Y^u_1 \Tilde{\Phi}_1 + \Y^u_2 \Tilde{\Phi}_2\right) u_R + \overline{\ell_L}\left(\Y^e_1 \Phi_1 + \Y^e_2 \Phi_2\right)e_R + \mathrm{H.c.} \, ,
	\label{eq:Lyuk2hdm}
	\end{align}
	where $q_L$ are the left-handed (LH) quark doublets, $d_R$ and $u_R$ are the down- and up-type right-handed (RH) quark singlets, respectively. The LH lepton doublets and the RH charged-lepton singlets are $\ell_L$ and $e_R$, respectively. As usual, we define $\Tilde{\Phi}_{a} = i \tau_2 \Phi_a^\ast$, with~$\tau_2$ being the complex Pauli matrix. As for $\Y^{d,u,e}_{1,2}$, these are $3 \times 3$ general complex Yukawa matrices for the down-type quarks, up-type quarks and charged leptons, respectively. We consider neutrinos to be Majorana particles with masses stemming from dimension-five Weinberg-like effective operators~\cite{Weinberg:1979sa}, which in the 2HDM read
	\begin{align}
	-\mathcal{L}_\mathrm{eff.} & = \frac{\boldsymbol{\kappa}_{a b}}{2 \Lambda} \left(\overline{\ell_L^c} \Tilde{\Phi}_a^\ast \right) \left(\Tilde{\Phi}_b^\dagger \ell_L \right) + \mathrm{H.c.} \;,\; a,b=1,2\,.
	\label{eq:Leff2hdm}
	\end{align}
	Summation over repeated indices is implicit, $\boldsymbol{\kappa}_{a b}$ are $3\times3$ symmetric dimensionless matrices in flavour space ($\boldsymbol{\kappa}_{12} = \boldsymbol{\kappa}_{21}$), and $\Lambda$ is a suppression mass scale.
	
	The EWSB is triggered when the neutral components of the Higgs doublets acquire non-zero VEVs,
	\begin{equation}
	\left<\phi^0_a\right> = \frac{v_a e^{i \varphi_a}}{\sqrt{2}} \; ,
	\end{equation} 
	where only the relative phase $\varphi = \varphi_2 - \varphi_1$ is physical. The down- and up-type quark mass matrices are
	\begin{equation}
	\mathbf{M}_d = \frac{v_1}{\sqrt{2}} \mathbf{Y}^d_1 + \frac{v_2 e^{i \varphi}}{\sqrt{2}} \mathbf{Y}^d_2 \; , \;  \mathbf{M}_u = \frac{v_1}{\sqrt{2}} \mathbf{Y}^u_1 + \frac{v_2 e^{-i \varphi}}{\sqrt{2}} \mathbf{Y}^u_2 \; ,
	\label{eq:quarksmass2hdm}
	\end{equation}
	which can be brought to the quark physical basis via the unitary transformations $d_{L,R} \rightarrow \mathbf{V}_{L,R}^d d_{L,R}$ and $u_{L,R} \rightarrow \mathbf{V}_{L,R}^u u_{L,R}$, such that
	\begin{align}
	\mathbf{V}_L^{d \dagger} \mathbf{M}_d \mathbf{V}_R^d = \mathbf{D}_d = \text{diag}(m_d,m_s,m_b) \; , \; \mathbf{V}_L^{u \dagger} \mathbf{M}_u \mathbf{V}_R^u = \mathbf{D}_u = \text{diag}(m_u,m_c,m_t) \; .
	\label{eq:massdiag}
	\end{align}
	Here, $m_{d,s,b}$ and $m_{u,c,t}$ denote the physical down- and up-type quark masses. The above unitary matrices are obtained by diagonalising the Hermitian matrices $\mathbf{H}_{d,u}= \mathbf{M}_{d,u} \mathbf{M}_{d,u}^\dagger$ and $\mathbf{H}_{d,u}^\prime= \mathbf{M}_{d,u}^\dagger \mathbf{M}_{d,u}$ as
	\begin{align}
	\mathbf{V}_L^{d \dagger} \mathbf{H}_d \mathbf{V}_L^d = \mathbf{V}_R^{d \dagger} \mathbf{H}_d^\prime \mathbf{V}_R^d = \mathbf{D}_d^2 \; , \;
	\mathbf{V}_L^{u \dagger} \mathbf{H}_u \mathbf{V}_L^u = \mathbf{V}_R^{u \dagger} \mathbf{H}_u^\prime \mathbf{V}_R^u = \mathbf{D}_u^2 \; ,
	\label{eq:hermitianmassdiag}
	\end{align}
	yielding the CKM quark mixing matrix $\mathbf{V}$,
	\begin{align}
	\mathbf{V} = \mathbf{V}_L^{u \dagger} \mathbf{V}_L^{d} \;,
	\label{eq:CKM}
	\end{align}
	appearing in the weak CC interactions. We adopt the standard parameterisation~\cite{ParticleDataGroup:2022pth}
	\begin{align}
	\mathbf{V}  = \begin{pmatrix}  c_{12}^q c_{13}^q & s_{12}^q c_{13}^q & s_{13}^q e^{-i\delta^q}  \\
	- s_{12}^q c_{23}^q - c_{12}^q s_{23}^q s_{13}^q e^{i\delta^q} & c_{12}^q c_{23}^q - s_{12}^q s_{23}^q s_{13}^q e^{i\delta^q} & s_{23}^q c_{13}^q  \\ 
	s_{12}^q s_{23}^q - c_{12}^q c_{23}^q s_{13}^q e^{i\delta^q} &- c_{12}^q s_{23}^q - s_{12}^q c_{23}^q s_{13}^q e^{i\delta^q} & c_{23}^q c_{13}^q \\ 
	\end{pmatrix} \; , 
	\label{eq:VCKMparam}
	\end{align}
	where $\theta_{ij}^q$ ($i<j=1,2,3$) are the three quark mixing angles with $c_{i j}^q \equiv \cos \theta_{i j}^q$,~$s_{i j}^q \equiv \sin \theta_{i j}^q$, and $\delta^q$ is the CKM CP-violating phase.
	
	For the lepton sector, the charged-lepton and effective neutrino mass matrices, generated upon EWSB, are given~by
	\begin{equation}
	\mathbf{M}_e = \frac{v_1}{\sqrt{2}} \mathbf{Y}^e_1 + \frac{v_2 e^{i \varphi}}{\sqrt{2}} \mathbf{Y}^e_2 \; , \;  \mathbf{M}_\nu = \frac{1}{2 \Lambda} \left(\frac{v_1^2}{2} \boldsymbol{\kappa}_{11}  + v_1 v_2 e^{i \varphi}\boldsymbol{\kappa}_{1 2} + \frac{v_2^2 e^{i 2 \varphi}}{2}  \boldsymbol{\kappa}_{2 2}\right) \; ,
	\label{eq:leptonmass2hdm}
	\end{equation}
	which can be diagonalised through the rotations $e_{L,R} \rightarrow \mathbf{U}_{L,R}^e e_{L,R}$ and $\nu_{L} \rightarrow \mathbf{U}_\nu \nu_{L}$, such that
	\begin{align}
	\mathbf{U}_L^{e \dagger} \mathbf{M}_e \mathbf{U}_R^e = \mathbf{D}_e = \text{diag}(m_e,m_\mu,m_\tau) \; , \; \mathbf{U}^{T}_\nu \mathbf{M}_\nu \mathbf{U}_\nu = \mathbf{D}_\nu = \text{diag}(m_1,m_2,m_3) \; ,
	\label{eq:leptonmassdiag}
	\end{align}
	where $m_{e,\mu,\tau}$ and $m_{1,2,3}$ are the physical charged-lepton and effective neutrino masses, respectively. The charged-lepton unitary rotations are obtained by diagonalising the Hermitian matrices $\mathbf{H}_{e}= \mathbf{M}_{e} \mathbf{M}_{e}^\dagger$ and $\mathbf{H}_{e}^\prime= \mathbf{M}_{e}^\dagger \mathbf{M}_{e}$, while the neutrino unitary rotation is obtained by diagonalising  $\mathbf{H}_{\nu}= \mathbf{M}_{\nu}^\dagger \mathbf{M}_{\nu}$, as follows,
	\begin{align}
	\mathbf{U}_L^{e \dagger} \mathbf{H}_e \mathbf{U}_L^e = \mathbf{U}_R^{e \dagger} \mathbf{H}_e^\prime \mathbf{U}_R^e = \mathbf{D}_e^2 \; , \;
	\mathbf{U}_\nu^{\dagger} \mathbf{H}_\nu \mathbf{U}_\nu = \mathbf{D}_\nu^2 \; .
	\label{eq:leptonhermitianmassdiag}
	\end{align}
	The above procedure results in the PMNS lepton mixing matrix $\mathbf{U}$, 
	\begin{align}
	\mathbf{U} = \mathbf{U}_L^{e \dagger} \mathbf{U}_\nu \; ,
	\label{eq:PMNS}
	\end{align}
	appearing in the weak CC interactions. For Majorana neutrinos, it can be parameterised as~\cite{Rodejohann:2011vc}
	\begin{align}
	\U=\begin{pmatrix}
	c_{12}^\ell c_{13}^\ell&s_{12}^\ell c_{13}^\ell&s_{13}^\ell\\
	-s_{12}^\ell c_{23}^\ell-c_{12}^\ell s_{23}^\ell s_{13}^\ell e^{i\delta^\ell}&c_{12}^\ell c_{23}^\ell-s_{12}^\ell s_{23}^\ell s_{13}^\ell e^{i\delta^\ell}&s_{23}^\ell c_{13}^\ell e^{i\delta^\ell}\\
	s_{12}^\ell s_{23}^\ell-c_{12}^\ell c_{23}^\ell s_{13}^\ell e^{i\delta^\ell}&-c_{12}^\ell s_{23}^\ell-s_{12}^\ell c_{23}^\ell s_{13}^\ell e^{i\delta^\ell}&c_{23}^\ell c_{13}^\ell e^{i\delta^\ell}
	\end{pmatrix} \begin{pmatrix}
	1&0&0\\
	0&e^{i\frac{\alpha_{21}}{2}}&0\\
	0&0&e^{i\frac{\alpha_{31}}{2}}
	\end{pmatrix}\;,
	\label{eq:UPMNSparam}
	\end{align}
	where $c_{ij}^\ell\equiv\cos\theta_{ij}^\ell$ and $s_{ij}^\ell\equiv\sin\theta_{ij}^\ell$ with $\theta_{ij}^\ell$ ($i<j=1,2,3$) being the lepton mixing angles, $\delta^\ell$ is the leptonic Dirac CP-violating phase and $\alpha_{21,31}$ are Majorana phases. 
	
	\section{Minimal 2HDMs for quarks and leptons with U(1) Abelian symmetries}
	\label{sec:texturesrealisable}
	
	In general, the mass matrices $\mathbf{M}_{d},\mathbf{M}_{u},\mathbf{M}_{e},\mathbf{M}_{\nu}$ are completely arbitrary having more independent parameters than the quark and lepton observables. Our aim is to reduce this arbitrariness by making use of Abelian symmetries that will impose restrictive flavour patterns and may lead to predictions for fermion masses, mixing angles and/or CP violation. To this end, we require the full Yukawa Lagrangian~\eqref{eq:Lyuk2hdm} and the effective Weinberg operator~\eqref{eq:Leff2hdm} to be invariant under the field transformations
	\begin{align}    
	\Phi \rightarrow \mathbf{S}_{\Phi} \Phi \; , \; q_L \rightarrow \mathbf{S}_{q} q_L \; , \; d_R \rightarrow \mathbf{S}_{d} d_R \; , \; u_R \rightarrow \mathbf{S}_{u} u_R \; , \; \ell_L \rightarrow \mathbf{S}_{\ell} \ell_L \; , \; e_R \rightarrow \mathbf{S}_{e} e_R \; ,
	\end{align}
	where $\Phi \equiv ( \Phi_1 \; \Phi_2)^T$ and the $\mathbf{S}_{\Phi,q,d,u,\ell,e}$ diagonal unitary matrices are given by~\cite{Ferreira:2010ir}:
		\begin{align}
		\mathbf{S}_\Phi &= \text{diag}\left(e^{i \theta_1} , e^{i \theta_2} \right) \; , \; \mathbf{S}_q = \text{diag}\left(e^{i \alpha_1} , e^{i \alpha_2} , e^{i \alpha_3} \right) \; , \; \mathbf{S}_d = \text{diag}\left(e^{i \beta_1} , e^{i \beta_2} , e^{i \beta_3} \right) \; , \nonumber \\ 
		\mathbf{S}_u &= \text{diag}\left(e^{i \gamma_1} , e^{i \gamma_2} , e^{i \gamma_3} \right) \; , \;
		\mathbf{S}_\ell = \text{diag}\left(e^{i \delta_1} , e^{i \delta_2} , e^{i \delta_3} \right) \; , \; \mathbf{S}_e = \text{diag}\left(e^{i \epsilon_1} , e^{i \epsilon_2} , e^{i \epsilon_3} \right) \; ,
		\label{eq:chargesfields}
		\end{align}
		with $\theta_i, \alpha_i, \beta_i, \gamma_i, \delta_i$ and $\epsilon_i$ being the continuous phases of a U(1) flavour symmetry. As a consequence, the following invariance conditions are imposed on the Yukawa matrices:
		\begin{align}
		\Y^d_a &= \mathbf{S}_{q} \Y^d_b \mathbf{S}_{d}^\dagger (\mathbf{S}_{\Phi}^\dagger)_{ba} \; , \; 
		\Y_u^a = \mathbf{S}_{q} \Y^u_b \mathbf{S}_{u}^\dagger (\mathbf{S}_{\Phi}^T)_{ba} \; , \; \Y^e_a = \mathbf{S}_{\ell} \Y^e_b \mathbf{S}_{e}^\dagger (\mathbf{S}_{\Phi}^\dagger)_{ba} \; , \; \boldsymbol{\kappa}_{ab} = \mathbf{S}_{\ell}^\dagger \boldsymbol{\kappa}_{cd} \mathbf{S}_{\ell}^\dagger (\mathbf{S}_{\Phi}^\dagger)_{ca} (\mathbf{S}_{\Phi}^\dagger)_{db} \; ,
		\label{eq:symmetrytrfs}
		\end{align}
		where summation over repeated indices is implicit. The above conditions can be written as
	\begin{align}
	(\mathbf{Y}_a^x)_{jk} = e^{i (\Theta_a^x)_{jk}} (\mathbf{Y}_a^x)_{jk} \; , \; (\boldsymbol{\kappa}_{ab})_{jk} = e^{i (\Theta_{ab}^\nu)_{jk}} (\boldsymbol{\kappa}_{ab})_{jk} \; ,
	\label{eq:phasesinv}
	\end{align}
	where $j,k=1,2,3$ are flavour indices, $x=d,u,e$ and
	\begin{align}
	(\Theta^d_a)_{ij}=\beta_j-\alpha_i+\theta_a \; , \; (\Theta^u_a)_{ij}=\gamma_j-\alpha_i-\theta_a \; , \; (\Theta^e_a)_{ij}=\epsilon_j-\delta_i+\theta_a \; , \; (\Theta^\nu_{ab})_{ij}=\delta_i+\delta_j+\theta_a + \theta_b \; .
	\label{eq:phaserel}
	\end{align}
	Eq.~\eqref{eq:phasesinv} defines U(1) phase relations which determine the presence or absence of zero entries in the Yukawa and mass matrices, defined in Eqs.~\eqref{eq:Lyuk2hdm} and~\eqref{eq:quarksmass2hdm} for quarks and Eqs.~\eqref{eq:Lyuk2hdm}, ~\eqref{eq:Leff2hdm} and~\eqref{eq:leptonmass2hdm} for leptons. Namely,
		\begin{align}
		(\mathbf{M}_x)_{ij} &= 0 \Leftrightarrow (\Theta^x_1)_{ij} \neq 0 \  \text{(mod $2\pi$)} \ \wedge \ (\Theta^x_2)_{ij} \neq 0 \ \text{(mod $2\pi$)} \; , \nonumber \\
		(\mathbf{M}_x)_{ij} & \neq 0 \Leftrightarrow (\Theta^x_1)_{ij} = 0 \ \text{(mod $2\pi$)} \ \vee \ (\Theta^x_2)_{ij} = 0 \ \text{(mod $2\pi$)} \; , \nonumber \\
		(\mathbf{M}_\nu)_{ij} &= 0 \Leftrightarrow (\Theta^\nu_{11})_{ij} \neq 0 \  \text{(mod $2\pi$)} \ \wedge \ (\Theta^\nu_{12})_{ij} \neq 0 \ \text{(mod $2\pi$)} \ \wedge \ (\Theta^\nu_{22})_{ij} \neq 0 \ \text{(mod $2\pi$)} \; , \nonumber \\
		(\mathbf{M}_\nu)_{ij} &\neq 0 \Leftrightarrow (\Theta^\nu_{11})_{ij} = 0 \  \text{(mod $2\pi$)} \ \vee \ (\Theta^\nu_{12})_{ij} = 0 \ \text{(mod $2\pi$)} \ \vee \ (\Theta^\nu_{22})_{ij} = 0 \ \text{(mod $2\pi$)} \; .
		\label{eq:canonicalcharges}
		\end{align}
		Instead of working with U(1) continuous phases, we choose a specific realization with integer charges $q_X^{j}$ with $X=\alpha,\beta,\gamma,\delta,\epsilon$ $(X=\theta)$ and $j=1,2,3$ ($j=1,2$) for fermion (scalar) fields. In this way, the continuos phase $X_j$ can be expressed as $X_j=q_X^j \zeta$ with $\zeta \in [0,2\pi[$.
		The particular case of $\zeta = 2\pi/N$, where $N=2,3,...$, corresponds to a discrete $\mathbb{Z}_N$ symmetry. The conditions \eqref{eq:canonicalcharges} can be translated into equivalent ones among the charges $q_X^{j}$ in a straightforward way.
	\begin{table}[t!]
		\renewcommand*{\arraystretch}{1.2}
		\begin{minipage}[b]{.4\textwidth}
			\centering
			\begin{tabular}{c c}
				\hline \hline
				Parameter & Best fit $\pm 1 \sigma$ \\
				\hline 
				$m_d (\times \; \text{MeV})$ \; \; & $4.67^{+0.48}_{-0.17}$ \\
				$m_s (\times \; \text{MeV})$ \; \;& $93.4^{+8.6}_{-3.4}$  \\
				$m_b (\times \; \text{GeV})$ \; \;& $4.18^{+0.03}_{-0.02}$ \\
				$m_u (\times \; \text{MeV})$ \; \;& $2.16^{+0.49}_{-0.26}$ \\
				$m_c (\times \; \text{GeV})$ \; \;& $1.27 \pm 0.02$  \\
				$m_t (\times \; \text{GeV})$\; \; & $172.69 \pm 0.30$ \\ 
				$\theta_{12}^q (^\circ)$ \; \;& $13.04\pm0.05$ \\
				$\theta_{23}^q (^\circ)$ \; \;& $2.38\pm0.06$ \\
				$\theta_{13}^q (^\circ)$ \; \;& $0.201\pm0.011$ \\
				$\delta^q (^\circ)$ \; \;& $68.75\pm4.5$ \\
				\hline \hline
			\end{tabular}
		\end{minipage}
		\begin{minipage}[b]{.55\textwidth}
			\centering
			\begin{tabular}{c c}  
				\hline \hline
				Parameter  & Best Fit $\pm 1 \sigma$ \\ \hline
				$m_e (\times \; \text{keV})$ \; \; & $510.99895000\pm 0.00000015$ \\
				$m_\mu (\times \; \text{MeV})$ \; \;& $105.6583755\pm 0.0000023$   \\
				$m_\tau (\times \; \text{GeV})$ \; \;& $1.77686\pm 0.00012$ \\
				$\Delta m_{21}^2 \left(\times 10^{-5} \ \text{eV}^2\right)$ \; \; & $7.50^{+0.22}_{-0.20}$  \\
				$\left|\Delta m_{31}^2\right| \left(\times 10^{-3}  \ \text{eV}^2\right) [\text{NO}]$ \; \; & $2.55^{+0.02}_{-0.03}$  \\
				$\left|\Delta m_{31}^2\right| \left(\times 10^{-3} \ \text{eV}^2\right) [\text{IO}]$ \; \; & $2.45^{+0.02}_{-0.03}$ \\
				$\theta_{12}^\ell (^\circ)$ \; \; & $34.3\pm1.0$ \\
				$\theta_{23}^\ell (^\circ) [\text{NO}]$ \; \; & $49.26\pm0.79$ \\
				$\theta_{23}^\ell (^\circ) [\text{IO}]$ \; \; & $49.46^{+0.60}_{-0.97}$ \\
				$\theta_{13}^\ell (^\circ) [\text{NO}]$ \; \; & $8.53^{+0.13}_{-0.12}$ \\
				$\theta_{13}^\ell (^\circ) [\text{IO}]$ \; \; & $8.58^{+0.12}_{-0.14}$ \\
				$\delta^\ell  (^\circ) [\text{NO}]$ \; \; & $194^{+24}_{-22}$ \\
				$\delta^\ell  (^\circ) [\text{IO}]$ \; \; & $284^{+26}_{-28}$ \\
				\hline \hline
			\end{tabular}
		\end{minipage}
		\caption{(Left) Current quark data: masses, mixing angles and Dirac CP phase~\cite{ParticleDataGroup:2022pth}. (Right) Current lepton data: charged-lepton 
			masses~\cite{ParticleDataGroup:2022pth}, neutrino mass-squared differences, mixing angles and Dirac CP phase, obtained from the global fit of neutrino oscillation data of Ref.~\cite{deSalas:2020pgw} (see also Refs.~\cite{Esteban:2020cvm} and~\cite{Capozzi:2021fjo}).}
		\label{tab:data}
	\end{table}

	To determine if a given set of mass matrices $\left(\mathbf{M}_{d},\mathbf{M}_{u},\mathbf{M}_{e},\mathbf{M}_{\nu}\right)$ are compatible with quark, charged-lepton and neutrino data within the 2HDM framework [see Eqs.~\eqref{eq:quarksmass2hdm} and~\eqref{eq:leptonmass2hdm}], we make use of a standard $\chi^2$-analysis, with the~function
	\begin{equation}
	\chi^2(x) = \sum_i \frac{\left[\mathcal{P}_i(x) - \mathcal{O}_i\right]^2}{\sigma_i^2} \; ,
	\label{eq:chi2}
	\end{equation}
	where $x$ denotes the input parameters, i.e., the matrix elements of $\mathbf{M}_{d}$, $\mathbf{M}_{u}$, $\mathbf{M}_{e}$ and $\mathbf{M}_{\nu}$; $\mathcal{P}_i(x)$ is the model prediction for a given observable with best-fit value $\mathcal{O}_i$, and $\sigma_i$ denotes its $1\sigma$ experimental uncertainty. In our search for viable sets $\left(\mathbf{M}_{d},\mathbf{M}_{u},\mathbf{M}_{e},\mathbf{M}_{\nu}\right)$, we use the current data reported in Table~\ref{tab:data} and require that the $\chi^2$-function is minimised with respect to ten observables in the quark sector: the six quark masses $m_{d,s,b}$ and $m_{u,c,t}$, as well as the CKM parameters -- the three mixing angles $\theta_{12,23,13}^q$ and the CP-violating phase $\delta^q$; and nine observables in the lepton sector: three charged-lepton masses $m_{e,\mu,\tau}$, the two neutrino mass-squared differences $\Delta m^2_{21}, \Delta m^2_{31}$, and the PMNS parameters -- the three mixing angles $\theta_{12,23,13}^\ell$ and the Dirac CP-violating phase $\delta^\ell$. Note that, for the neutrino oscillation parameters, $\chi^2(x)$ is computed using the one-dimensional profiles $\chi^2(\sin^2 \theta^\ell_{ij})$ and $\chi^2(\Delta m^2_{ij})$, and the two-dimensional~(2D) distribution $\chi^2(\delta^\ell,\sin^2 \theta^\ell_{23})$ for $\delta^\ell$ and $\theta_{23}^\ell$ given in Ref.~\cite{deSalas:2020pgw}. In our analysis, we consider a set of mass matrices $\left(\mathbf{M}_d, \mathbf{M}_u, \mathbf{M}_e, \mathbf{M}_\nu\right)$ to be compatible with data if the 
	observables in Table~\ref{tab:data} fall within the $1\sigma$ range at the $\chi^2$-function minimum.
	
	A given matrix set $\left(\mathbf{M}_{d},\mathbf{M}_{u},\mathbf{M}_{e},\mathbf{M}_{\nu}\right)$ is unique up to the following weak-basis permutations:
	\begin{align}
	\mathbf{M}_d \rightarrow \mathbf{P}_q^T \mathbf{M}_d \mathbf{P}_d\; , \;  \mathbf{M}_u \rightarrow \mathbf{P}_q^T \mathbf{M}_u \mathbf{P}_u \; , \; \mathbf{M}_e \rightarrow \mathbf{P}_\ell^T \mathbf{M}_e \mathbf{P}_e\; , \; \mathbf{M}_\nu \rightarrow \mathbf{P}_\ell^T \mathbf{M}_\nu \mathbf{P}_\nu\; , \;
	\end{align}
	where $\textbf{P}$ are the $3 \times 3$ permutation matrices:
	\begin{align}
	\mathbf{I} &= \begin{pmatrix}
	1 & 0 & 0 \\
	0 & 1 & 0 \\
	0 & 0 & 1 \\
	\end{pmatrix} \; , \; \mathbf{P}_{12} = \begin{pmatrix}
	0 & 1 & 0 \\
	1 & 0 & 0 \\
	0 & 0 & 1 \\
	\end{pmatrix} \; , \; \mathbf{P}_{13} = \begin{pmatrix}
	0 & 0 & 1 \\
	0 & 1 & 0 \\
	1 & 0 & 0 \\
	\end{pmatrix} \; , \nonumber \\
	\mathbf{P}_{23} &= \begin{pmatrix}
	1 & 0 & 0 \\
	0 & 0 & 1 \\
	0 & 1 & 0 \\
	\end{pmatrix} \; , \; \mathbf{P}_{123} = \begin{pmatrix}
	0 & 0 & 1 \\
	1 & 0 & 0 \\
	0 & 1 & 0 \\
	\end{pmatrix} \; , \; \mathbf{P}_{321} = \begin{pmatrix}
	0 & 1 & 0 \\
	0 & 0 & 1 \\
	1 & 0 & 0 \\
	\end{pmatrix} \; .
	\label{eq:permutationmatrices}
	\end{align}
	Note that these permutations do not affect the mass matrix diagonalisation procedure outlined in Section~\ref{sec:framework} for extracting the mass, mixing and CP-violating observables. Thus, our search for compatible matrix sets is simplified, since only the physically-equivalent sets, the so-called equivalence classes~\cite{Ludl:2014axa,Ludl:2015lta,GonzalezFelipe:2014zjk,Correia:2019vbn,Camara:2020efq}, need to be considered.
	
	We will identify the minimal quark and lepton models featuring Abelian symmetries that impose the maximally-restrictive flavour patterns in the set of mass matrices $\left(\mathbf{M}_{d},\mathbf{M}_{u},\mathbf{M}_{e},\mathbf{M}_{\nu}\right)$. By maximally restrictive we mean that the maximal number of vanishing entries in the mass matrices is imposed in such a way that the resulting flavour patterns are compatible with the current quark and lepton data given in Table~\ref{tab:data}. We employ the methodology described in Refs.~\cite{GonzalezFelipe:2016tkv,Correia:2019vbn,Camara:2020efq}, namely, we use the canonical method~\cite{Ferreira:2010ir,Serodio:2013gka} to identify the realisable flavour patterns for the set $\left(\mathbf{M}_{d},\mathbf{M}_{u},\mathbf{M}_{e},\mathbf{M}_{\nu}\right)$ and the corresponding field transformation charges by making use of Eq.~\eqref{eq:canonicalcharges}, followed by the Smith Normal Form method~\cite{Ivanov:2011ae,Ivanov:2013bka} that ascertains the minimal Abelian group under which the flavour patterns are symmetric. In our framework, the minimal Abelian group that establishes the patterns corresponds to a single global U$(1)$ (which can be discretised as commented below). In Table~\ref{tab:charges}, we show the U$(1)$ Abelian charges that realise the minimal quark and lepton flavour patterns presented in Table~\ref{tab:Matrices}. Next we briefly comment on the results.
	\begin{table}[t!]
		\renewcommand*{\arraystretch}{1.4}
		\begin{minipage}[b]{0.40\textwidth}
			\centering
			\begin{tabular}{lccc}
				\hline
				\hline
				($\mathbf{M}_d$,$\mathbf{M}_u$) & 
				$q_\alpha^j$ & 
				$q_\beta^j$
				& $q_\gamma^j$
				\\ \hline
				$(4_{3}^d,\mathbf{P}_{12}{5}_{1}^u\mathbf{P}_{23})$             &      $(0,1,2)$    &    $(2,1,0)$  &  $(3,2,0)$      \\
				$(4_{3}^d,\mathbf{P}_{123}5_{1}^u \mathbf{P}_{12})$             &      $(0,1,2)$    &    $(2,1,0)$  &  $(3,0,1)$  \\
				$(5_{1}^d,\mathbf{P}_{12}{4}_{3}^u)$          &      $(0,-1,1)$    &    $(1,-2,0)$  &  $(2,1,0)$      \\
				$(5_{1}^d,\mathbf{P}_{321}4_{3}^u\mathbf{P}_{23})$           &      $(0,-1,1)$    &    $(1,-2,0)$  &  $(-1,1,0)$ \\
				\hline \hline
			\end{tabular}
		\end{minipage}
		\begin{minipage}[b]{0.4\textwidth}
			\centering
			\begin{tabular}{ccc}
				\hline
				\hline
				($\mathbf{M}_{e}$,$\mathbf{M}_\nu$)      & 
				$q_\delta^j$
				& $q_\epsilon^j$ 
				\\ \hline
				$(5_{1}^{e},2_{3}^{\nu})$             &      $\left(-1,-3,1\right)$    &    $\left(1,-5,-1\right)$     
				\\
				$(5_{1}^{e},2_{7}^{\nu})$             &      $(-1,-2,0)$    &    $(0,-3,-1)$        
				\\
				$(5_{1}^{e},2_{10}^{\nu})$          &      $(0,-1,1)$    &    $(1,-2,0)$            \\ \hline \hline
			\end{tabular}
		\end{minipage}
		\caption{
			Maximally-restrictive quark matrix pairs ($\mathbf{M}_d$,$\mathbf{M}_u$) and lepton matrix pairs ($\mathbf{M}_e$,$\mathbf{M}_\nu$) for both NO and IO, compatible with data at~$1 \sigma$ CL (see Table~\ref{tab:data}) and realisable by Abelian flavour symmetries. The U(1) flavour charges $q_X^j$ with $X=\alpha,\beta,\gamma,\delta,\epsilon$ and $j=1,2,3$ for the various fermion fields are shown. In all cases $q_\theta^1=0$ and $q_\theta^2=1$, except for $(5_{1}^{e},2_{3}^{\nu})$ where $q_\theta^2=2$.} 
		\label{tab:charges}
	\end{table}
	\begin{table}[t!]
		\renewcommand*{\arraystretch}{1.2}
		\begin{minipage}[b]{0.45\textwidth}
			\centering
			\begin{tabular}{rcc}
				\hline
				\hline
				Texture Decomposition  & $\mathbf{Y}_1^{d}$ & $\mathbf{Y}_2^{d}$ 
				\\
				\hline
				\\[-10pt] 
				$4_{3}^{d}\sim \begin{pmatrix}
				0 & 0 & \times \\
				0 & \times & \times \\
				\times & \times & 0\\
				\end{pmatrix}$ \; \;    &    
				$\begin{pmatrix}
				0 & 0 & \times \\
				0 & \times & 0 \\
				\times & 0 & 0 \\
				\end{pmatrix}$ &
				$\begin{pmatrix}
				0 & 0 & 0 \\
				0 & 0 & \times \\
				0 & \times & 0 \\
				\end{pmatrix}$ 
				\\
				[20pt]
				$5_{1}^{d}\sim \begin{pmatrix}
				0 & 0 & \times \\
				0 & \times & 0 \\
				\times & 0 & \times\\
				\end{pmatrix}$ \; \;   
				&   
				$\begin{pmatrix}
				0 & 0 & \times \\
				0 & 0 & 0 \\
				\times & 0 & 0 \\
				\end{pmatrix}$ &
				$\begin{pmatrix}
				0 & 0 & 0 \\
				0 & \times & 0 \\
				0 & 0 & \times \\
				\end{pmatrix}$
				\\
				[19pt]
				\hline
				\hline
				Texture Decomposition & $\mathbf{Y}_1^{u}$ & $\mathbf{Y}_2^{u}$ \\
				\hline
				\\[-10pt] 
				$\mathbf{P}_{12}{5}_{1}^u\mathbf{P}_{23}\sim \begin{pmatrix}
				0 & 0 & \times \\
				0 & \bullet & 0 \\
				\times & \times & 0 \\
				\end{pmatrix}$ \; \; 
				&    
				$\begin{pmatrix}
				0 & 0 & \times \\
				0 & 0 & 0 \\
				0 & \times & 0 \\
				\end{pmatrix}$ &
				$\begin{pmatrix}
				0 & 0 & 0 \\
				0 & \bullet & 0 \\
				\times & 0 & 0 \\
				\end{pmatrix}$ 
				\\
				[20pt]
				$\mathbf{P}_{123}5_{1}^u \mathbf{P}_{12}\sim \begin{pmatrix}
				0 & \times & \bullet \\
				0 & 0 & \times \\
				\times & 0 & 0\\
				\end{pmatrix}$ \; \; 
				&    
				$\begin{pmatrix}
				0 & \times & 0 \\
				0 & 0 & \times \\
				0 & 0 & 0 \\
				\end{pmatrix}$ &
				$\begin{pmatrix}
				0 & 0 & \bullet \\
				0 & 0 & 0 \\
				\times & 0 & 0 \\
				\end{pmatrix}$ 
				\\
				[20pt]
				$\mathbf{P}_{12}{4}_{3}^u\sim \begin{pmatrix}
				0 & \bullet & \times \\
				0 & 0 & \times \\
				\times & \times & 0\\
				\end{pmatrix}$ \; \;     
				&    
				$\begin{pmatrix}
				0 & 0 & \times \\
				0 & 0 & 0 \\
				0 & \times & 0 \\
				\end{pmatrix}$ &
				$\begin{pmatrix}
				0 & \bullet & 0 \\
				0 & 0 & \times \\
				\times & 0 & 0 \\
				\end{pmatrix}$ 
				\\
				[20pt]
				$\mathbf{P}_{321}4_{3}^u\mathbf{P}_{23}\sim \begin{pmatrix}
				0 & \bullet & \times \\
				\times & 0 & \times \\
				0 & \times & 0\\
				\end{pmatrix}$ \; \; 
				&           
				$\begin{pmatrix}
				0 & 0 & \times \\
				\times & 0 & 0 \\
				0 & \times & 0 \\
				\end{pmatrix}$ &
				$\begin{pmatrix}
				0 & \bullet & 0 \\
				0 & 0 & \times \\
				0 & 0 & 0 \\
				\end{pmatrix}$ 
				\\
				[19pt]
				\hline
				\hline
			\end{tabular}
		\end{minipage}%
		\hspace{0.5mm}
		\begin{minipage}[b]{0.5\textwidth}
			\centering
			\begin{tabular}{rccc}
				\hline
				\hline
				Texture Decomposition & $\mathbf{Y}_1^{e}$ & $\mathbf{Y}_2^{e}$
				\\
				\hline
				\\[-10pt]
				$5_{1}^{e}\sim \begin{pmatrix}
				0 & 0 & \times \\
				0 & \times & 0 \\
				\times & 0 & \times\\
				\end{pmatrix}$ \; \;   
				&
				$\begin{pmatrix} 0 & 0 & \times \\ 0 & 0 & 0 \\ \times & 0 & 0 \end{pmatrix}$ & $\begin{pmatrix} 0 & 0 & 0 \\ 0 & \times & 0 \\ 0 & 0 & \times \end{pmatrix}$
				\\
				[19pt]
				\hline
				\hline
				& $\boldsymbol{\kappa}_{11}$ & $\boldsymbol{\kappa}_{12}$ & $\boldsymbol{\kappa}_{22}$
				\\
				\hline
				\\[-10pt]
				$2_{3}^{\nu}\sim \begin{pmatrix}
				\times & \times & \bullet \\
				\cdot & 0 & \bullet \\
				\cdot & \cdot & 0\\
				\end{pmatrix}$ \; \; 
				& 
				$\begin{pmatrix} 0 & 0 & \bullet \\ \cdot & 0 & 0 \\ \cdot & \cdot & 0 \end{pmatrix}$ & $\begin{pmatrix} \times & 0 & 0 \\ \cdot & 0 & \bullet \\ \cdot & \cdot & 0 \end{pmatrix}$ 
				& 
				$\begin{pmatrix} 0 & \times & 0 \\ \cdot & 0 & 0 \\ \cdot & \cdot & 0 \end{pmatrix}$
				\\[20pt]
				$2_{7}^{\nu}\sim \begin{pmatrix}
				\times & 0 & \bullet \\
				\cdot & 0 & \times \\
				\cdot & \cdot & \bullet\\
				\end{pmatrix}$ \; \;  
				& 
				$\begin{pmatrix} 0 & 0 & 0 \\ \cdot & 0 & 0 \\ \cdot & \cdot & \bullet \end{pmatrix}$
				& 
				$\begin{pmatrix} 0 & 0 & \bullet \\ \cdot & 0 & 0 \\ \cdot & \cdot & 0 \end{pmatrix}$ & $\begin{pmatrix} \times & 0 & 0 \\ \cdot & 0 & \times \\ \cdot & \cdot & 0 \end{pmatrix}$
				\\[20pt]
				$2_{10}^{\nu}\sim \begin{pmatrix}
				\times & \bullet & 0 \\
				\cdot & \times & \bullet \\
				\cdot & \cdot & 0\\
				\end{pmatrix}$ \; \;    
				& $\begin{pmatrix} \times & 0 & 0 \\ \cdot & 0 & \bullet \\ \cdot & \cdot & 0 \end{pmatrix}$ & $\begin{pmatrix} 0 & \bullet & 0 \\ \cdot & 0 & 0 \\ \cdot & \cdot & 0 \end{pmatrix}$ & 
				$\begin{pmatrix} 0 & 0 & 0 \\ \cdot & \times & 0 \\ \cdot & \cdot & 0 \end{pmatrix}$ 
				\\
				[19pt]
				\hline
				\hline
			\end{tabular}
		\end{minipage}
		\caption{Realisable decomposition into Yukawa matrices of the quark (left) and lepton (right) mass matrices for the texture pairs of Table~\ref{tab:charges}. A matrix entry ``$0$" denotes a texture zero, ``$\times$" and ``$\bullet$" denote a real positive and complex entry, respectively. The symmetric character of the Majorana matrix is marked by a ``$\cdot$".} 
		\label{tab:Matrices}
	\end{table}
	\begin{itemize}
		\item \textbf{Minimal flavour patterns for quarks:} The maximally-restrictive textures in general quark mass matrices were derived in Ref.~\cite{Ludl:2015lta} with a less constraining definition of compatibility. Therefore, it suffices to examine whether such matrices can be realised by Abelian flavour symmetries, being simultaneously compatible with data, according to the definition provided in this paper. This procedure led to four models, denoted in the notation of Ref.~\cite{Ludl:2015lta} as $G_9:(4_{3}^d,\mathbf{P}_{12}{5}_{1}^u\mathbf{P}_{23})$, $G_{12}:(4_{3}^d,\mathbf{P}_{123}5_{1}^u \mathbf{P}_{12})$, $G_{20}:(5_{1}^d,\mathbf{P}_{12}{4}_{3}^u)$ and $G_{23}:(5_{1}^d,\mathbf{P}_{321}4_{3}^u\mathbf{P}_{23})$, where to distinguish between cases we explicitly show the permutation matrices $\mathbf{P}$ [see Eq.~\eqref{eq:permutationmatrices}]. The results are shown in more detail in Table~\ref{tab:charges} for the U(1) flavour charges and in Table~\ref{tab:Matrices} for the matrix structures. It is worth noting that the maximally restrictive ($\mathbf{M}_d$, $\mathbf{M}_u$) pairs contain a total of ten independent parameters that match the number of observables. Additionally, in all cases, the minimal set of discrete charges corresponds to a $\mathbb{Z}_4$ symmetry (see Ref.~\cite{Ferreira:2010ir}).
		
		\item \textbf{Minimal flavour patterns for leptons:} To identify the realisable maximally-restrictive matrix pairs ($\mathbf{M}_{e}$, $\mathbf{M}_\nu$), we examine the equivalence classes with the highest number of zeros derived in Ref.~\cite{Ludl:2014axa}. We then check whether they can be realised by Abelian symmetries and whether they are compatible with the current lepton data given in Table~\ref{tab:data}. If none of the equivalence classes under consideration passes this test, we add a non-zero entry to the mass matrix pairs and repeat the process until compatibility is achieved. This methodology, which is similar to that of previous works~\cite{GonzalezFelipe:2016tkv,Correia:2019vbn,Camara:2020efq}, led to three different models. The U(1) flavour charges for each model are shown in Table \ref{tab:charges}, while their respective matrix structures can be found in Table \ref{tab:Matrices}. As it turns out, ($\mathbf{M}_{e}$, $\mathbf{M}_\nu$) contain only ten parameters, two less than the number of lepton observables. These observables are those listed in Table~\ref{tab:data}, together with the lightest neutrino mass $m_\text{lightest}$ and two Majorana phases $\alpha_{21,31}$. Consequently, these symmetry-constrained models provide predictions for the neutrino sector. Moreover, the minimal set of discrete charges in all three cases corresponds to  a $\mathbb{Z}_5$ symmetry.
		
	\end{itemize}
	We conclude this section by remarking that the minimal flavour models found compatible with data, four for quarks and three for leptons, will result in a total of twelve combined models. In what follows, we will investigate in Section~\ref{sec:neutrino} the predictions for the neutrino sector and thoroughly study the phenomenological properties of these models in Section~\ref{sec:pheno}.
	
	\section{Lepton sector predictions}
	\label{sec:neutrino}
	
	In the previous section, we have identified which of the maximally-restrictive texture sets for the lepton sector can be realised through Abelian symmetries in the 2HDM context. For the three identified cases, namely $(5_{1}^{e},2_{3}^{\nu})$, $(5_{1}^{e},2_{7}^{\nu})$ and $(5_{1}^{e},2_{10}^{\nu})$, shown in Table~\ref{tab:Matrices}, the charged-lepton mass matrix has the same texture structure $5_{1}^{e}$, which can be parameterised as
	\begin{equation}
	5_1^{e} : \mathbf{M}_{e} = \begin{pmatrix} 
	0 & 0 & a_1\\ 
	0 & a_3 & 0 \\
	a_2 & 0 & a_4     
	\end{pmatrix}, \ a_1^2 = \frac{m^2_{e_{2}} m^2_{e_{3}}}{a_2^2}, \ a_3^2 =  m^2_{e_{1}}, \ a_4^2 = \frac{(a_2^2-m^2_{e_{2}}) (m^2_{e_{3}}-a_2^2)}{a_2^2}, \ m_{e_{2}}<a_2<m_{e_{3}},
	\end{equation}
	where $a_i$ can always be made real by phase field redefinitions, and $m_{e_{i}} (i=1,2,3)$ are the charged-lepton masses. Note that the charged-lepton state $e_1$ is decoupled from the remaining ones. The unitary matrices $\mathbf{U}_{L,R}^{e \prime}$ that diagonalise the Hermitian matrices $\mathbf{H}_{e} = \mathbf{M}_{e} \mathbf{M}_{e}^{\dagger}$ and $\mathbf{H}_{e}^\prime = \mathbf{M}_{e}^{\dagger} \mathbf{M}_{e} $ are given by
	\begin{equation}
	\mathbf{H}_{e} = \begin{pmatrix} 
	a_1^2 & 0 & a_1 a_4\\ 
	0 & a_3^2 & 0 \\
	a_1 a_4 & 0 & a_2^2 + a_4^2     
	\end{pmatrix}\;,\; \mathbf{U}_{L}^{e\prime} = \begin{pmatrix} 
	c_L & 0 & s_L\\ 
	0 & 1 & 0 \\
	- s_L & 0 & c_L
	\end{pmatrix} \;,\;
	\mathbf{H}_{e}^\prime = \begin{pmatrix} 
	a_2^2 & 0 & a_2 a_4\\ 
	0 & a_3^2 & 0 \\
	a_2 a_4 & 0 & a_1^2 + a_4^2     
	\end{pmatrix}\;,\; \ \mathbf{U}_{R}^{e\prime} = \begin{pmatrix} 
	c_R & 0 & s_R\\ 
	0 & 1 & 0 \\
	- s_R & 0 & c_R
	\end{pmatrix} \, ,
	\end{equation}
	using the compact notation $c_{L,R}\equiv \cos \theta_{L,R}$ and $s_{L,R}\equiv \sin \theta_{L,R}$ with the angles $\theta_{L,R}$ given by
	\begin{equation}
	\tan \left(2 \theta_{L}\right) = \frac{2 m_{e_{2}} m_{e_{3}} \sqrt{(a_2^2-m_{e_{2}}^2)(m_{e_{3}}^2-a_2^2)}}{a_2^2(m_{e_{2}}^2+m_{e_{3}}^2)-2 m_{e_{2}}^2 m_{e_{3}}^2} \;,\;
	\tan \left(2 \theta_{R}\right) = \frac{2 \sqrt{(a_2^2-m_{e_{2}}^2)(m_{e_{3}}^2-a_2^2)}}{m_{e_{2}}^2+m_{e_{3}}^2-2 a_2^2}.
	\end{equation}
	We consider the three distinct cases of $5_1^{e_1}$ textures with $e_{1}=e,\mu,\tau$, labelled as $5_1^{e,\mu,\tau}$. Note that while performing the numerical fits in Section~\ref{sec:texturesrealisable} our aim was to identify the best-fit model, requiring compatibility with data at $1\sigma$ confidence level. As it turned out, in some $e_{1}$-decoupled cases, the best-fit value exceeded this threshold and thus such texture pairs were not presented. In this section, nevertheless, we proceed with their analysis for the sake of completeness. 
	
	Since after the diagonalisation of the charged-lepton mass matrix the unitary matrices $\mathbf{U}_{L,R}^{e}$ are such that the correct mass ordering is obtained as in Eq.~\eqref{eq:leptonmassdiag}, we have
	\begin{equation}
	5_1^{e}: \mathbf{U}_{L,R}^{e}=\mathbf{U}_{L,R}^{e\prime} \mathbf{P}_{12}, \quad 5_1^{\mu}: \mathbf{U}_{L,R}^{e}=\mathbf{U}_{L,R}^{e\prime}, \quad 5_1^{\tau}: \mathbf{U}_{L,R}^{e}=\mathbf{U}_{L,R}^{e\prime} \mathbf{P}_{23},
	\label{eq:VLdecoupled}
	\end{equation}
	with the permutation matrices $\mathbf{P}_{12,23}$ defined in Eq.~\eqref{eq:permutationmatrices}. Regarding the neutrino mass matrix textures identified in Table~\ref{tab:Matrices}, the $2_{3,7,10}^\nu$ effective neutrino mass matrices can be parameterised as follows:
	\begin{align}
	2_3^e \ &\text{:} \ \mathbf{M}_{\nu} = \begin{pmatrix} 
	b_1  & b_2 & b_3 e^{i \xi_1}\smallskip\\
	\cdot & 0 &b_4 e^{i \xi_2}\smallskip\\
	\cdot & \cdot & 0
	\end{pmatrix}\Rightarrow \U_L^{eT}\,\M_{\nu} \U_L^e = \begin{pmatrix} 
	0  & b_2 c_L - b_4 s_L e^{i \xi_2} & b_2 s_L + b_4 c_L e^{i \xi_2}\smallskip\\
	\cdot & b_1 c_L^2 - b_3 \sin (2\theta_L) e^{i \xi_1} &b_1 c_L s_L + b_3 \cos (2\theta_L) e^{i \xi_1}\smallskip\\
	\cdot & \cdot & b_1 s_L^2 + b_3 \sin (2\theta_L) e^{i \xi_1}
	\end{pmatrix} \; ,
	\label{eq:Meff23} \\
	2_7^e \ &\text{:} \ \mathbf{M}_{\nu} = \begin{pmatrix} 
	b_1  & 0 & b_2 e^{i \xi_1}\smallskip\\
	\cdot & 0 &b_3 \smallskip\\
	\cdot & \cdot & b_4 e^{i \xi_2}
	\end{pmatrix} \nonumber \\
	& \Rightarrow \U_L^{eT}\,\M_{\nu} \U_L^e = \begin{pmatrix} 
	0  & - b_3 s_L& b_3 c_L\smallskip\\
	\cdot & b_1 c_L^2 - b_2 \sin (2 \theta_L) e^{i \xi_1} + b_4 s_L^2 e^{i \xi_2} & \left(b_1 - b_4 e^{i \xi_2}\right) c_L s_L + b_2 \cos (2 \theta_L) e^{i \xi_1} \smallskip\\
	\cdot & \cdot & b_1 s_L^2 + b_2 \sin (2 \theta_L) e^{i \xi_1} + b_4 c_L^2 e^{i \xi_2}
	\end{pmatrix} \; ,
	\label{eq:Meff27} \\
	2_{10}^e \ &\text{:} \ \mathbf{M}_{\nu} = \begin{pmatrix} 
	b_1  & b_2 e^{i \xi_1}& 0 \smallskip\\
	\cdot & b_3  & b_4 e^{i \xi_2}\smallskip\\
	\cdot & \cdot & 0
	\end{pmatrix} \Rightarrow \U_L^{eT}\,\M_{\nu} \U_L^e = \begin{pmatrix} 
	b_3  & b_2 c_L e^{i \xi_1} - b_4 s_L e^{i \xi_2} & b_2 s_L e^{i \xi_1} + b_4 c_L e^{i \xi_2} \smallskip\\
	\cdot & b_1 c_L^2 & b_1 c_L s_L \smallskip\\
	\cdot & \cdot & b_1 s_L^2
	\end{pmatrix} \; ,
	\label{eq:Meff210}
	\end{align}
	with all parameters real. Note also that we have performed the rotation to the charged-lepton mass basis with the unitary matrix $\mathbf{U}_{L}^e$ given in Eq.~\eqref{eq:VLdecoupled}, for the $5_1^{e}$ decoupled case. For $5_1^\mu$ and $5_1^{\tau}$, the permutation matrices $\mathbf{P}_{12}$ and $\mathbf{P}_{12}\mathbf{P}_{23}$ of Eq.~\eqref{eq:permutationmatrices} were applied on the left and right. In what follows, we denote each decoupled case as $2_{3,7,10}^{e,\mu,\tau}\,$. 

	\begin{figure}[t!]
		\centering
		\includegraphics[scale=0.30]{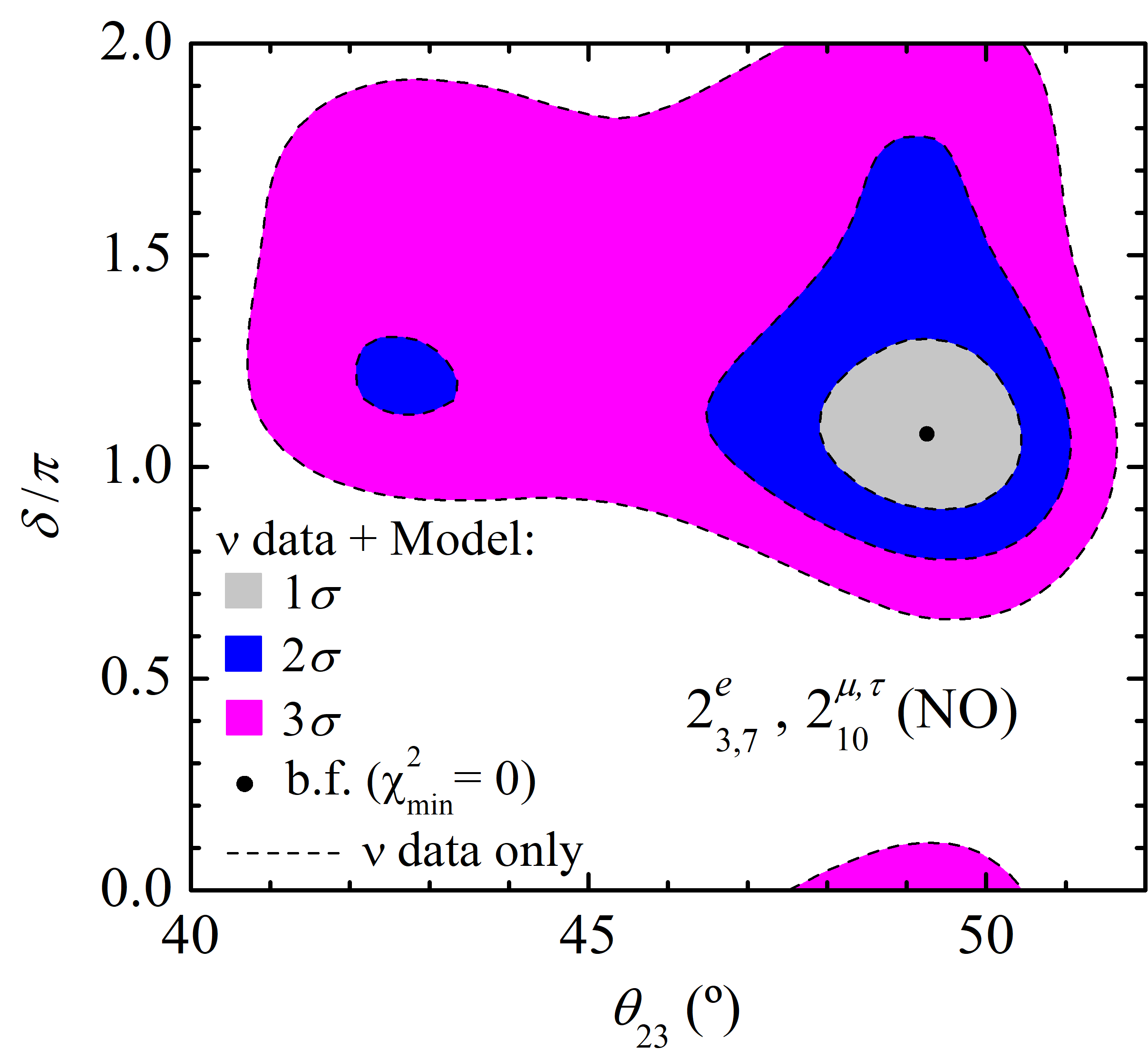} \hspace{0.2cm} 
		\includegraphics[scale=0.30,trim={0cm 0cm 0cm 0cm},clip]{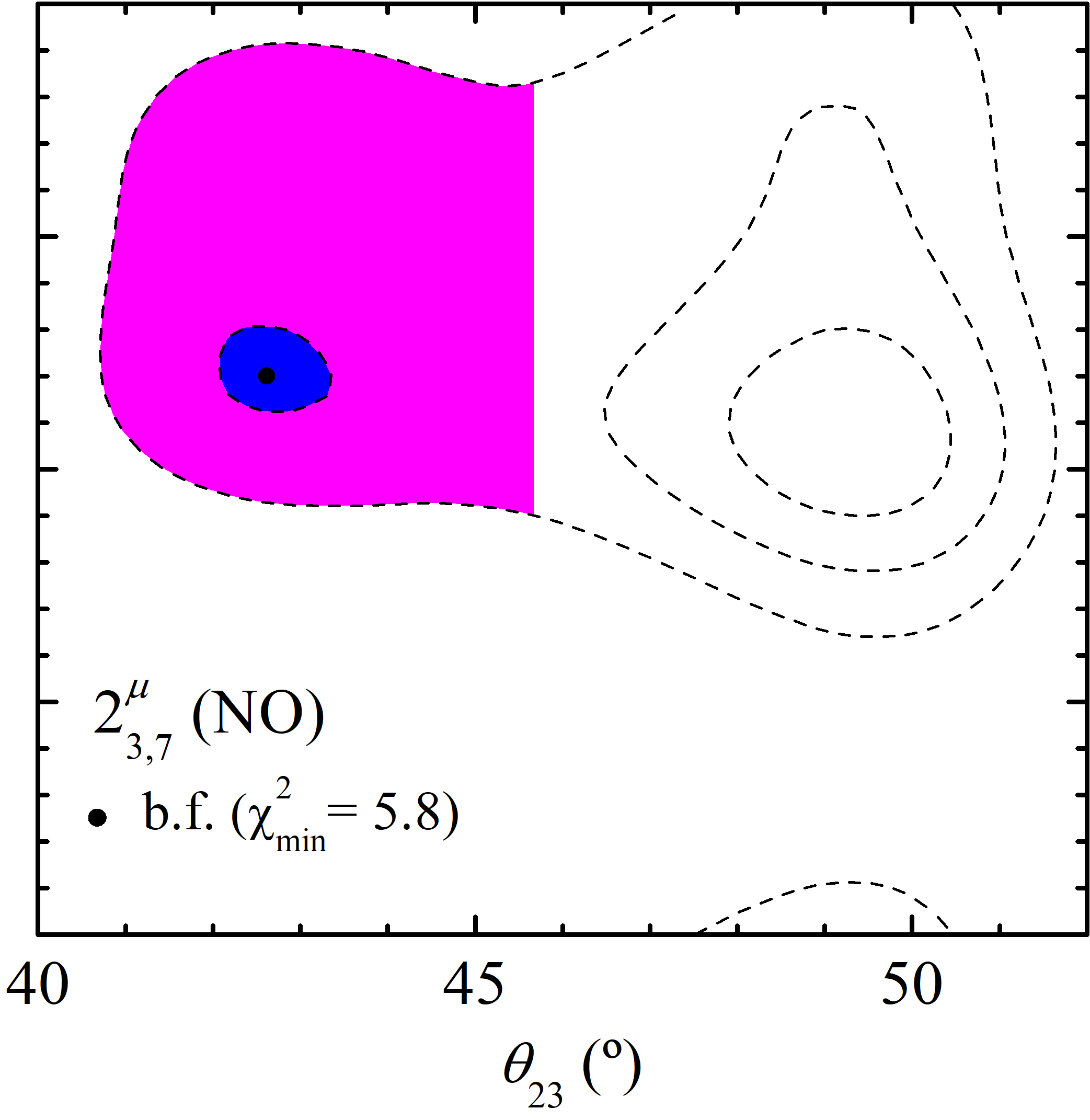}  
		\hspace{0.2cm} 
		\includegraphics[scale=0.30,trim={0cm 0cm 0cm 0cm},clip]{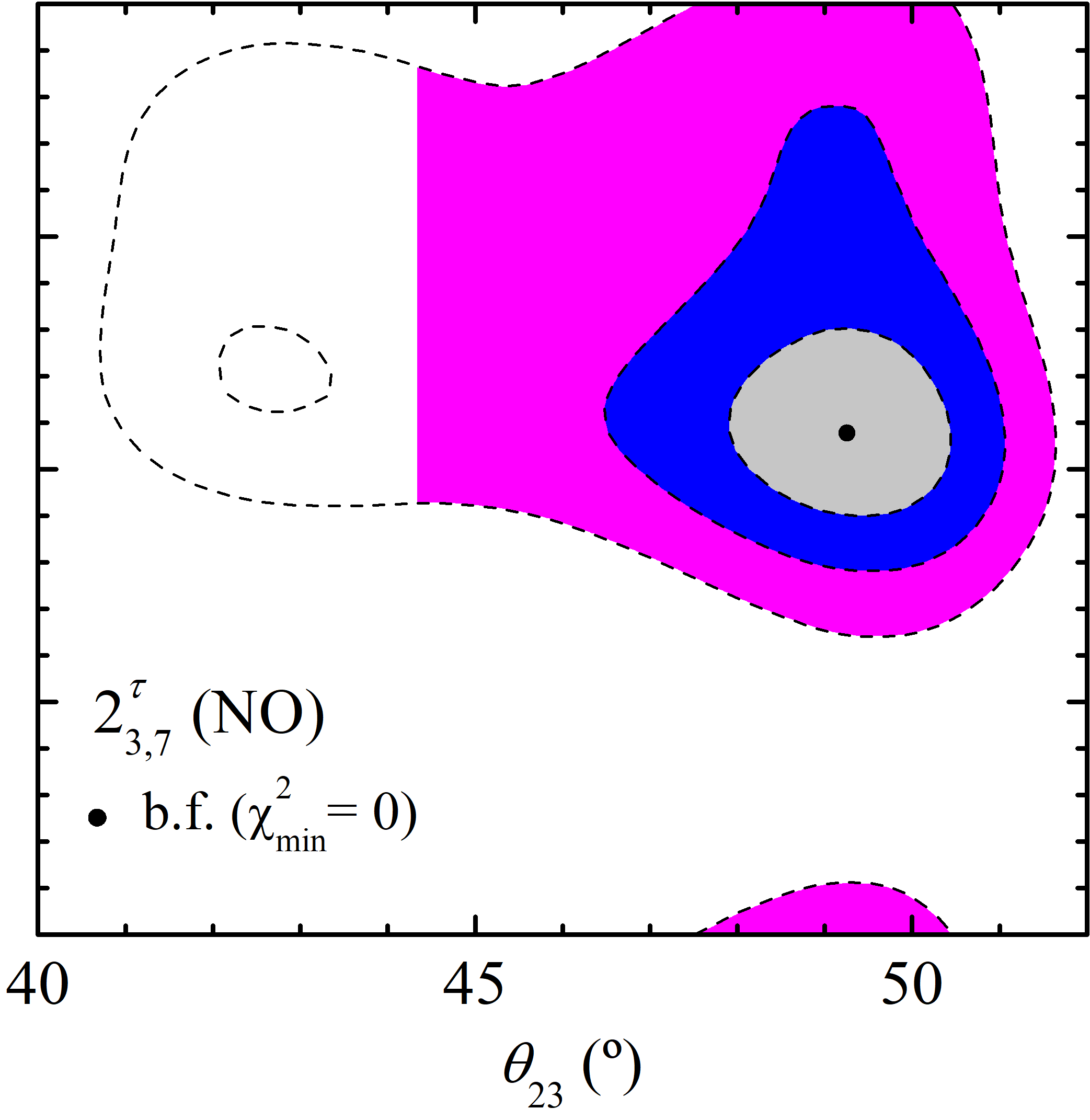} \\
		\vspace{+0.2cm} \includegraphics[scale=0.30]{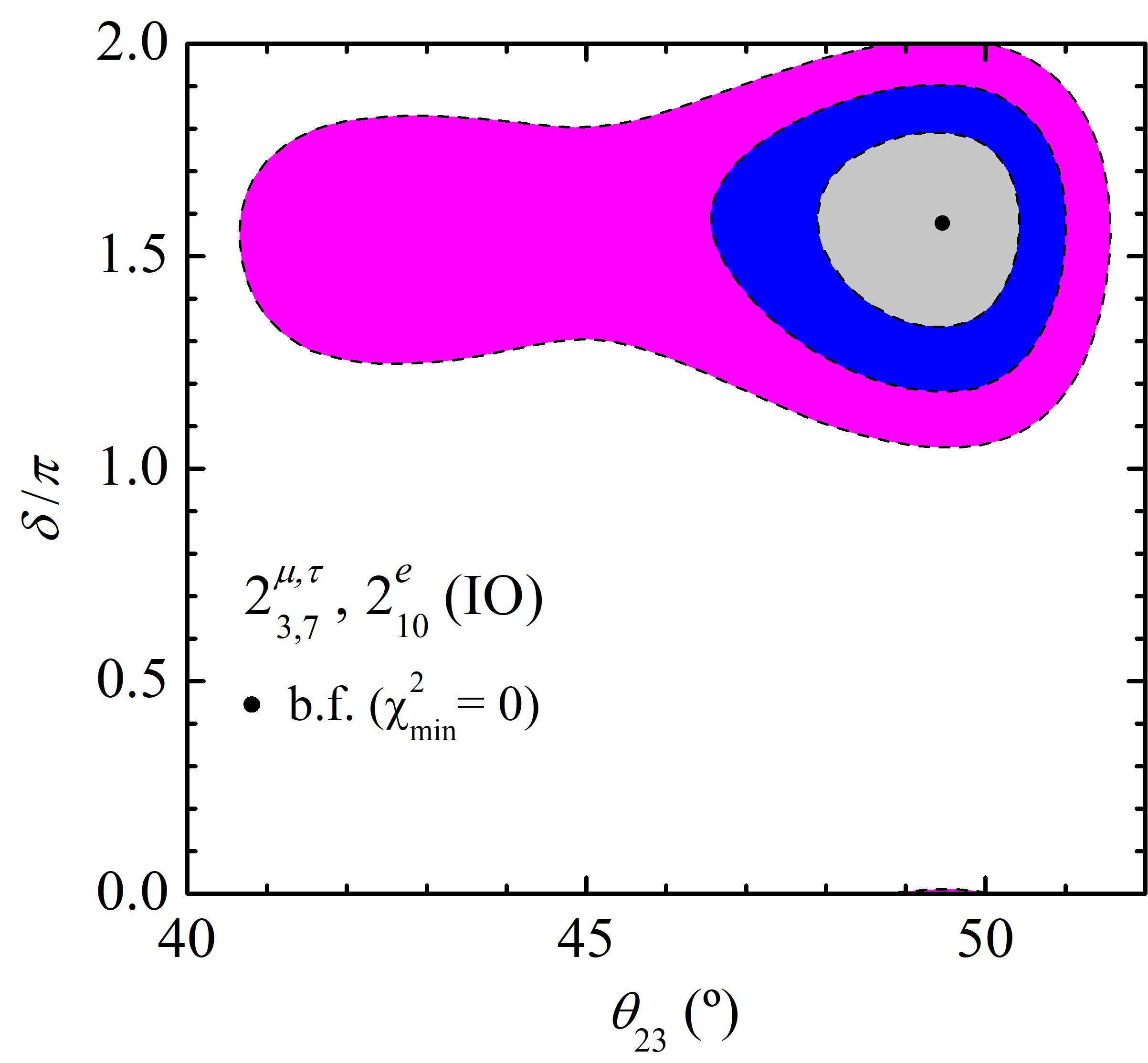} \hspace{0.2cm} 
		\includegraphics[scale=0.30,trim={0cm 0cm 0cm 0cm},clip]{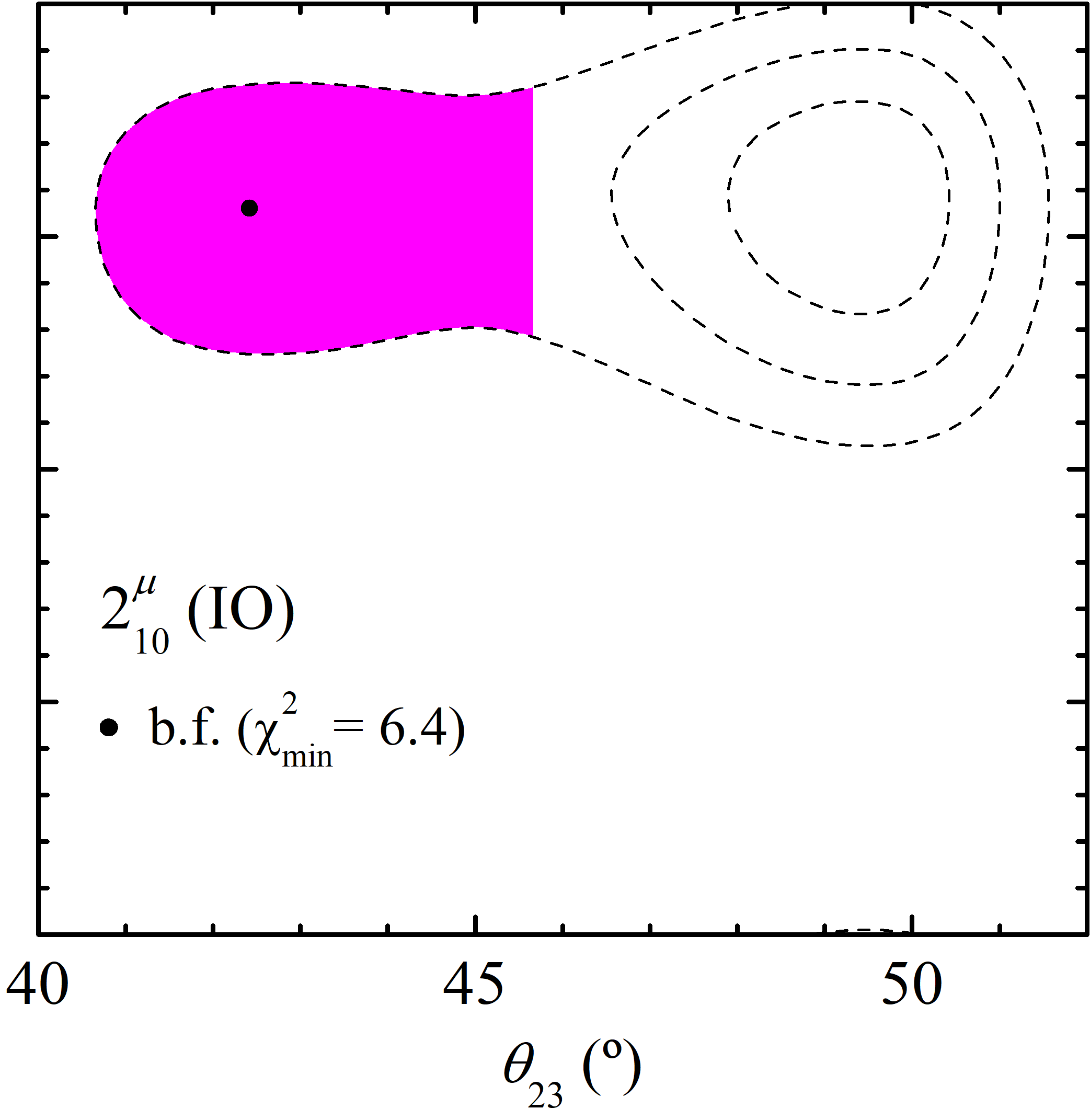}
		\hspace{0.2cm}
		\includegraphics[scale=0.30,trim={0cm 0cm 0cm 0cm},clip]{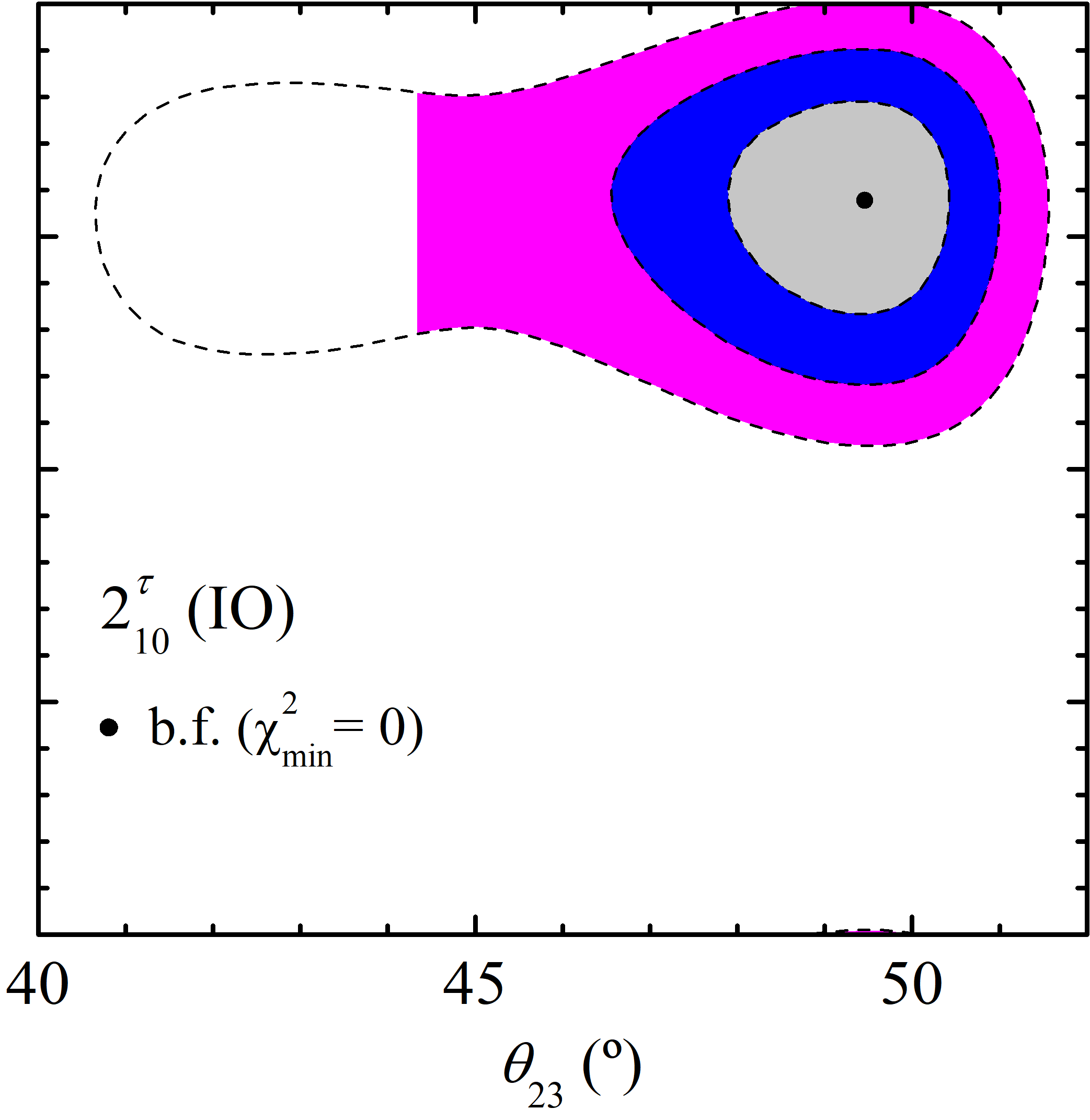}
		\caption{Allowed regions at 1$\sigma$, 2$\sigma$ and 3$\sigma$ level (in grey, blue and magenta, respectively) in the plane ($\theta_{23}$,$\delta$), for the cases $2_{3,7,10}$ discussed in the main text and for both NO (upper plots) and IO (lower plots) neutrino masses. The black dot marks the best-fit (b.f.) value for each case, while the dashed contours correspond to the $\chi^2$ contours at 1$\sigma$, 2$\sigma$ and 3$\sigma$, allowed by the global fit of neutrino oscillation data (see Table~\ref{tab:data}).}
		\label{fig:predictionst23delta}
	\end{figure}
	\begin{figure}[t!]
		\centering
		\includegraphics[scale=0.3]{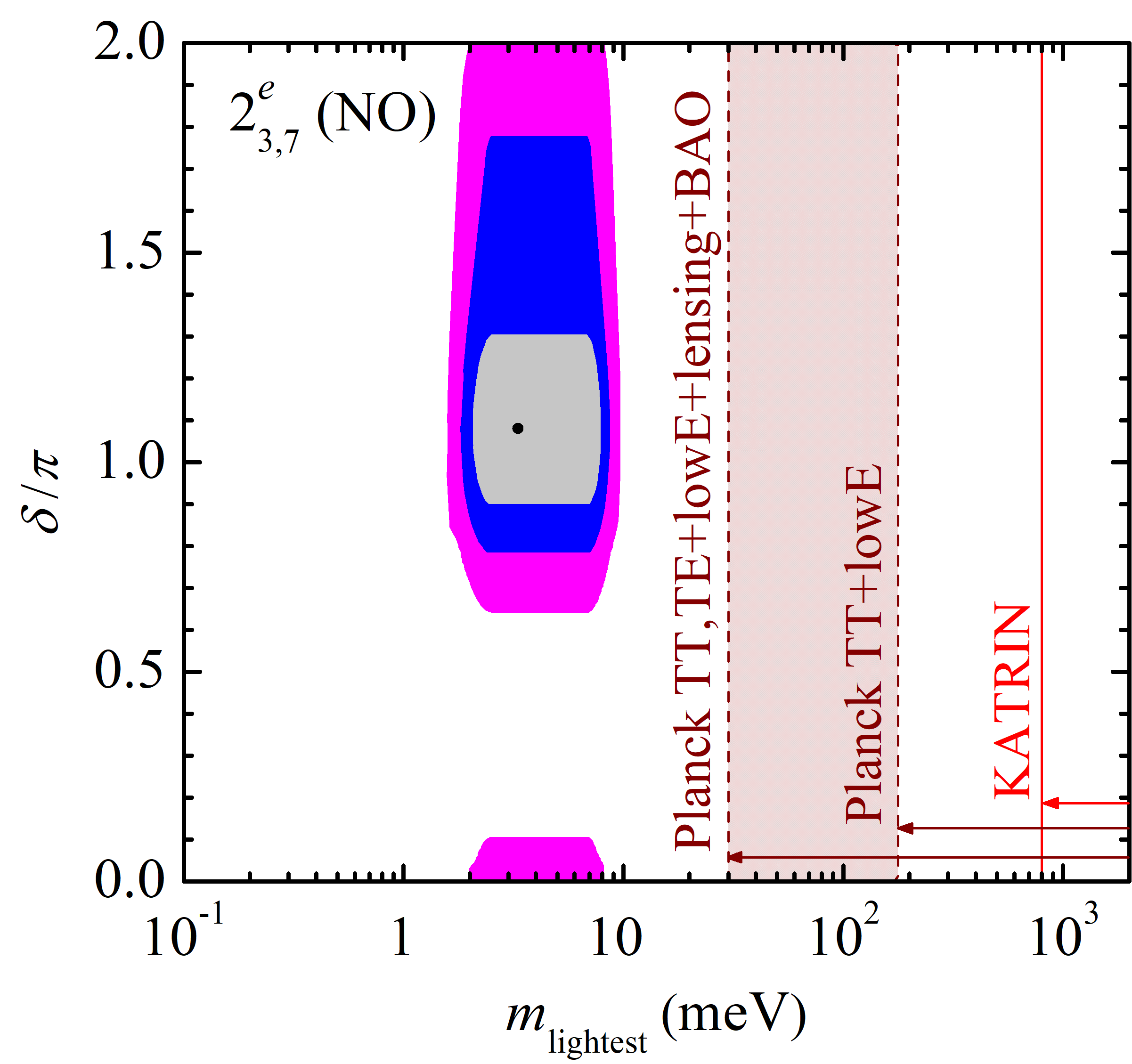}
		\includegraphics[scale=0.096]{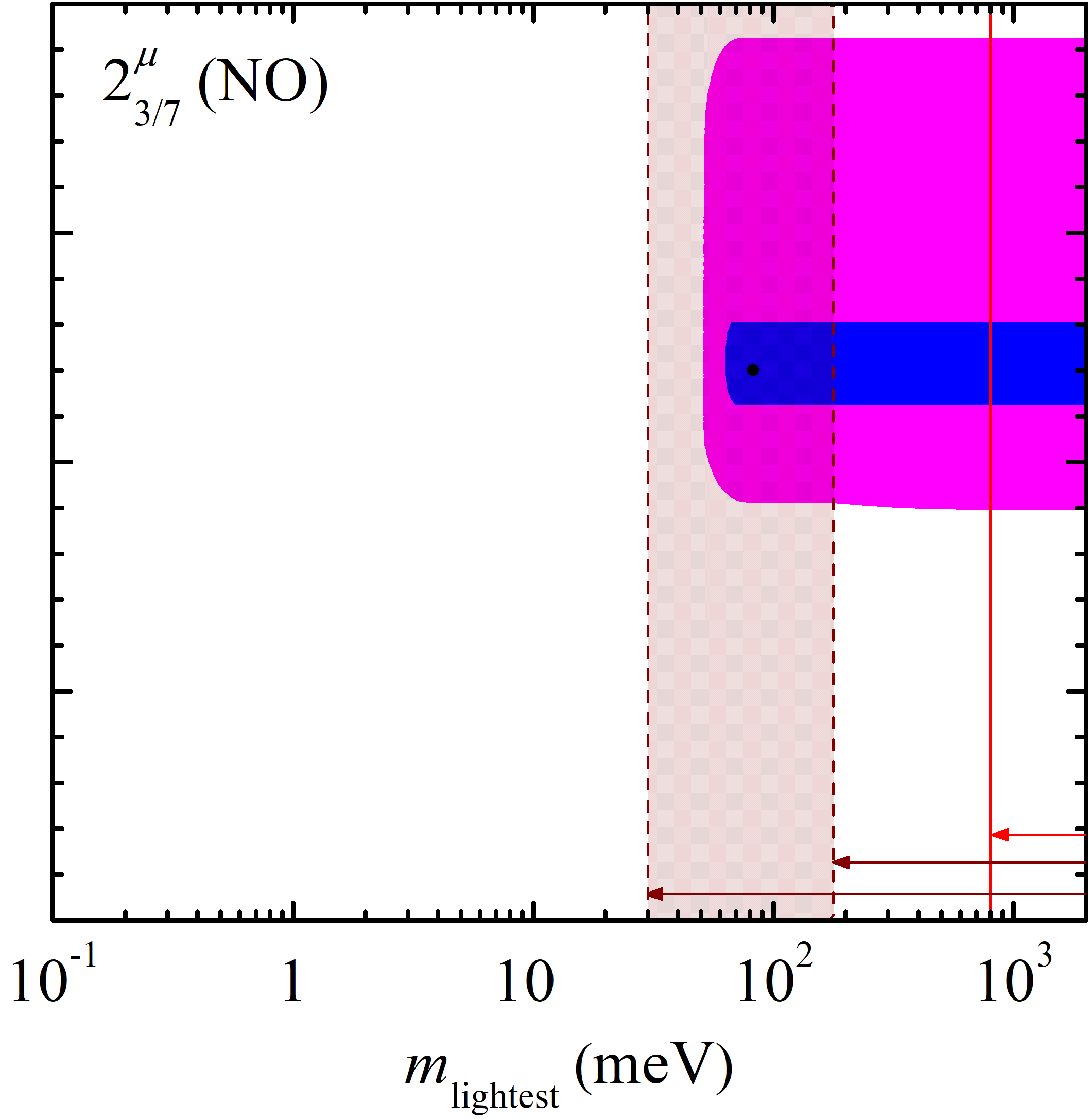}  
		\includegraphics[scale=0.30]{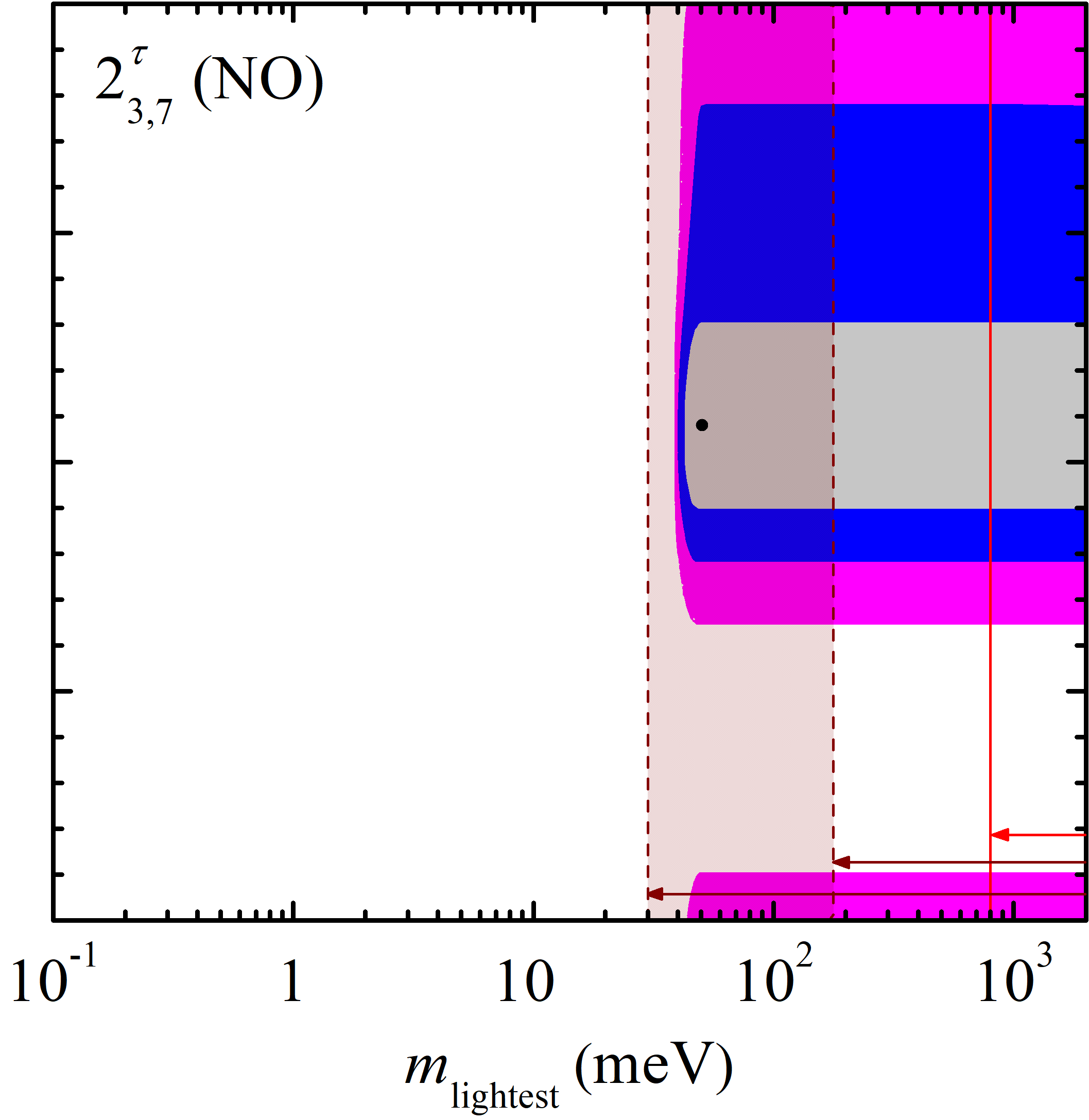} \\
		\vspace{+0.5cm}
		\includegraphics[scale=0.3]{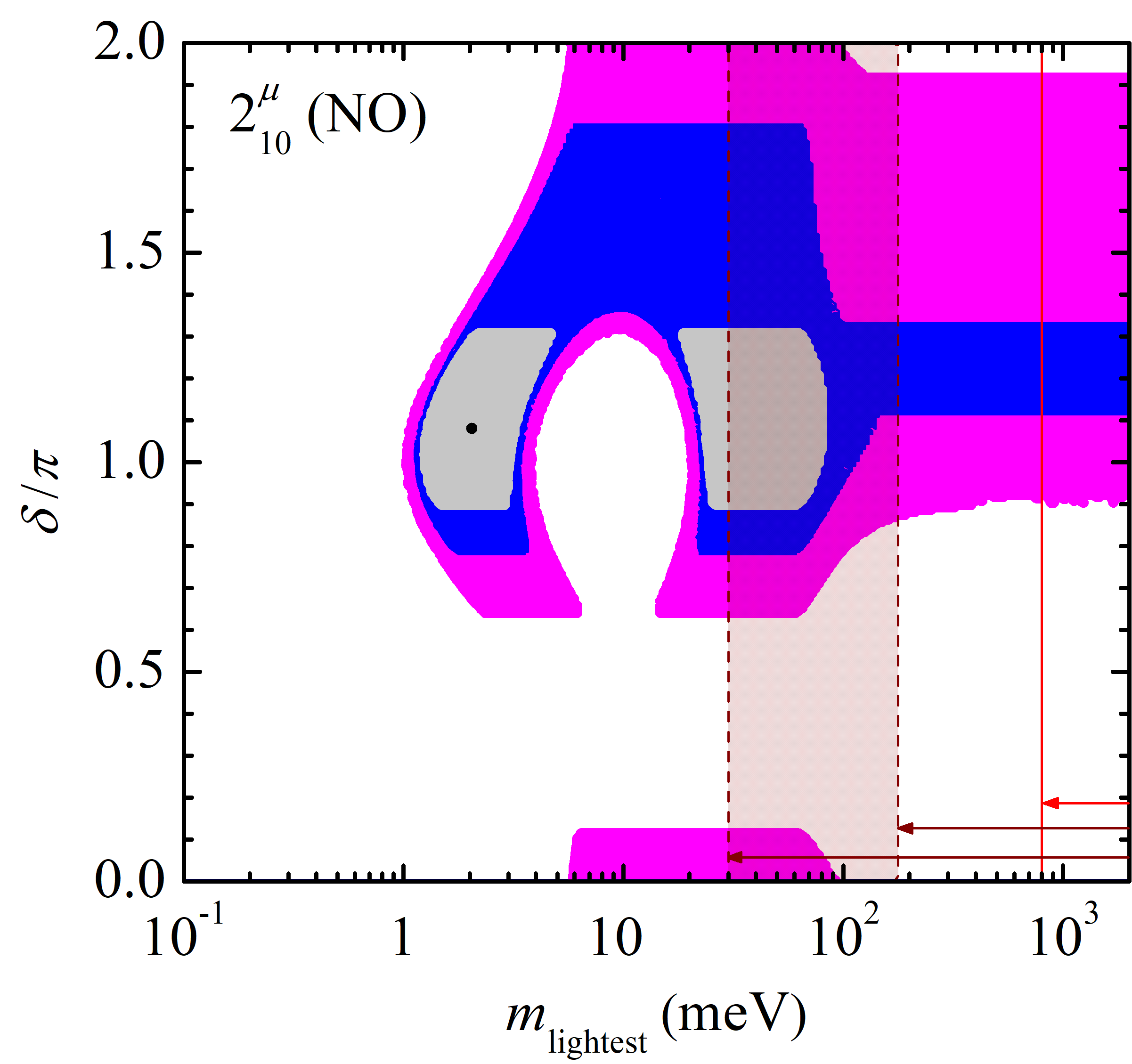} 
		\includegraphics[scale=0.096]{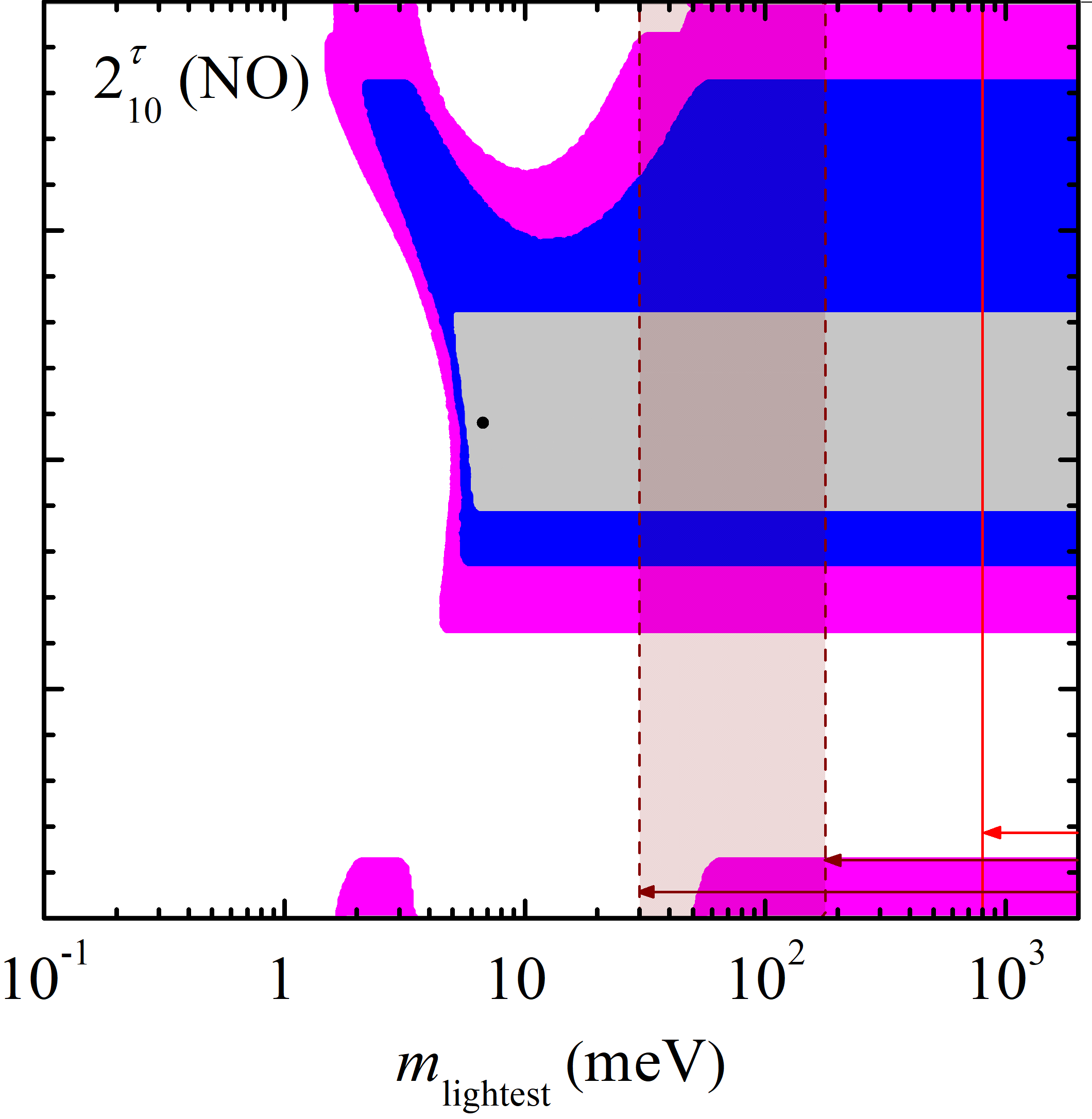}
		\caption{Allowed regions in the plane ($m_\text{lightest}$,$\delta$) for the cases $2_{3,7,10}$ discussed in the main text for NO neutrino masses. The colour scheme is the same as in Fig.~\ref{fig:predictionst23delta}. The vertical red line corresponds to the upper limit for $m_\text{lightest}$ extracted from the KATRIN bound on $m_\beta$. The brown shaded vertical region shows the $m_\text{lightest}$ upper-limit interval obtained from the Planck cosmological bounds on $\sum_k m_k$. }
		\label{fig:predictionsmlightestdeltaNO} 
	\end{figure}
	\begin{figure}[t!]
		\centering
		\includegraphics[scale=0.3]{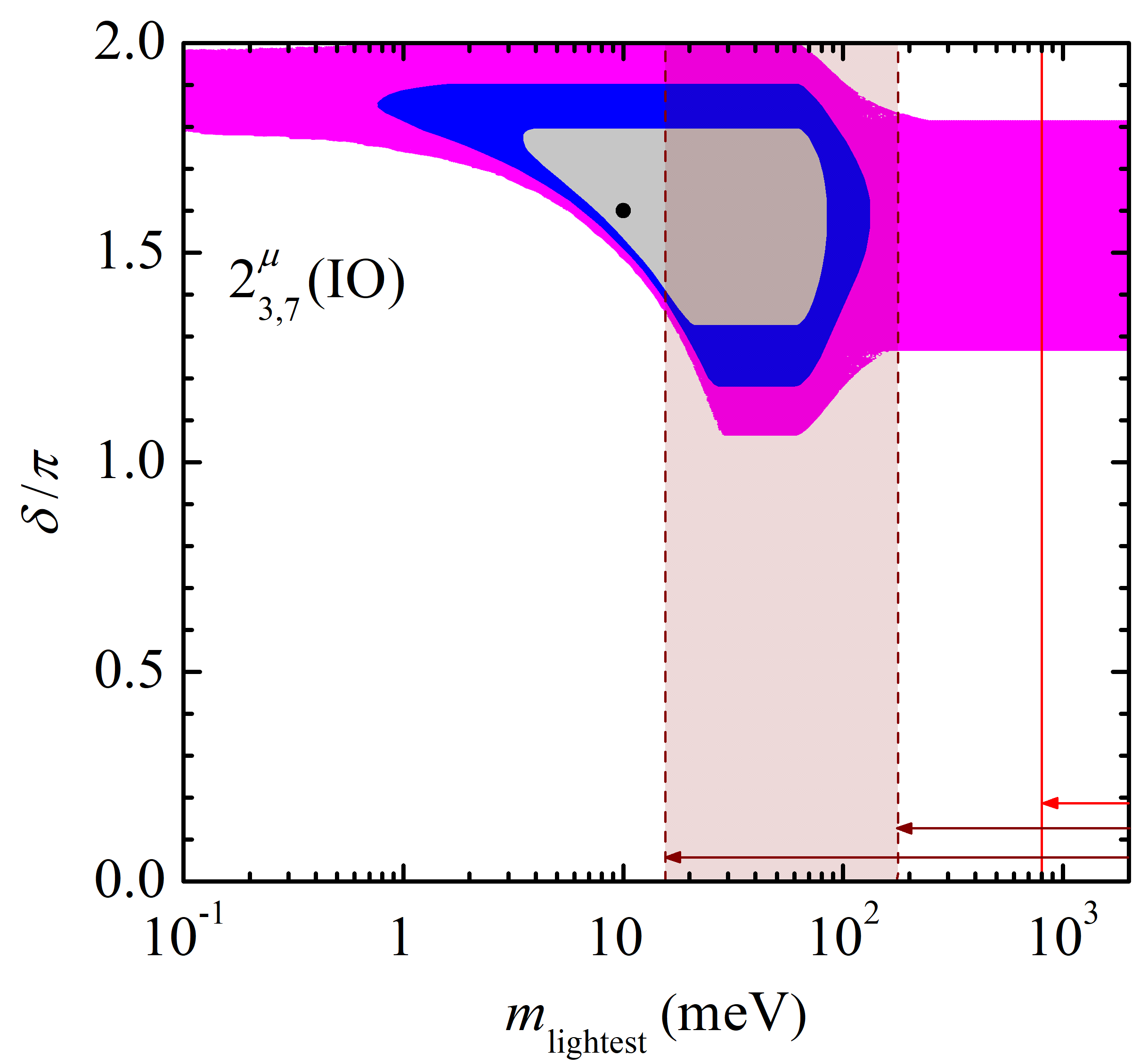}
		\includegraphics[scale=0.30,trim={0cm 0cm 0cm 0cm},clip]{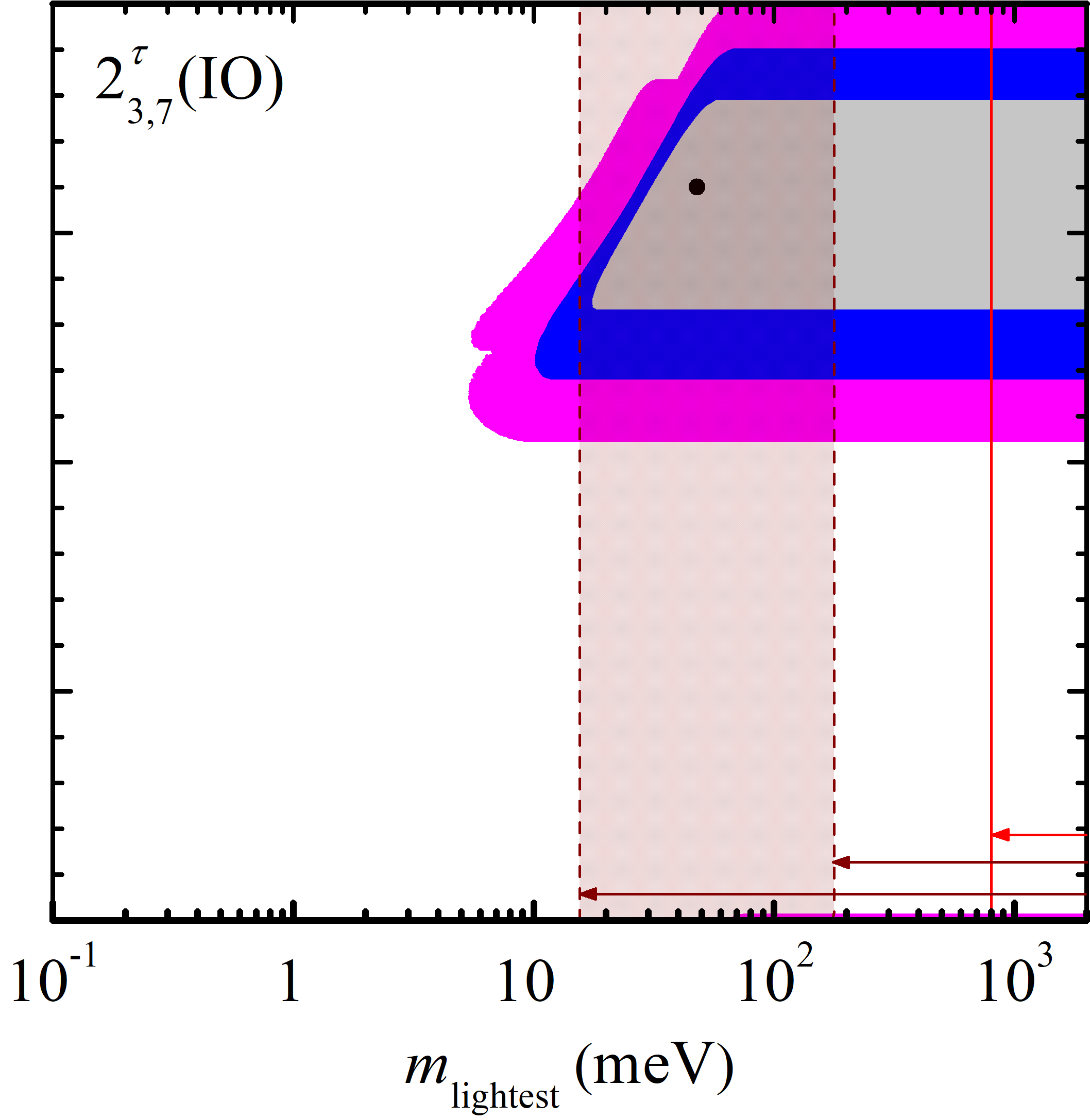} \\ 
		\vspace{+0.5cm} \includegraphics[scale=0.30,trim={0cm 0cm 0cm 0cm},clip]{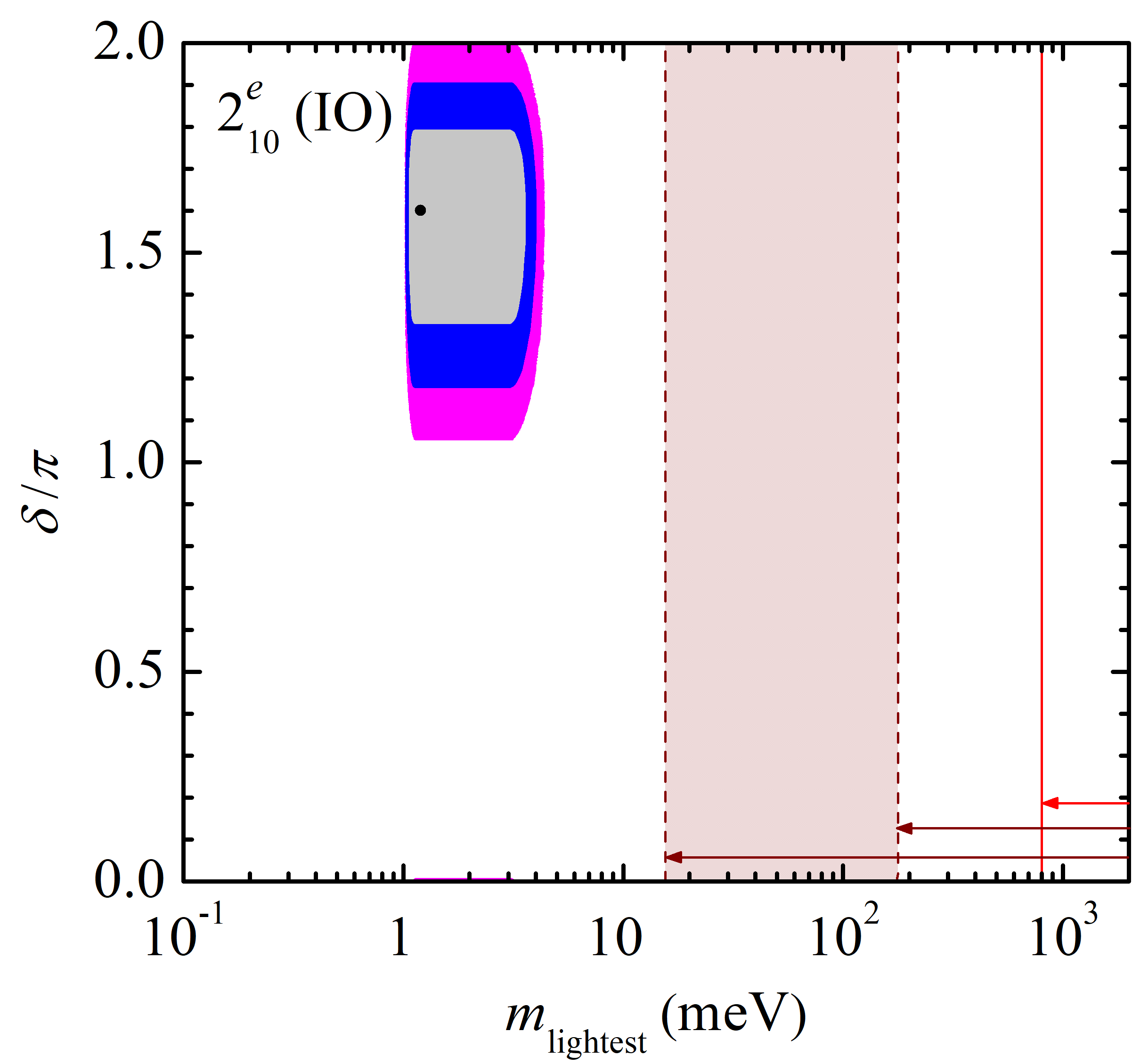} 
		\includegraphics[scale=0.30]{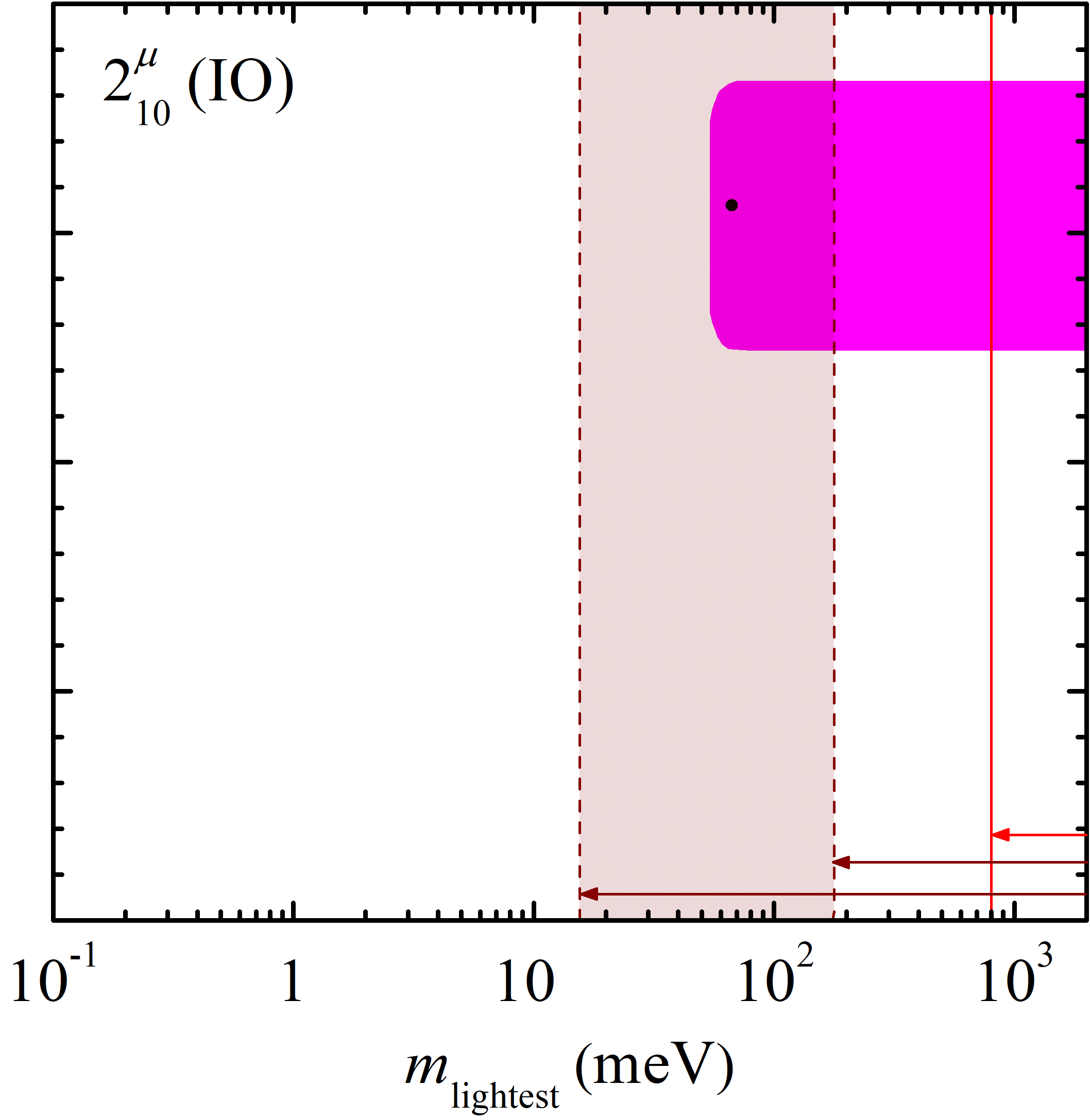} 
		\hspace{+0.1cm}\includegraphics[scale=0.30]{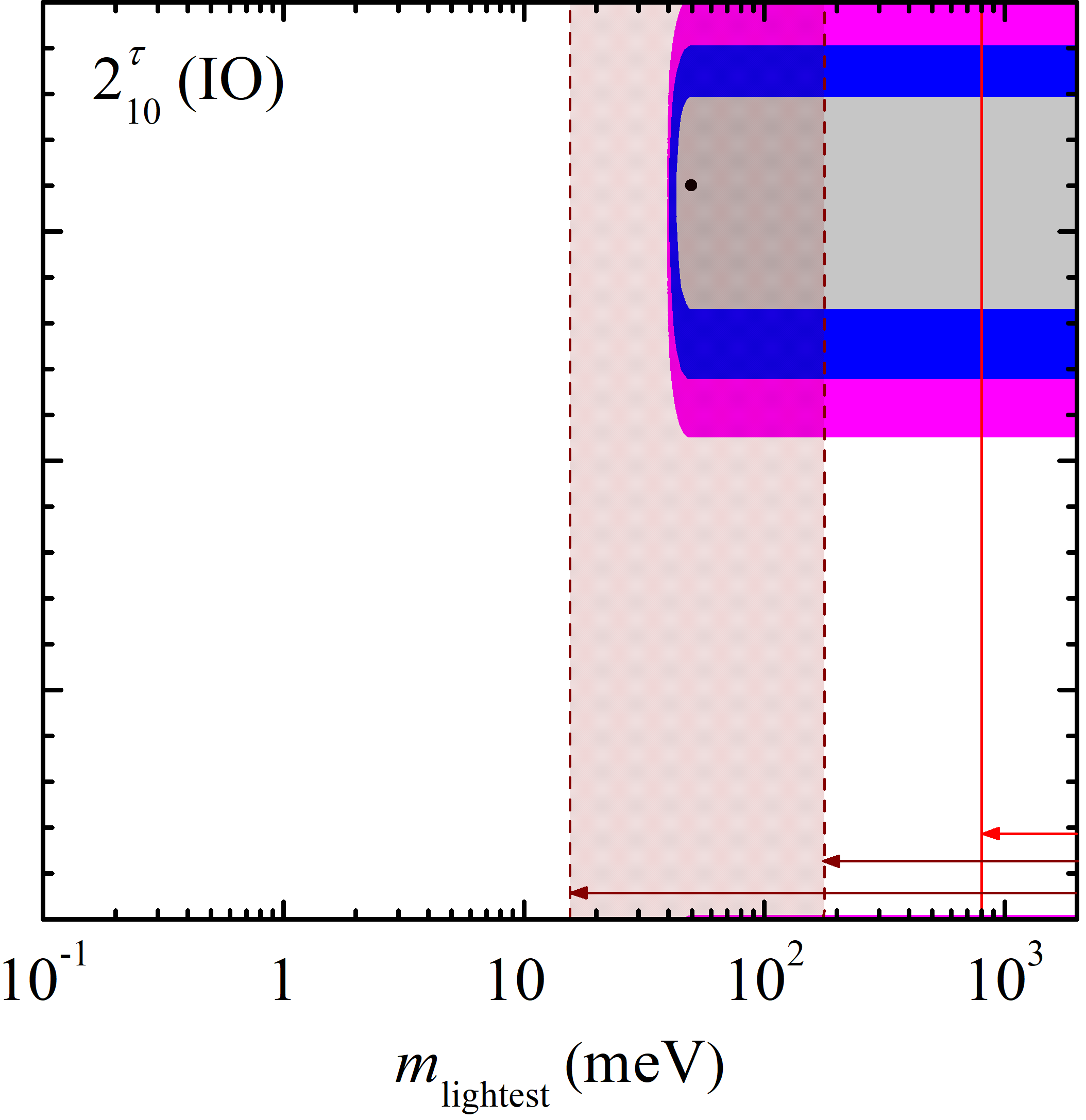}
		\caption{Allowed regions in the plane ($m_\text{lightest}$,$\delta$) for the cases $2_{3,7,10}$ discussed in the main text for IO neutrino masses. The colour scheme is the same as in Figs.~\ref{fig:predictionst23delta} and ~\ref{fig:predictionsmlightestdeltaNO}.}
		\label{fig:predictionsmlightestdeltaIO} 
	\end{figure}
	\begin{figure}[t!]
		\centering
		\includegraphics[scale=0.63]{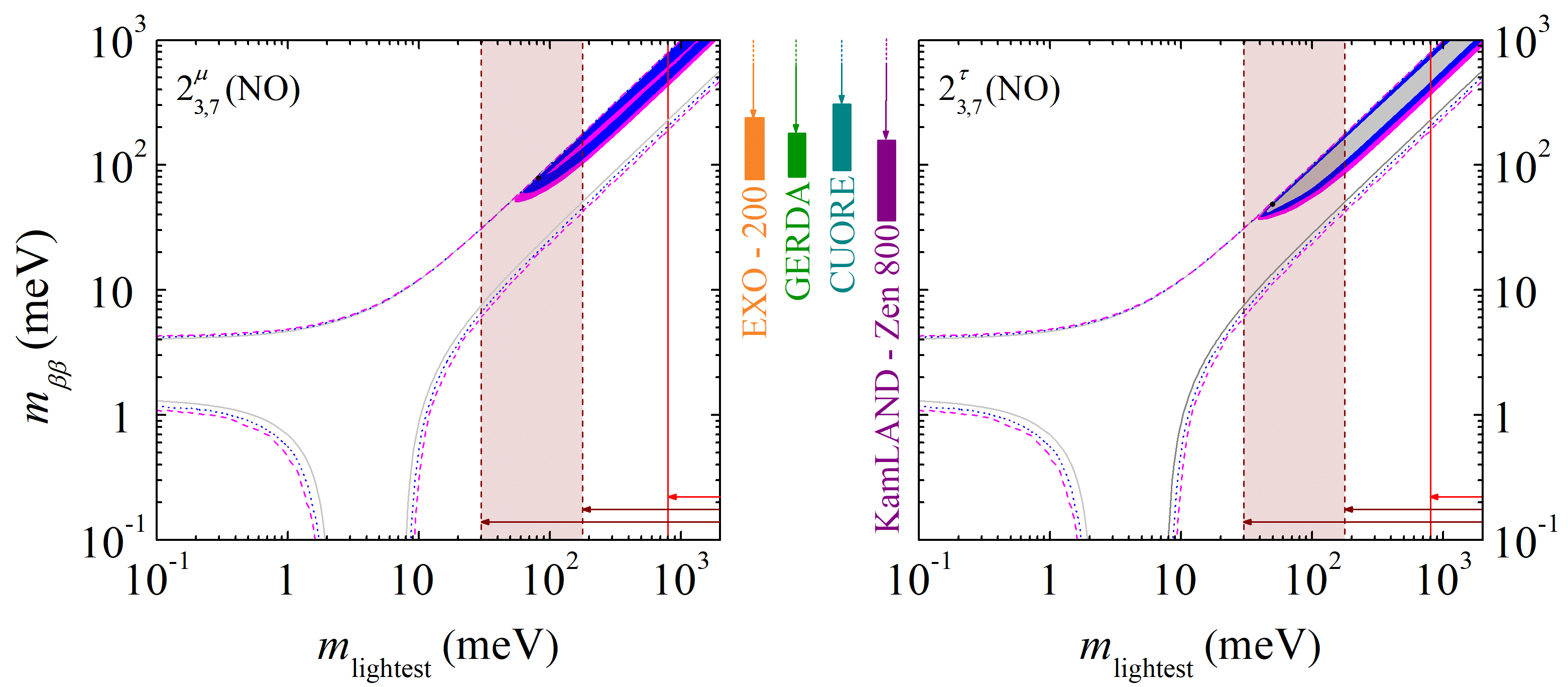} \\
		\includegraphics[scale=0.2]{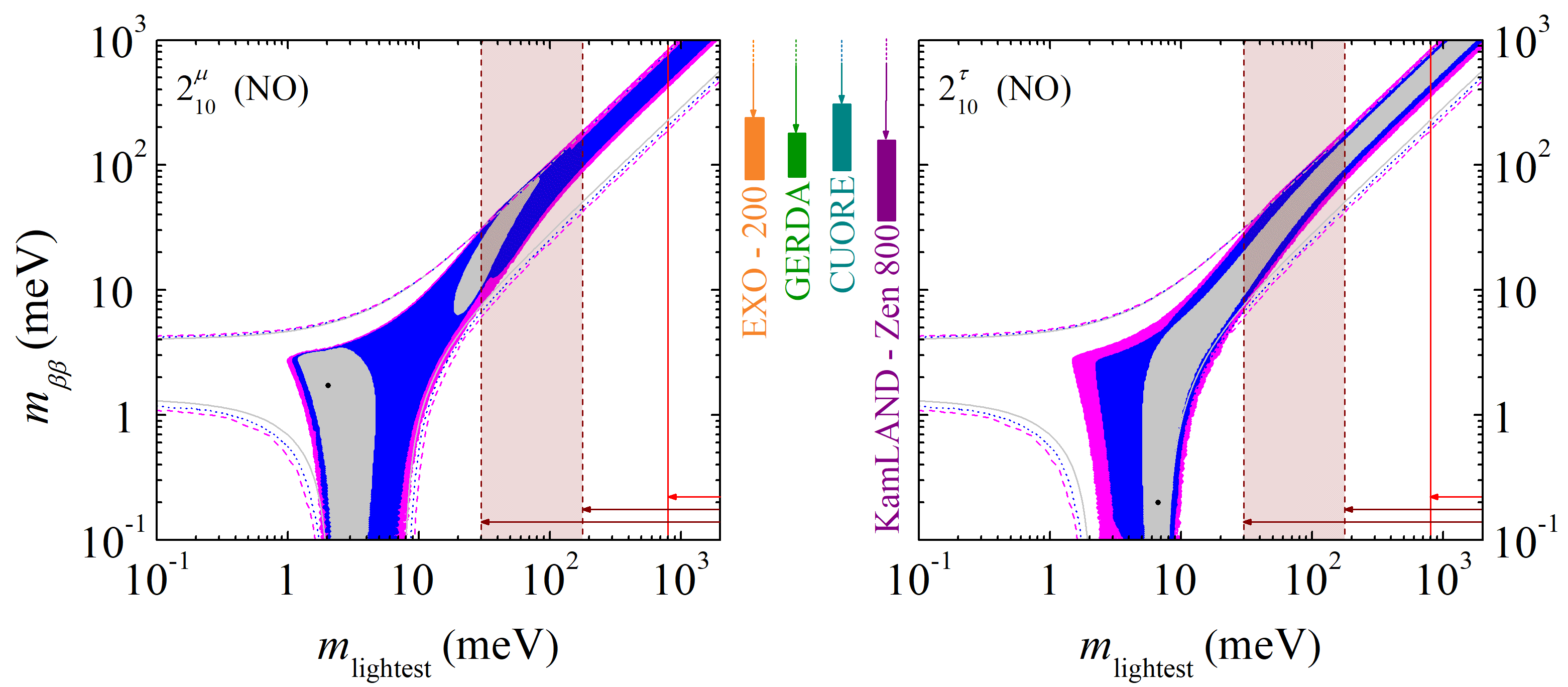} 
		\caption{Allowed regions in the ($m_\text{lightest}$,$m_{\beta\beta}$) plane for the cases $2_{3,7,10}$ discussed in the main text for NO neutrino masses. The colour scheme is the same as in Figs.~\ref{fig:predictionst23delta} and ~\ref{fig:predictionsmlightestdeltaNO}. The dashed contours delimit the 1$\sigma$, 2$\sigma$ and 3$\sigma$ regions allowed by neutrino oscillation data only, without considering any extra constraint on $\Mnu$. The coloured vertical bars correspond to the upper-bound ranges on $m_{\beta\beta}$ from EXO-200~\cite{EXO-200:2019rkq}, GERDA~\cite{GERDA:2020xhi}, CUORE~\cite{CUORE:2021mvw} and KamLAND-Zen 800~\cite{KamLAND-Zen:2022tow}.}
		\label{fig:predictionsmlightestmbetabetaNO} 
	\end{figure}
	\begin{figure}[t!]
		\centering
		\includegraphics[scale=0.65]{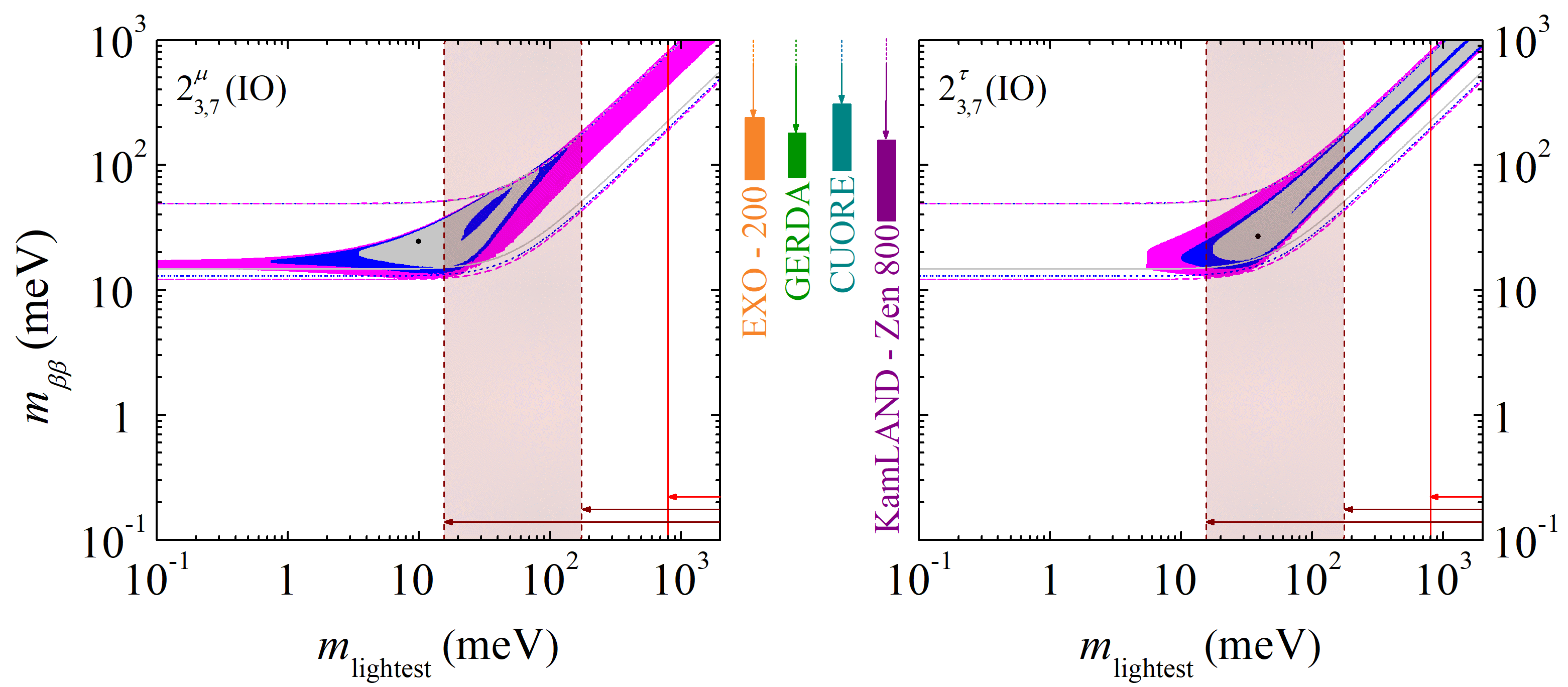} \\
		\includegraphics[scale=0.65]{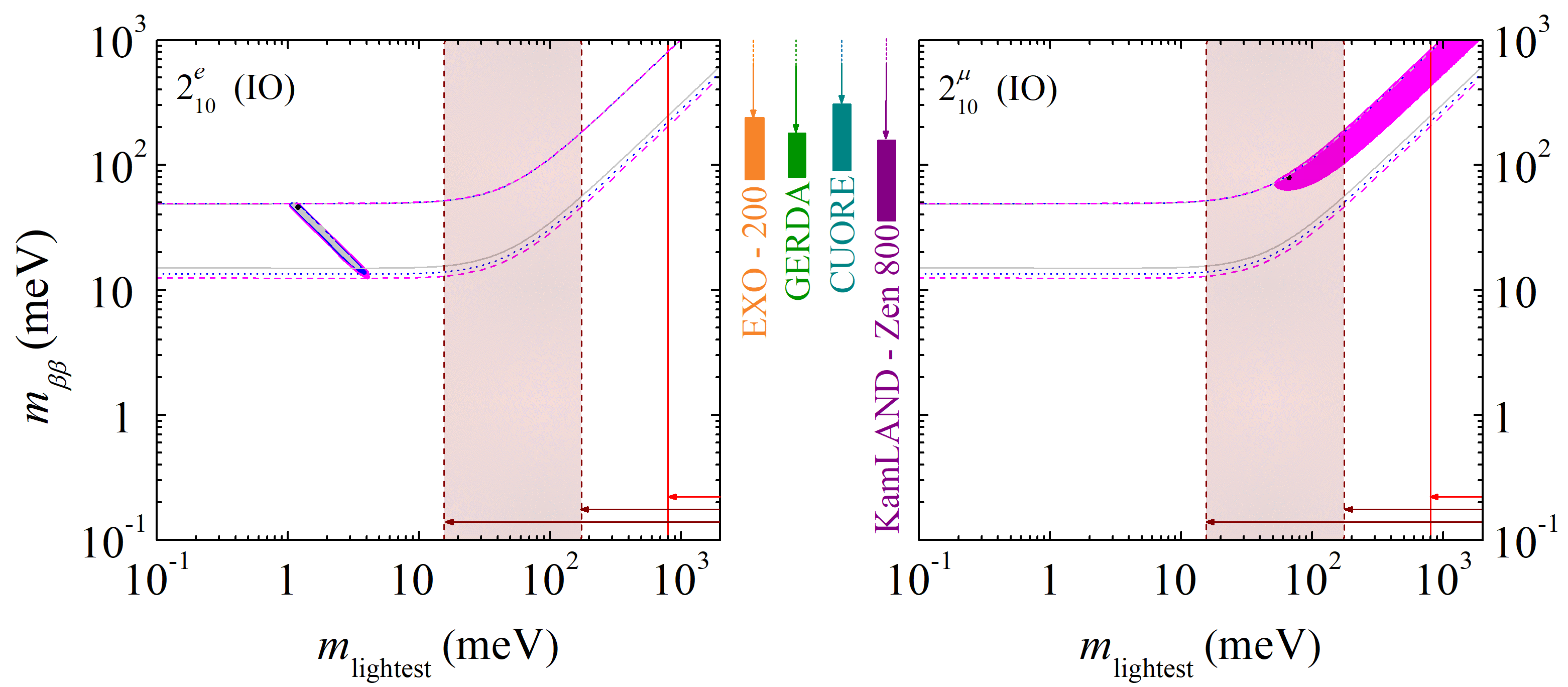}  \\
		\includegraphics[scale=0.65]{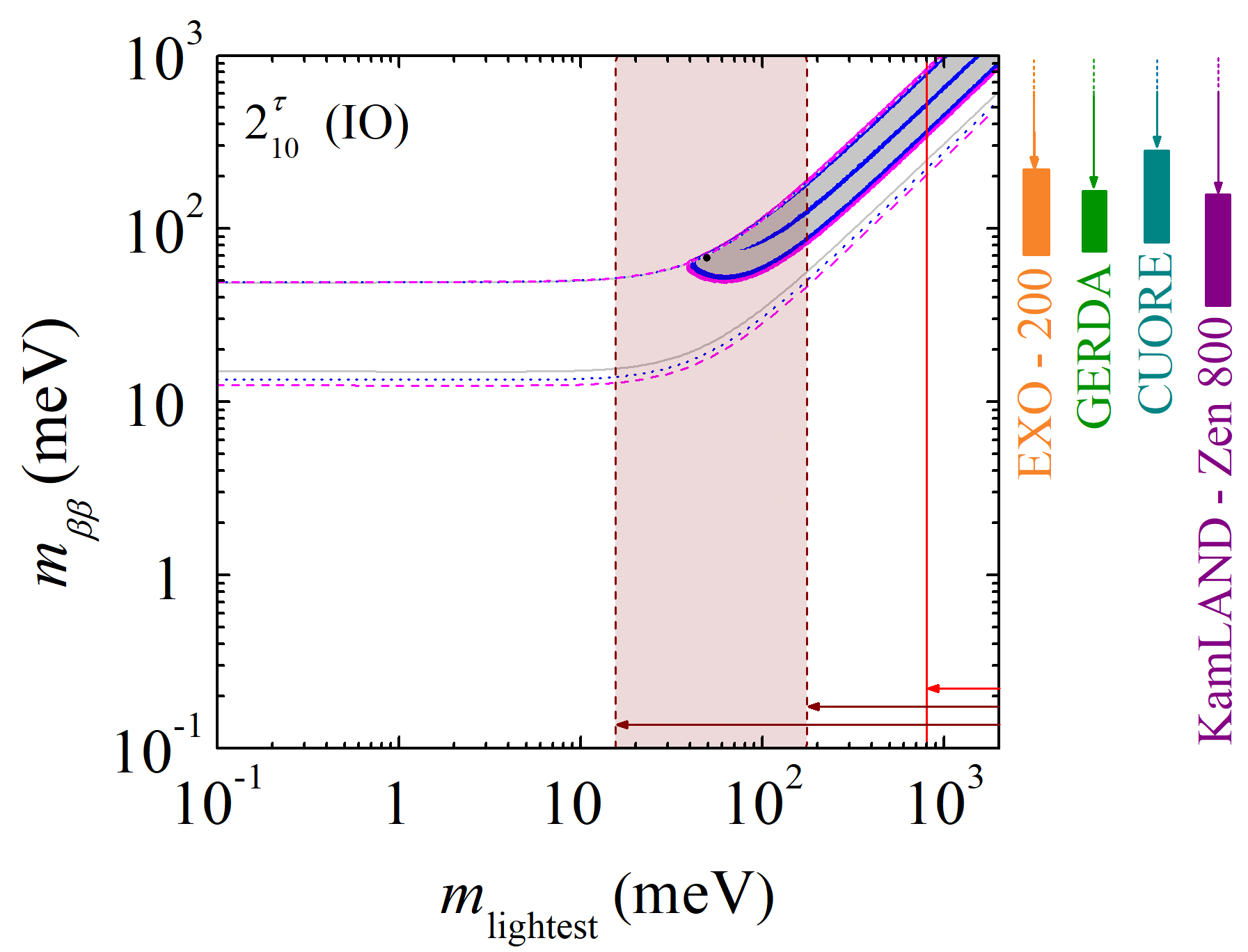} 
		\caption{Allowed regions in the ($m_\text{lightest}$,$m_{\beta\beta}$) plane for the cases $2_{3,7,10}$ discussed in the main text for IO neutrino masses. The colour scheme is the same as in Figs.~\ref{fig:predictionst23delta}, ~\ref{fig:predictionsmlightestdeltaNO}, ~\ref{fig:predictionsmlightestdeltaIO} and~\ref{fig:predictionsmlightestmbetabetaNO}.}
		\label{fig:predictionsmlightestmbetabetaIO}
	\end{figure}
	The independent parameters in the lepton sector: $a_{1,3,4},\theta_L$ for charged leptons and $b_{1,2,3,4}, \xi_{1,2}$ for neutrinos, amount to a total number of ten, which should be compared to twelve lepton sector observables: $m_{e,\mu,\tau}$, $m_{1,2,3}$, $\theta_{12,23,13}^\ell$ and $\delta^\ell,\alpha_{21,31}$ [see Eqs.~\eqref{eq:leptonmassdiag} and~\eqref{eq:UPMNSparam}, and Table~\ref{tab:data}]~\footnote{For simplicity, in this section we omit the superscript $\ell$ used for the neutrino oscillation parameters.}. Thus, the identified minimal Abelian flavour symmetries impose mass matrix patterns that lead to predictive scenarios for the lepton sector. To analyse their predictions we need to match the above effective neutrino mass matrices to the one defined in terms of the physical low-energy parameters according to Eqs.~\eqref{eq:leptonmassdiag}-\eqref{eq:UPMNSparam}, namely,
	\begin{align}
	\Mnuh=\U^\ast\, \text{diag}(m_1,m_2,m_3)\,\U^\dagger,
	\label{eq:mnureconstruction}
	\end{align} 
	being $m_i$ the neutrino masses and $\U$ the lepton mixing matrix of Eq.~\eqref{eq:UPMNSparam}. The matching between $\mathbf{M}_{\nu}$ and $\Mnuh$ is done by performing proper rephasings of the lepton fields and imposing the texture-zero conditions stemming from Eqs.~\eqref{eq:Meff23},~\eqref{eq:Meff27} and~\eqref{eq:Meff210}. This allows us to check whether those conditions are compatible with current neutrino data and possibly select regions of the physical parameter space preferred by the model. The conditions are
	\begin{align}
	2_{3,7}^e & : \ (\Mnuh)_{11} = 0 , \ 2_{3,7}^\mu  : \ (\Mnuh)_{22} = 0 , \ 2_{3,7}^\tau : \ (\Mnuh)_{33} = 0 \; ; \label{eq:cond2327} \\
	2_{10}^e  & : \ (\Mnuh)_{22} (\Mnuh)_{33} - (\Mnuh)_{23}^2 = 0 , \ 2_{10}^\mu  : \ (\Mnuh)_{11} (\Mnuh)_{33} - (\Mnuh)_{13}^2 = 0 , \ 2_{10}^\tau  : \ (\Mnuh)_{11} (\Mnuh)_{22} - (\Mnuh)_{12}^2 = 0 \;.\label{eq:cond210}
	\end{align}
	The above conditions of $\Mnu$ are tested by performing a $\chi^2$-analysis based on the minimisation of the $\chi^2$ function defined in Eq.~\eqref{eq:chi2}. As mentioned before, for the neutrino oscillation parameters this function is computed using the one-dimensional profiles $\chi^2(\sin^2\theta_{ij})$ and $\chi^2(\Delta m^2_{ij})$, and the 2D distribution $\chi^2(\delta,\sin^2\theta_{23})$ for $\delta$ and $\theta_{23}$ given in Refs.~\cite{deSalas:2020pgw}. The constraints on $\Mnu$ are incorporated using the Lagrange multiplier method~\cite{Alcaide:2018vni}. For a fixed pair of parameters, the 2D 1$\sigma$, 2$\sigma$ and 3$\sigma$ compatibility regions were obtained by requiring $\chi^2\leq 2.30$, $6.18$ and $11.83$, respectively. We remark that the compatibility with neutrino oscillation data of the neutrino mass matrices for the $2_{3,7}^{e,\mu,\tau}$ cases, leading to the conditions of Eq.~\eqref{eq:cond2327}, has been previously analysed in Ref.~\cite{Barreiros:2020gxu}. In this work, we update the analysis in light of the latest global fits of neutrino data and current bounds on the absolute neutrino mass scale and $0\nu\beta\beta$ decay.
	
	Our results are presented in Figs.~\ref{fig:predictionst23delta}--\ref{fig:predictionsmlightestmbetabetaIO}. We show the 1$\sigma$, 2$\sigma$ and 3$\sigma$ allowed regions (in grey, blue and magenta, respectively) in the planes ($\theta_{23}$,$\delta$), ($m_\text{lightest}$,$\delta$) and ($m_\text{lightest}$,$m_{\beta\beta}$) for the compatible cases $2_{3,7,10}^{e,\mu,\tau}$, for both NO and IO neutrino masses. In Fig.~\ref{fig:predictionst23delta}, the 1$\sigma$, 2$\sigma$ and 3$\sigma$ contours of the regions allowed by neutrino oscillation data~\cite{deSalas:2020pgw} (dashed curves) together with the best-fit (b.f.) point for each case and its corresponding minimum value of $\chi^2$ are shown. Note that the cases $2_{10}^{e}$ for NO and $2_{3,7}^{e}$ for IO are not compatible and therefore are not shown in the figures. For all the remaining cases the $\chi^2_\text{min}$ that best fits the data is $\chi^2_\text{min}=0$ for all cases, except for $2_{3,7}^{\mu}$ for NO and $2_{10}^{\mu}$ for IO which have $\chi^2_\text{min}=5.8$ and $\chi^2_\text{min}=6.4$, being therefore compatible with data only at the 2$\sigma$ and 3$\sigma$ level, respectively.~\footnote{We have not included in our $\chi^2$ calculation the $2.5\sigma$ data preference for NO (see Ref.~\cite{deSalas:2020pgw}).} Below we discuss the results in detail:
	
	\begin{itemize}
		
		\item \textbf{Dirac CP phase in terms of atmospheric mixing angle:} From the plots in Fig.~\ref{fig:predictionst23delta}, we conclude that the cases $2_{3,7}^{e},2_{10}^{\mu,\tau}$ for NO (upper left plot) and $2_{3,7}^{\mu,\tau},2_{10}^{e}$ for IO (lower left plot) do not lead to any constraints on $\theta_{23}$ and $\delta$ plane. However, for NO, $2_{3,7}^{\mu}$ and $2_{3,7}^{\tau}$ (upper middle and right plots) select the first and second octant for the atmospheric $\theta_{23}$ mixing angle, respectively. Thus, $2_{3,7}^{\mu}$ for NO is only compatible at the 2$\sigma$ level. For IO, $2_{10}^{\mu}$ and $2_{10}^{\tau}$ textures (lower middle and right plots) select the first and second octant for $\theta_{23}$, respectively. Thus, $2_{10}^{\mu}$ is only compatible at the 3$\sigma$ level for IO. The aforementioned $\theta_{23}$ octant preference of these textures will be tested by improving the sensitivity on $\delta$ and $\theta_{23}$ with upcoming experiments like DUNE~\cite{DUNE:2015lol} or T2HK~\cite{Hyper-KamiokandeWorkingGroup:2014czz}.
		
		\item \textbf{Dirac CP phase versus lightest neutrino mass:} For NO and IO, two of the three neutrino masses may be expressed in terms of the lightest neutrino mass $m_{\text{lightest}}$ (corresponding to $m_1$ and $m_3$ for NO and IO, respectively) and the measured neutrino mass-squared differences $\dmsol=m_2^2-m_1^2$ and $\dmatm=m_3^2-m_1^2$ as
		\begin{align}
		\text{NO:}&\quad m_2=\sqrt{m_1^2+\dmsol},\quad m_3=\sqrt{m_1^2+\dmatm}\;,\\
		\text{IO:} &\quad m_1=\sqrt{m_3^2+|\dmatm|},\quad m_2=\sqrt{m_3^2+\dmsol+|\dmatm|}\;.
		\end{align}

		The constraints of our model on the lightest neutrino mass are presented in Figs.~\ref{fig:predictionsmlightestdeltaNO} and~\ref{fig:predictionsmlightestdeltaIO}, for NO and IO neutrino masses, respectively. We have not included the $m_\text{lightest}$ bounds from cosmology or $\beta$-decay experiments in our $\chi^2$ analysis; instead, we explicitly present in the figures the upper limit on $m_\text{lightest}$ obtained from the KATRIN experiment, being the current bound $m_\beta<0.8$~eV (at 90\% CL)~\cite{KATRIN:2021uub} (red vertical line), and the upper-bound range on $m_\text{lightest}$ calculated from Planck limits on the sum of neutrino masses (brown shaded vertical band). The left (right) vertical dashed brown line refers to the less (most) conservative 95\% CL bound, corresponding to $\sum_k m_k<0.12$~eV ($0.54$~eV), where Planck TT,TE+lowE+lensing+BAO (Planck TT+lowE) data was used~\cite{Planck:2018vyg}. Focusing on Fig.~\ref{fig:predictionsmlightestdeltaNO} for NO, one can see that all models predict a 3$\sigma$ lower limit for $m_\text{lightest}$, while for $2_{3,7}^{e}$ (upper left plot) an upper bound is also established where $m_\text{lightest}$ is constrained to the range~$\sim [1.5,9]$~meV (3$\sigma$), well below the current limits from cosmology and KATRIN. The remaining $2_{3,7}^{\mu,\tau}$ (upper middle and right plots) cases establish a lower limit on $m_\text{lightest} \sim 40$~meV (3$\sigma$) which is now being probed by cosmological observations, and lying above (below) the less (most) conservative Planck bound. Regarding $2_{10}^{\mu}$ and $2_{10}^{\tau}$ (bottom left and right plots), these cases establish a lower limit on $m_\text{lightest} \sim 1$~meV and $\sim 1.8$~meV (3$\sigma$), respectively, well below the current limits from Planck and KATRIN. Turning our attention to Fig.~\ref{fig:predictionsmlightestdeltaIO} for IO, we remark that all models predict a 3$\sigma$ lower limit for $m_\text{lightest}$ except $2_{3,7}^{\mu}$ (upper left plot) which allows for a vanishing $m_{\text{lightest}}$. However, for $2_{10}^{e}$ (lower left plot), an upper bound is also established where $m_\text{lightest}$ is constrained at 3$\sigma$ in the interval $\sim [1,5]$~meV, which is below current limits from Planck and KATRIN. The remaining $2_{10}^{\mu,\tau}$ (lower middle and right plots) cases establish a lower limit on $m_\text{lightest}$ around $\sim 50$~meV (3$\sigma$), which is now being probed by cosmological observations, lying within the brown shaded region. As for the $2_{3,7}^{\tau}$ textures (upper right plot), the lower limit $m_\text{lightest}\sim 5$~meV at 3$\sigma$ is established, below the Planck lower limit, in a very small region of parameter space. Overall, from these figures it is clear that for NO the $2_{3,7}$ models are more constraining than the $2_{10}$ models while for IO the opposite is true.
		
		\newpage
		
		\item \textbf{Effective Majorana mass in terms of the lightest neutrino mass:} We now analyse the constraints on $0\nu\beta\beta$ decay rate by presenting the allowed regions for the effective neutrino mass parameter $m_{\beta\beta}$. This parameter is of utmost importance since it can be (in principle) extracted from $0\nu\beta\beta$ experimental data, which could elucidate on the Majorana character of neutrinos~\cite{Schechter:1981bd}. Using the parameterisation given in Eq.~\eqref{eq:UPMNSparam}, $m_{\beta\beta}$ reads, for NO and IO,
		\begin{align}
		{\rm NO}: \;m_{\beta\beta}&=\left| c_{12}^2 c_{13}^2  \,m_{\rm lightest} + s_{12}^2 c_{13}^2  \,\sqrt{m_{\rm lightest}^2+\dmsol} \,e^{-i \alpha_{21}} +s_{13}^2  \, \sqrt{m_{\rm lightest}^2+\dmatm}  \,e^{-i \alpha_{31}}\right| \,,\\
		{\rm IO}: \;m_{\beta\beta}&=\left| c_{12}^2 c_{13}^2  \,\sqrt{m_{\rm lightest}^2+\lvert\dmatm\rvert}  + s_{12}^2 c_{13}^2  \,\sqrt{m_{\rm lightest}^2+\dmsol+|\dmatm|}  \,e^{-i \alpha_{21}}\right.\nonumber \\
		&\left.+s_{13}^2  \, m_{\rm lightest}  \,e^{-i \alpha_{31}}
		\right| \,.
		\end{align}

		\clearpage
		
		The constraints of our model on the effective Majorana mass parameter are presented in Figs.~\ref{fig:predictionsmlightestmbetabetaNO} and~\ref{fig:predictionsmlightestmbetabetaIO}, for NO and IO neutrino masses, respectively. The regions in the ($m_\text{lightest}$,$m_{\beta\beta}$)-plane, allowed by neutrino oscillation data at the 1$\sigma$, 2$\sigma$ and 3$\sigma$ level without imposing any constraint on $\Mnu$, are shown by the dashed contours. Besides the current limits on $m_\text{lightest}$, previously presented in Figs.~\ref{fig:predictionsmlightestdeltaNO} and~\ref{fig:predictionsmlightestdeltaIO}, we also show the upper-limit ranges on $m_{\beta\beta}$ imposed by the $0\nu\beta\beta$ experiments EXO-200~\cite{EXO-200:2019rkq}, GERDA~\cite{GERDA:2020xhi}, CUORE~\cite{CUORE:2021mvw} and KamLAND-Zen 800~\cite{KamLAND-Zen:2022tow}. These ranges are indicated by coloured vertical bars, which reflect the uncertainty in theoretical calculations of the nuclear matrix elements. Note that $(\Mnuh)_{11}=0$ [see Eqs.~\eqref{eq:cond2327},~\eqref{eq:Meff23} and~\eqref{eq:Meff27}] for $2_{3,7}^e$ implying $m_{\beta\beta}=0$, thus we do not show this case in the aforementioned figures. This scenario is still viable for NO but is not compatible for IO neutrino masses.
		From the results of Fig.~\ref{fig:predictionsmlightestmbetabetaNO} for NO, one can see that for $2_{3,7}^{\mu,\tau}$ (upper plots) the lower bounds on $m_{\beta\beta}$ are within the sensitivity of current $0\nu\beta\beta$ decay experiments while being simultaneously in tension with the cosmological constraints on $m_{\text{lightest}}$. 
		In fact, the current KamLAND-Zen 800 result strongly disfavours both of these cases and one expects them to be probed  
		by future projects such as KamLAND2-Zen~\cite{KamLAND-Zen:2016pfg}, AMORE II~\cite{Lee:2020rjh}, CUPID~\cite{CUPID:2015yfg}, LEGEND~\cite{LEGEND:2017cdu}, nEXO\cite{nEXO:2017nam}, PandaX-III~\cite{Chen:2016qcd} or SNO+ I~\cite{SNO:2015wyx}. In contrast, for $2_{10}^{\mu,\tau}$ (lower plots), a wide region of parameter space containing even the b.f. point lies in the well known NO ``cancellation region" within the interval $m_{\text{lightest}} \sim [1,10]$ meV. This region cannot be probed experimentally due to the very suppressed values of $m_{\beta\beta}$. In Fig.~\ref{fig:predictionsmlightestmbetabetaNO} the results are shown for IO. We remark that for $2_{10}^{\mu,\tau}$ (middle right and lower plots) the lower bounds on $m_{\beta\beta}$ are within the sensitivity of current $0\nu\beta\beta$ decay experiments, with the KamLAND-Zen 800 bound strongly disfavouring these cases, which are also in tension with Planck data (brown shaded region). These are the most constrained models for IO. The remaining cases exhibit regions of the parameter space not currently constrained by $0\nu\beta\beta$ experiments and well below the cosmological bounds on the lightest neutrino mass. However, in contrast to NO, there is no $m_{\beta \beta}$ suppression region for IO. Therefore, the $0\nu\beta\beta$ experiments mentioned above could potentially rule out an IO neutrino mass spectrum. In summary, from these figures we conclude that for NO (IO) the $2_{3,7}$ ($2_{10}$) cases are more restrictive than the $2_{10}$ ($2_{3,7}$) cases.
		
	\end{itemize}
	%
	
	\section{Flavour phenomenology}
	\label{sec:pheno}
	
	Within the framework of the 2HDM, the presence of new charged and neutral scalars that interact with SM fermions introduces NP contributions in flavour processes. In Appendix~\ref{sec:interactions} we present the scalar-fermion interactions in the mass-eigenstate basis where it is clear that in our models there will be tree-level FCNCs which are controlled by the following matrices,
		\begin{align}
		\textbf{N}_d &= \textbf{V}_L^{d \dagger}\textbf{N}^0_d \mathbf{V}_R^d \; ,\;
		\textbf{N}^0_d = \frac{v}{\sqrt{2}} \left(s_\beta\Y_1^d - c_\beta\Y_2^d e^{i\varphi}\right) \; , \nonumber \\
		\textbf{N}_u &= \textbf{V}_L^{u \dagger}\textbf{N}^0_u \mathbf{V}_R^u \; ,\;
		\textbf{N}^0_u = \frac{v}{\sqrt{2}} \left(s_\beta\Y_1^u - c_\beta\Y_2^u e^{-i\varphi}\right) \; , \nonumber \\
		\textbf{N}_e &= \textbf{U}_L^{e \dagger}\textbf{N}^0_e \mathbf{U}_R^e \; ,\;
		\textbf{N}^0_e = \frac{v}{\sqrt{2}} \left(s_\beta\Y_1^e - c_\beta\Y_2^e e^{i\varphi}\right) \; ,
		\label{eq:NMatrices}
		\end{align}
		with the rotation angles $\beta$ and $\alpha$ defined in Eqs.~\eqref{eq:Higgsbasis} and~\eqref{eq:tan_alpha}, respectively. The unitary matrices $\mathbf{V}_{L,R}^{d,u}$ are defined in Eqs.~\eqref{eq:massdiag} and~\eqref{eq:hermitianmassdiag}, leading to the CKM quark mixing matrix $\mathbf{V}$ of Eq.~\eqref{eq:CKM}, while $\mathbf{U}_{L,R}^{e}$ and $\mathbf{U}^{\nu}$ are defined in Eqs.~\eqref{eq:leptonmassdiag} and~\eqref{eq:leptonhermitianmassdiag}, leading to the PMNS lepton mixing matrix $\mathbf{U}$ of Eq.~\eqref{eq:PMNS}. However, the Abelian flavour symmetries considered here provide, in some cases, a natural mechanism to control such dangerous FCNCs effects. In fact, note that the mass matrices labelled ``$5$" in Table~\ref{tab:Matrices} exhibit an isolated non-vanishing entry in a given row and column, which coincides with the mass of a specific fermion [see e.g. Eq.~\eqref{eq:VLdecoupled}]. This leads to zero entries in the $\mathbf{N}_x$ (with $x=d,u,e$) matrices of Eq.~\eqref{eq:NMatrices}, which control the strength of FCNCs,
	\begin{align}
	5^{d,u,e} &: \mathbf{N}_{d,u,e} \sim \begin{pmatrix}
	\times & 0 & 0 \\
	0 & \times & \times \\
	0 & \times & \times \\
	\end{pmatrix} \; , \;
	5^{s,c,\mu} : \mathbf{N}_{s,c,\mu} \sim \begin{pmatrix}
	\times & 0 & \times  \\
	0 & \times & 0 \\
	\times & 0 & \times \\
	\end{pmatrix} \; , \;
	5^{b,t,\tau} : \mathbf{N}_{b,t,\tau} \sim \begin{pmatrix}
	\times & \times & 0  \\
	\times & \times & 0 \\
	0 & 0 & \times \\
	\end{pmatrix} \; ,
	\label{eq:Nudtextures}
	\end{align}
	where the decoupled state is identified by a superscript. On the other hand, the mass matrices labelled ``$4$'' lead to $\mathbf{N}_x$ matrices without zero entries. In cases where the Abelian flavour symmetry leads to a decoupled state, there will be processes without NP contributions, depending of the flavour transition involved. In Tables~\ref{tab:LFV}--\ref{tab:SemiLeptonicDecays}, we list the flavour constraints considered in our numerical analysis in order to test the compatibility of the U(1) models given in Table~\ref{tab:charges}. In the last column of those tables we indicate the states that, when decoupled, automatically satisfy the bounds, since there is no NP contribution for the corresponding process. 
	
	\vspace{+0.2cm}
	
	The flavour constraints under consideration include:
	\begin{itemize}

		\begin{table}[t!]
			\renewcommand*{\arraystretch}{1.2}
			\centering
			\begin{tabular}{llcc}
				\hline \hline 
				Observable \ & Constraint & Ref. \ & Decoupled state \\
				\hline
				$|g_{\mu}/g_e|-1$ & $0.0019 \pm 0.0014$ & HFLAV~\cite{HFLAV:2022pwe} & -
				\\
				$|g^S_{RR,\mu e}|$ & $<0.035$ & TWIST~\cite{TWIST:2011aa} & -
				\\
				$|g^S_{RR,\tau e}|$ & $<0.70$ & PDG~\cite{ParticleDataGroup:2022pth} & - 
				\\
				$|g^S_{RR,\tau \mu}|$ & $<0.72$ & PDG~\cite{ParticleDataGroup:2022pth} & -
				\\
				$\operatorname{BR}(\tau^- \rightarrow e^- e^+ e^-)$ & $<2.7 \times 10^{-8}$ & 
				Belle~\cite{Hayasaka:2010np} & $e,\tau$ 
				\\
				$\operatorname{BR}(\tau^- \rightarrow \mu^- \mu^+ \mu^-)$ & $<2.1 \times 10^{-8}$ & 
				Belle~\cite{Hayasaka:2010np} & $\mu,\tau$ 
				\\
				$\operatorname{BR}(\tau^- \rightarrow e^- \mu^+ e^-)$  & $<1.5 \times 10^{-8}$ & 
				Belle~\cite{Hayasaka:2010np} & $e,\mu,\tau$  
				\\
				$\operatorname{BR}(\tau^- \rightarrow e^- e^+ \mu^-)$ & $<1.8 \times 10^{-8}$ & 
				Belle~\cite{Hayasaka:2010np} & $\mu,\tau$  
				\\
				$\operatorname{BR}(\tau^- \rightarrow \mu^- e^+ \mu^-)$ & $<1.7 \times 10^{-8}$ & Belle~\cite{Hayasaka:2010np} & $e,\mu,\tau$ 
				\\
				$\operatorname{BR}(\tau^- \rightarrow \mu^- \mu^+ e^-)$ & $<2.7 \times 10^{-8}$ & Belle~\cite{Hayasaka:2010np} & $e,\tau$ 
				\\
				$\operatorname{BR}(\mu^- \rightarrow e^- e^+ e^-)$ & $<1.0 \times 10^{-12}$ & SINDRUM~\cite{SINDRUM:1987nra} & $e,\mu$ 
				\\
				$\operatorname{BR}(\mu \rightarrow e \gamma)$ & $<4.2 \times 10^{-13}$ & MEG \cite{MEG:2016leq} & $e,\mu$ 
				\\
				$\operatorname{BR}(\tau \rightarrow e \gamma)$ & $<3.3 \times 10^{-8}$ & BaBar~\cite{BaBar:2009hkt} & $e,\tau$ 
				\\
				$\operatorname{BR}(\tau \rightarrow \mu \gamma)$  & $<4.2 \times 10^{-8}$ & Belle \cite{Belle:2021ysv} & $\mu,\tau$  
				\\
				\hline \hline
			\end{tabular}
			\caption{
				Current experimental bounds on lepton universality observables and BRs of cLFV processes. When one of the lepton states indicated in the last column is decoupled, there are no NP contributions to the corresponding process in the first column (see text for details).}
			\label{tab:LFV}
		\end{table}
		\item \textbf{Lepton universality and cLFV:} Table~\ref{tab:LFV} shows the current bounds for several cLFV processes. For the 2HDM, the NP contributions can be computed as discussed in Refs.~\cite{Botella:2014ska,Correia:2019vbn}. The constraints on lepton universality are encoded in the parameters $|g_{\mu}/g_e|-1$, $|g^S_{RR,\mu e}|$, $|g^S_{RR,\tau e}|$ and $|g^S_{RR,\tau \mu}|$. For 3-body cLFV decays $e_\alpha^{-} \rightarrow e_\beta^{-} e_\gamma^{+} e_\delta^{-}$, mediated by neutral scalars $H$ and $I$ at tree-level, the branching ratio~(BR) is
		\begin{align}
		\operatorname{BR}\left(e_\alpha^{-} \rightarrow e_\beta^{-} e_\gamma^{+} e_\delta^{-}\right)& = \; \frac{1}{\left(1+\delta_{\beta \delta}\right)} \frac{m_\alpha^5 G_{\mathrm{F}}}{3 \times 2^{12} \pi^3 \Gamma_{{\alpha}}}\times \bigg[
		\left| g_{L L}^{\alpha \beta, \gamma \delta}\right|^2+\left|g_{L L}^{\alpha \delta, \gamma \beta}\right|^2+\left|g_{R R}^{\alpha \beta, \gamma \delta}\right|^2+\left|g_{R R}^{\alpha \delta, \gamma \beta}\right|^2  \label{eq:Gamma3BodyLFV}
		\\
		& + \left|g_{L R}^{\alpha \beta , \gamma \delta}\right|^2+\left|g_{L R}^{\alpha \delta, \gamma \beta}\right|^2+\left|g_{R L}^{\alpha \beta, \gamma \delta}\right|^2+\left|g_{R L}^{\alpha \delta, \gamma \beta}\right|^2 
		-\operatorname{Re}\left(g_{L L}^{\alpha \beta, \gamma \delta} g_{L L}^{\alpha \delta, \gamma \beta *}+g_{R R}^{\alpha \beta, \gamma \delta} g_{R R}^{\alpha \delta , \gamma \beta *}\right) 
		\bigg] \; , \nonumber
		\end{align}
		where $\Gamma_{{\alpha}}$ is the total decay width of $e_{\alpha}$, namely $\Gamma_\mu = 3.0\times 10^{-19}\text{ GeV}$ and $\Gamma_\tau= 2.3
		\times 10^{-12}\text{ GeV}$~\cite{ParticleDataGroup:2022pth}, and
		\begin{align}
		g_{L L}^{\alpha \beta , \gamma \delta} & =\left(\mathbf{N}_e^{\dagger}\right)_{\beta \alpha}\left(\mathbf{N}_e^{\dagger}\right)_{\delta \gamma}\left(\frac{1}{m_H^2}-\frac{1}{m_I^2}\right) \; , \quad
		g_{R L}^{\alpha \beta , \gamma \delta} =\left(\mathbf{N}_e\right)_{\beta \alpha}\left(\mathbf{N}_e^{\dagger}\right)_{\delta \gamma}\left(\frac{1}{m_H^2}+\frac{1}{m_I^2}\right)\; , \nonumber \\
		g_{L R}^{\alpha \beta , \gamma \delta} & =\left(\mathbf{N}_e^{\dagger}\right)_{\beta \alpha}\left(\mathbf{N}_e\right)_{\delta \gamma}\left(\frac{1}{m_H^2}+\frac{1}{m_I^2}\right)\; , \quad
		g_{R R}^{\alpha \beta , \gamma \delta}  =\left(\mathbf{N}_e\right)_{\beta \alpha}\left(\mathbf{N}_e\right)_{\delta \gamma}\left(\frac{1}{m_H^2}-\frac{1}{m_I^2}\right)\; .
		\label{eq:Gamma3BodyLFV_aux}
		\end{align}
		Finally, we also consider the radiative cLFV decay $e_\alpha \rightarrow e_\beta \gamma$. Neglecting the contributions proportional to neutrino masses and the subleading terms suppressed by $m_{e}^2 / m_{H, I}^2$, the BR at one loop is
		\begin{align}
		\operatorname{BR}\left(e_\alpha \rightarrow e_\beta \gamma\right)=\frac{\alpha_e m_\alpha^5 G_{\mathrm{F}}^2}{128 \pi^4 \Gamma_{\alpha}}\left(\left|\mathcal{A}_L\right|^2+\left|\mathcal{A}_R\right|^2\right) \; ,
		\label{eq:RadiativeLFV}
		\end{align}
		where $\alpha_e=e^2 /(4 \pi)$. The amplitudes $\mathcal{A}_{L, R}$ are given by
		\begin{align}
		\mathcal{A}_L= & \left(\mathbf{N}_e^{\dagger} \mathbf{N}_e\right)_{\beta \alpha} \frac{1}{12}\left(\frac{1}{m_H^2}+\frac{1}{m_I^2}-\frac{1}{m_{H^\pm}^2}\right) - \frac{\left(\mathbf{N}_e^{\dagger}\right)_{\beta i}\left(\mathbf{N}_e^{\dagger}\right)_{i \alpha}}{2 m_\alpha / m_i} \left\{\frac{1}{m_H^2} \left[\frac{3}{2}+\ln \left(\frac{m_i^2}{m_H^2}\right)\right] - \frac{1}{m_I^2} \left[\frac{3}{2}+\ln \left(\frac{m_i^2}{m_I^2}\right)\right] \right\} \; , \nonumber \\
		\mathcal{A}_R= & \left(\mathbf{N}_e \mathbf{N}_e^{\dagger}\right)_{\beta \alpha} \frac{1}{12}\left(\frac{1}{m_H^2}+\frac{1}{m_I^2}\right) - \frac{\left(\mathbf{N}_e\right)_{\beta i}\left(\mathbf{N}_e\right)_{i \alpha}}{2 m_\alpha / m_i} \left\{\frac{1}{m_H^2} \left[\frac{3}{2}+\ln \left(\frac{m_i^2}{m_H^2}\right)\right] - \frac{1}{m_I^2} \left[\frac{3}{2}+\ln \left(\frac{m_i^2}{m_I^2}\right)\right] \right\} \; ,
		\label{eq:RadiativeLFV_aux}
		\end{align}
		where a sum over $i=e, \mu, \tau$ is implicit and chirallity-suppressed terms $\propto m_\beta / m_\alpha$ have been neglected.
		
		From Eqs.~\eqref{eq:Gamma3BodyLFV}--\eqref{eq:RadiativeLFV_aux}, we can easily deduce that most 3-body and radiative cLFV processes are forbidden at the one-loop level for the $5_1^{e}$ structures (see Table~\ref{tab:LFV}). For example, the decay $\mu \rightarrow e \gamma$ is forbidden if $\mathbf{M}_e \sim 5_1^e\left(5_1^\mu\right)$, since the electron (muon) is decoupled and, consequently, $(\mathbf{N}_e)_{\mu e}=(\mathbf{N}_e)_{e \mu}=0$, i.e. $\mu-e$ transitions are not possible. For the 3-body decays $e_\alpha^{-} \rightarrow e_\beta^{-} e_\gamma^{+} e_\delta^{-}$, at least two of these processes are disallowed for every $5_1^{e}$ case.

		\begin{table}[t!]
			\renewcommand*{\arraystretch}{1.2}
			\centering
			\begin{tabular}{cccc}
				\hline \hline
				Observable \ & Constraint & Ref. \ & Decoupled state \\
				\hline
				$\operatorname{BR}\left(\overline{B} \rightarrow X_s \gamma \right)$  & $(3.49 \pm  0.19) \times 10^{-4} $   & PDG~\cite{ParticleDataGroup:2022pth} & - \\
				$|\varepsilon_K|$ &  $(2.228 \pm  0.011) \times 10^{-3}$  & PDG~\cite{ParticleDataGroup:2022pth} & $u,d,s$ \\
				$\Delta m_{K}^{\text{NP}}$ & $<3.484 \times 10^{-15}$ GeV  & PDG~\cite{ParticleDataGroup:2022pth} 
				& $d,s$    \\
				$\Delta m_{B_d}$ & $(3.334 \pm 0.013)  \times 10^{-13}$ GeV   & PDG~\cite{ParticleDataGroup:2022pth}  
				& $d,b$ \\
				$\Delta m_{B_s}$ &  $(1.1693 \pm 0.0004)  \times 10^{-11}$ GeV  & PDG~\cite{ParticleDataGroup:2022pth} & $s,b$    \\
				$\Delta m_{D}^{\text{NP}}$ & $< 6.56 \times 10^{-15}$ GeV & PDG~\cite{ParticleDataGroup:2022pth} & $u,c$    \\
				\hline
				\hline
			\end{tabular}
			\caption{Present experimental constraints on BR of $\overline{B} \rightarrow X_s \gamma$, CP violation in the neutral Kaon system via the $\varepsilon_K$ parameter and mass differences of $K^0,B^0_{d,s},D^0$ meson-antimeson systems. When one of the quark states indicated in the last column is decoupled, there are no NP contributions to the corresponding observable in the first column (see text for details).}
			\label{tab:mesonOscillations}
		\end{table}
		\item \textbf{$\overline{B} \rightarrow X_s \gamma$ and neutral meson–antimeson observables:} Table~\ref{tab:mesonOscillations} shows the constraints for the quark sector. These include the radiative $\overline{B} \rightarrow X_s \gamma$ decay, CP violation in the neutral Kaon system via the $\varepsilon_K$ parameter and the mass differences of the meson-antimeson systems $K^0$, $B^0_{d}$, $B^0_{s}$ and $D^0$. In particular, for $\Delta m_K$ and $\Delta m_D$ only NP contributions are considered, as highlighted in Ref.~\cite{Ferreira:2019aps}. To understand the impact of flavour symmetries, consider the NP contribution to the matrix element contributing to the $\overline{K}_0 \rightarrow K_0$ transition~\cite{Ferreira:2019aps}:
		\begin{align}
		M_{21}^{\text{NP}} = \;
		& \dfrac{f_k^2 m_K}{96 v ^2} \bigg\{ \left[ (\mathbf{N}_{d}^*)^2_{ds} 
		+ (\mathbf{N}_{d})^2_{sd} \right] \dfrac{10 m_k^2}{(m_s + m_d)^2} \bigg( \dfrac{1}{m_I^2} - \dfrac{c_{\beta-\alpha}^2}{m_h^2} - \dfrac{s_{\beta-\alpha}^2}{m_H^2} \bigg) \nonumber
		\\
		& + 4 (\mathbf{N}_{d}^*)_{ds} (\mathbf{N}_{d})_{sd}
		\bigg [ 1 + \dfrac{6 m_K^2}{(m_s + m_d)^2}\bigg ]  \bigg( \dfrac{1}{m_I^2} + \dfrac{c_{\beta-\alpha}^2}{m_h^2} + \dfrac{s_{\beta-\alpha}^2}{m_H^2} \bigg)  
		\bigg\} \; .
		\label{eq:M21NP}
		\end{align}
		If one of the constituent quarks in $K^0$, either $d$ or $s$, is decoupled, then $(\mathbf{N}_{d})_{sd} = (\mathbf{N}_{d})_{ds} = 0$. This implies $M_{21}^{\text{NP}}=0$ and $\Delta m_{K}^{\text{NP}} =2|M_{21}^{\text{NP}}| =0$. Furthermore, the $\varepsilon_K$ parameter is given by~\cite{Ferreira:2019aps}:
		\begin{align}
		\varepsilon_K =
		- \frac{\text{Im}(M_{21} {\lambda_u^*}^2)}{\sqrt{2} \Delta m_K |\lambda_u|^2}
		=
		\varepsilon_K^{\text{SM}}
		- \frac{\text{Im}(M_{21}^{\text{NP}} {\lambda_u^*}^2)}{\sqrt{2} \Delta m_K |\lambda_u|^2} \; ,
		\label{eq:epsilonK}
		\end{align}
		where we use the experimental value for $\Delta m_K$ in the denominator and $\lambda_u =  \mathbf{V}_{11}^* \mathbf{V}_{12} $. Thus, the condition $M_{21}^{\text{NP}}=0$ also implies $\varepsilon_K = \varepsilon_K^{\text{SM}}$, which is in agreement with the current experimental bound. Hence, the first two conditions in Table~\ref{tab:mesonOscillations} are  satisfied automatically. A similar argument applies to $\Delta m_{B_{d,s}}$ and $\Delta m_{D}^{\text{NP}}$, i.e. these observables have no NP contributions if one of the meson constituent quarks is decoupled. On the contrary, the $\overline{B} \rightarrow X_s \gamma$ constraint cannot be automatically fulfilled by a decoupled state, as its BR depends not only on the $\mathbf{N}_x$ matrices but also on terms involving $\mathbf{V}^\dagger \mathbf{N}_u$ and $\mathbf{N}_d^\dagger \mathbf{V}^\dagger$.
		
		If $u$ is decoupled, $\varepsilon_K$ is automatically within the allowed range. To understand this we note that, in our models, $\mathbf{M}_d$ can be made real by appropriate rephasing of the fields as shown in Table~\ref{tab:Matrices}. Hence, in this case, $\mathbf{H}_{d}= \mathbf{M}_{d} \mathbf{M}_{d}^\dagger$ and $\mathbf{H}_{d}^\prime= \mathbf{M}_{d}^\dagger \mathbf{M}_{d} $ are also real. Since their eigenvalues are also real, as indicated in Eq.~\eqref{eq:hermitianmassdiag}, their eigenvectors $\mathbf{V}_L^d$ and $\mathbf{V}_R^d$ are also real. Thus, from Eq.~\eqref{eq:NMatrices}, we conclude that $\mathbf{N}_d$ is real and so is the matrix $M_{21}^{\text{NP}}$ [see Eq.~\eqref{eq:M21NP}]. 
		
		For the cases $\mathbf{P}_{12}{5}_{1}^u \mathbf{P}_{23}$ and $\mathbf{P}_{123}{5}_{1}^u \mathbf{P}_{12}$, we obtain
		\begin{align}
		\mathbf{V}_L^u = 
		\begin{pmatrix}
		1 & 0     & 0     \\
		0 & \cdot & \cdot \\
		0 & \cdot & \cdot
		\end{pmatrix} ,
		\quad
		\mathbf{V}_L^u = 
		\begin{pmatrix}
		0 & \cdot & \cdot \\
		0 & \cdot & \cdot \\
		1 & 0     & 0
		\end{pmatrix},
		\label{eq:eigenvectorsUdecoupled}
		\end{align}
		respectively. By using these matrices and taking into account that $\mathbf{V}_L^d$ is real, it follows that the first row of the CKM matrix is real [see Eq.~\eqref{eq:CKM}]. Consequently, $\lambda_u$ is real, along with $M_{21}^{\text{NP}}$, implying $\varepsilon_K= \varepsilon_K^{\text{SM}}$ [see Eq.~\eqref{eq:epsilonK}].

		\begin{table}[t!]
			\renewcommand*{\arraystretch}{1.2}
			\centering
			\begin{tabular}{lccc}
				\hline \hline
				Observable \ & Constraint & Ref. \ & Decoupled state \\
				\hline
				Br($B_d \rightarrow e^+ e^-$)   & $<2.5 \times 10^{-9}$ & LHCb~\cite{LHCb:2020pcv} & $d,b$ 
				\\
				Br($B_d \rightarrow \mu^+ \mu^-$)   & $<2.0 \times 10^{-10}$ & PDG~\cite{ParticleDataGroup:2022pth} & $d,b$ 
				\\
				Br($B_d \rightarrow \tau^+ \tau^-$)   & $<2.1 \times 10^{-3}$ & LHCb~\cite{LHCb:2017myy} & $d,b$
				\\
				Br($B_d \rightarrow \tau^{\pm} e^\mp$)   & $<1.6 \times 10^{-5}$ & Belle~\cite{Belle:2021rod} & $e,\tau,d,b$ 
				\\
				Br($B_d \rightarrow \tau^\pm \mu^\mp$)   & $<1.4 \times 10^{-5}$ & LHCb~\cite{LHCb:2019ujz} & $\mu,\tau,d,b$
				\\
				Br($B_s \rightarrow e^+ e^-$)   & $<9.4 \times 10^{-9}$ & LHCb~\cite{LHCb:2020pcv}  & $s,b$
				\\ 
				Br($B_s \rightarrow \mu^+ \mu^-$)   & $(3.01\pm 0.35)\times 10^{-9} $ & PDG~\cite{ParticleDataGroup:2022pth}  & $s,b$  
				\\
				Br($B_s \rightarrow \mu^\pm e^\mp $)   & $<5.4 \times 10^{-9}$ & LHCb~\cite{LHCb:2017hag} & $e,\mu,s,b$ 
				\\
				Br($D \rightarrow e^+ e^-$)   & $<7.9 \times 10^{-8}$ & Belle~\cite{Belle:2010ouj}  & $u,c$ 
				\\
				Br($D \rightarrow \mu^+ \mu^-$)   & $<6.2 \times 10^{-9}$ & LHCb~\cite{LHCb:2013jyo} & $u,c$ 
				\\
				Br($D \rightarrow \mu^\pm e^\mp$)   & $<1.3 \times 10^{-8}$ & LHCb~\cite{LHCb:2015pce} & $e,\mu,u,c$ 
				\\
				\hline
				\hline
			\end{tabular}
			\caption{Current experimental bounds on the BR of rare leptonic neutral meson decays. In the last column the decoupled state prediction automatically satisfies the constraint (see text for details).} 
			\label{tab:LeptonicDecays}
		\end{table}
		\item \textbf{Leptonic neutral meson decays and semileptonic processes:} 
		Tables~\ref{tab:LeptonicDecays} and~\ref{tab:SemiLeptonicDecays} show constraints involving both the quark and lepton sectors, namely, leptonic neutral meson decays and semileptonic processes, respectively. In the notation of Ref.~\cite{Crivellin:2013wna}, the NP Wilson coefficients relating to the leptonic decays are given by
		\begin{align}
		C_S^{q_f q_i}=-\frac{\sqrt{2}\, \pi^2}{G_F M_W^2}\left(c_{L R}^{f i, \alpha \beta}+c_{L L}^{f i, \alpha \beta}\right), 
		& \quad
		C_P^{q_f q_i}=-\frac{\sqrt{2}\, \pi^2}{G_F M_W^2}\left(c_{L R}^{f i, \alpha \beta}-c_{L L}^{f i, \alpha \beta}\right), \\
		C_S^{\prime q_f q_i}=-\frac{\sqrt{2}\, \pi^2}{G_F M_W^2}\left(c_{R R}^{f i, \alpha \beta}+c_{R L}^{f i, \alpha \beta}\right), 
		&
		\quad
		C_P^{\prime q_f q_i}=-\frac{\sqrt{2}\, \pi^2}{G_F M_W^2}\left(c_{R R}^{f i, \alpha \beta}-c_{R L}^{f i, \alpha \beta}\right),
		\label{eq:LeptonicWilson}
		\end{align}
		where for a down-type (up-type) meson,
		\begin{equation}
		\begin{array}{ll}
		c_{L L}^{f i, \alpha \beta}=-\dfrac{(\mathbf{N}_{d,u}^{\dagger})_{fi}(\mathbf{N}_{e}^{\dagger})_{\alpha \beta}}{m_H^2}
		\pm
		\dfrac{(\mathbf{N}_{d,u}^{\dagger})_{fi}(\mathbf{N}_{e}^{\dagger})_{\alpha \beta}}{m_I^2}, 
		& 
		c_{R L}^{f i, \alpha \beta}=-\dfrac{(\mathbf{N}_{d,u})_{fi}(\mathbf{N}_{\ell}^{\dagger})_{\alpha \beta}}{m_H^2}
		\mp
		\dfrac{(\mathbf{N}_{d,u})_{fi}(\mathbf{N}_{e}^{\dagger})_{\alpha \beta}}{m_I^2},
		\\ 
		c_{L R}^{f i , \alpha \beta}=
		-\dfrac{(\mathbf{N}_{d,u}^{\dagger})_{fi}(\mathbf{N}_{e})_{\alpha \beta}}{m_H^2}
		\mp
		\dfrac{(\mathbf{N}_{d,u}^{\dagger})_{fi}(\mathbf{N}_{e})_{\alpha \beta}}{m_I^2}, 
		& 
		c_{R R}^{f i, \alpha \beta}=
		-\dfrac{(\mathbf{N}_{d,u})_{fi}(\mathbf{N}_{e})_{\alpha \beta}}{m_H^2}
		\pm
		\dfrac{\left(\mathbf{N}_{d,u}\right)_{fi}(\mathbf{N}_{e})_{\alpha \beta}}{m_I^2} .
		\end{array}
		\label{eq:LeptonicWilson_aux}
		\end{equation}
		Consider, for instance, the process $B_s (s\overline{b}) \rightarrow \mu^\pm e^\mp$. In this example, if $s$, $b$, $\mu$ or $e$ are decoupled, then all NP Wilson coefficients are zero [see Eqs.~\eqref{eq:Nudtextures}, \eqref{eq:LeptonicWilson} and \eqref{eq:LeptonicWilson_aux}], i.e. there are no NP contributions. A similar analysis applies to the other leptonic neutral meson decays. However, this feature cannot be extended to semileptonic processes, as their NP Wilson coefficients do not depend directly on the $\mathbf{N}_x$ matrices, being given by
		\begin{equation}
		\begin{array}{ll}
		\dfrac{C_R^{u_f d_i, \ell_\alpha \nu_n}}{C_{\text{SM}}} = -\dfrac{(\mathbf{V} \mathbf{N}_d)_{fi} ( \mathbf{N}_e^{\dagger} \mathbf{U} )_{\alpha n}}{m_{H^{ \pm}}^2 \mathbf{V}_{fi} \mathbf{U}_{\alpha n}}, 
		& 
		\dfrac{C_L^{u_f d_i, \ell_\alpha \nu_n}}{C_{\text{SM}}} = \dfrac{(\mathbf{N}_u^\dagger \mathbf{V} )_{fi} ( \mathbf{N}_e^{\dagger} \mathbf{U} )_{\alpha n}}{m_{H^{ \pm}}^2 \mathbf{V}_{fi} \mathbf{U}_{\alpha n}}.
		\end{array}
		\label{eq:LeptonicWilson_not}
		\end{equation}

		Using the Wilson coefficients of Eqs.~\eqref{eq:LeptonicWilson_aux} and \eqref{eq:LeptonicWilson_not}, we calculate the BRs for leptonic neutral meson decays and semileptonic processes by employing the methodology outlined in Refs.\cite{Crivellin:2013wna, Botella:2014ska}. 
		
	\end{itemize}
	\begin{table}[t!]
		\renewcommand*{\arraystretch}{1.2}
		\centering
		\begin{tabular}{lcc}
			\hline \hline
			Observable \ & Constraint & Ref. \\
			\hline
			Br($B^+ \rightarrow e^+ \nu$)   & $<9.8\times 10^{-7}$ & Belle~\cite{Belle:2006tbq}
			\\ 
			Br($B^+ \rightarrow \mu^+ \nu$)   & $<8.6\times 10^{-7}$ & Belle~\cite{Belle:2019iji}
			\\ 
			Br($B^+ \rightarrow \tau^+ \nu$)   & $(1.09 \pm 0.24)\times 10^{-4}$  & PDG~\cite{ParticleDataGroup:2022pth}
			\\
			Br($D_s^+ \rightarrow e^+ \nu$)   & $<8.3\times 10^{-5}$  & Belle~\cite{Belle:2013isi}\\  
			Br($D_s^+ \rightarrow \mu^+ \nu$)   & $(5.43 \pm 0.15)\times 10^{-3}$   & PDG~\cite{ParticleDataGroup:2022pth}
			\\ 
			Br($D_s^+ \rightarrow \tau^+ \nu$)   & $(5.32 \pm 0.11)\times 10^{-2}$   & PDG~\cite{ParticleDataGroup:2022pth}
			\\
			Br($D^+ \rightarrow e^+ \nu$)   & $< 8.8 \times 10^{-6}$ & CLEO~\cite{CLEO:2008ffk} \\  
			Br($D^+ \rightarrow \mu^+ \nu$)   & $(3.74 \pm 0.17)\times 10^{-4}$  & PDG~\cite{ParticleDataGroup:2022pth}
			\\ 
			Br($D^+ \rightarrow \tau^+ \nu$)   & $(1.20\pm 0.27)\times 10^{-3}$   & BESIII~\cite{BESIII:2019vhn}
			\\
			$\dfrac{\text{Br}\left(\pi^{+} \rightarrow e^{+} \nu\right)}{\text{Br}\left(\pi^{+} \rightarrow \mu^{+} \nu\right)}$   & $(1.230\pm 0.004)\times 10^{-4}$ & PDG~\cite{ParticleDataGroup:2022pth}
			\\
			$\dfrac{\text{Br}\left(K^{+} \rightarrow e^{+} \nu\right)}{\text{Br}\left(K^{+} \rightarrow \mu^{+} \nu\right)}$   & $(2.488\pm 0.009)\times 10^{-5}$ & PDG~\cite{ParticleDataGroup:2022pth}
			\\
			$\dfrac{\text{Br}\left(\tau^{-} \rightarrow \pi^{-} \nu\right)}{\text{Br}\left(\pi^{+} \rightarrow \mu^{+} \nu\right)}$   & $(10.82\pm 0.05)\times 10^{-2}$ & PDG~\cite{ParticleDataGroup:2022pth}
			\\
			$\dfrac{\text{Br}\left(\tau^{-} \rightarrow K^{-} \nu\right)}{\text{Br}\left(K^{+} \rightarrow \mu^{+} \nu\right)}$  & $(1.095 \pm 0.016)\times 10^{-2}$ & PDG~\cite{ParticleDataGroup:2022pth}
			\\
			\hline
			\hline
		\end{tabular}
		\caption{Current experimental bounds on the BRs of semileptonic processes.} 
		\label{tab:SemiLeptonicDecays}
	\end{table}
	%
	
	\subsection{Numerical procedure and constraints}
	\label{sec:num}
	
	As already discussed, the Abelian flavour symmetries impose specific flavour structures that, in some cases, lead to controlled FCNCs. To investigate their phenomenology and explore the parameter space, we perform a numerical analysis considering all relevant constraints, which we now describe. Since the NP contributions to flavour processes stem from the additional scalar degrees of freedom of the 2HDM, we will consider as input parameters the angles $\alpha,\beta$ and the masses $m_{h,H,I,H^\pm}$ defined in Appendix~\ref{sec:scalar}. We will restrict our analysis to the so-called alignment limit, which corresponds to $\beta-\alpha = \pi/2$~\cite{Branco:2011iw}. In such a case, the new CP-even scalar $H^0$ aligns with the SM Higgs boson~$h$ for which $m_{h}= 125.25$ GeV~\cite{ParticleDataGroup:2022pth}. The remaining parameters vary within the intervals $10^{-2} \lesssim \tan \beta \equiv v_2/v_1 \lesssim 10^3$ and $10^2 \text{ GeV} \lesssim m_{H,I,H^\pm} \lesssim 10^{3} \text{ TeV}$, where the $100$~GeV lower mass limit is motivated by the $H^\pm$ mass constraint found in Ref.~\cite{Pierce:2007ut}. 
	
	The parameters of the scalar potential defined in Eq.~\eqref{eq:Vpotential2HDM} can be reconstructed in terms of the input variables:
	\begin{align}
	\mu_{12}^2 & = - m_I^2 \sin \beta \cos \beta \; ,\;
	\lambda_1 = \frac{m_h^2+(m_H^2-m_I^2)\tan^2 \beta}{v^2} \; ,\;
	\lambda_2 = \frac{m_h^2+(m_H^2-m_I^2)\cot^2 \beta}{v^2} \; ,\nonumber \\
	\lambda_3 &= \frac{2m_{H^{\pm}}^2+m_h^2-m_H^2-m_I^2}{v^2} \; ,\;
	\lambda_4 = \frac{2 \left(m_I^2 - m_{H^\pm}^2 \right)}{v^2} \; .
	\label{Eq:PotParameters}
	\end{align}
	As for the Yukawa couplings, which determine the scalar couplings to fermions, we extract them using the best-fit results on the mass matrices from Sections~\ref{sec:texturesrealisable} and~\ref{sec:neutrino}. In all cases, the scalar potential parameters, as well as the Yukawa couplings, must obey the following theoretical constraints:
	\begin{itemize}
		
		\item \textbf{Vacuum stability:} For the potential to be bounded from below the quartic parameters must satisfied~\cite{Branco:2011iw}
		\begin{equation}
		\lambda_1 > 0 \; , \; \lambda_2 > 0 \; , \; \lambda_3 + \sqrt{\lambda_1 \lambda_2} > 0 \; , \; \lambda_3 + \lambda_4 + \sqrt{\lambda_1 \lambda_2} > 0 \; ,
		\label{eq:BFB}
		\end{equation}
		as well as the condition \cite{Barroso:2012mj,Ivanov:2015nea}
		\begin{equation}
		-\frac{2 \mu_{12}^2}{v^2 c_\beta s_\beta} > \lambda_3 + \lambda_4 - \sqrt{\lambda_1 \lambda_2}  \; .
		\label{eq:BFBextra}
		\end{equation}

		\item \textbf{Perturbative unitarity:} The quartic scalar parameters are constrained by demanding perturbative unitarity in scalar-scalar scattering at tree level~\cite{PhysRevLett.38.883, PhysRevD.16.1519}, resulting in the following conditions~\cite{Branco:2011iw}:
		\begin{align}
		|\lambda_1| \; , \; |\lambda_2|\; , \; |\lambda_3|\; , \;  |\lambda_3 + \lambda_4|\; , \; |\lambda_3 - \lambda_4|\; , \; |\lambda_3 + 2\lambda_4| & \leq 8 \pi \; , \nonumber \\
		\left| \frac{3}{2}(\lambda_1 + \lambda_2) \pm \sqrt{\frac{9}{4}\left(\lambda_1 - \lambda_2\right)^2 + \left(2\lambda_3 - \lambda_4\right)^2} \right| & \leq 8 \pi \; , \nonumber \\
		\left| \frac{1}{2}(\lambda_1 + \lambda_2) \pm \frac{1}{2}\sqrt{\left(\lambda_1 - \lambda_2\right)^2 + 4\lambda_4^2} \right| & \leq 8 \pi \; .
		\label{eq:PertUnit}
		\end{align}

		\item \textbf{Quartic scalar couplings and Yukawa coupling perturbativity:} Quartic scalar and Yukawa couplings must fulfil perturbativity bounds, considered here to be
		\begin{align}
		|\lambda_1| \; , \; |\lambda_2|\; , \; |\lambda_3|\; , \;  |\lambda_4| \; , \; \left| \left(\Y_a^x\right)_{ij} \right| & \leq \sqrt{4 \pi} \;\; , a=1,2\;,\; x=e,d,u\,.
		\label{eq:perturbativityYuk}
		\end{align}

	\end{itemize}

	Regarding the experimental constraints, we consider the following: 
	\begin{itemize}
		
		\item \textbf{Electroweak precision observables~(EWPO):} The presently allowed ranges for the oblique $S$, $T$ and $U$ parameters at $95 \%$ confidence level (CL)~\cite{ParticleDataGroup:2022pth} are
		\begin{equation}
		S=-0.02 \pm 0.10 \; , \; T=0.03\pm0.12 \; , \; U=0.01\pm0.11 \; ,
		\end{equation}
		with correlation coefficients of $+0.92$ ($-0.80$) [$-0.93$] between $S$ and $T$ ($S$ and $U$) [$T$ and $U$]. To compute them in our 2HDM framework, we follow Refs.~\cite{Grimus:2007if,Grimus:2008nb} where the analytical expressions for the $S$, $T$ and $U$ were derived for a generic extension of the SM featuring arbitrary number of Higgs doublets and singlets. As for the measurement of the decay $Z \rightarrow b\overline{b}$, we follow Ref.~\cite{Ferreira:2019aps}, where the Lagrangian for the $Zb\overline{b}$ vertex is written as
		\begin{align}
		\mathcal{L}_{Z b \bar{b}} = -\frac{e Z_\mu}{s_W c_W} \bar{b} \gamma^\mu \left( \bar{g}_b^L P_L + \bar{g}_b^R P_R \right) b,
		\end{align}
		and it is required that the $H^\pm$ contribution added to the SM one does not deviate from the SM prediction by more than $2 \sigma$, viz. $2\left(\bar{g}_b^L\right)^2 + 2\left(\bar{g}_b^R\right)^2 = 0.36782 \pm 0.00143$.
		
		\item \textbf{SM Higgs boson:} As mentioned above, we work in the alignment limit, so that $H^0$ aligns with the SM Higgs boson $h$. Thus, the $h$ couplings to the gauge bosons and fermions are SM-like, i.e. within their experimental values. However, the charged scalar $H^\pm$ contributes at one-loop level to the radiative $h\rightarrow \gamma \gamma$ and $h\rightarrow Z \gamma$ decays. We computed analytically their decay widths following Refs.~\cite{Djouadi:2005gj,Posch:2010hx,Fontes:2014xva,Aguilar-Saavedra:2023tql}. To check if the set of generated points is compatible with the experimentally measured properties of the SM-like Higgs boson we use \texttt{HiggsSignals}~\cite{Bechtle:2013xfa,Bechtle:2020uwn}, which is part of \texttt{HiggsTools}~\cite{Bahl:2022igd}. For a given model, this tool computes global $\chi^2$ values taking into account Higgs measurements performed at the LHC. We define $\Delta \chi_{125}^2=\chi^2-\chi_{\mathrm{SM}}^2$, where $\chi_{\mathrm{SM}}^2=151.6$, is the $\chi^2$ value of the SM provided by \texttt{HiggsSignals}, for 159 degrees of freedom. This represents an agreement within a $37 \%$ CL. In our work we require an agreement at the $95 \%$ CL, that is, $\Delta \chi_{125}^2 \leq 37$~\cite{Aguilar-Saavedra:2023tql}.
		
		\item \textbf{New scalar searches:} The parameter space of our models is further constrained by searches for new scalars conducted at LEP and LHC. This analysis is assisted by \texttt{HiggsBounds}~\cite{Bechtle:2008jh,Bechtle:2011sb,Bechtle:2013wla,Bechtle:2015pma,Bechtle:2020pkv}, which is also integrated into \texttt{HiggsTools}. To use \texttt{HiggsBounds}, we input the couplings of new scalars with gauge bosons and fermions of the same flavour. However, neutral scalar flavour-violating decays into two fermions need to be provided. Namely, 
		\begin{align}
		\Gamma(\tilde{H}_k \rightarrow  f_\alpha \overline{f_\beta}) 
		& =
		n_c \sqrt{\left[m_k^2-(m_\alpha-m_\beta)^2\right] \left[m_k^2 - (m_\alpha+m_\beta)^2\right]} \nonumber
		\\
		& \times
		\frac{\left(\left|{a_k^{\alpha \beta}}\right|^2+\left|{b_k^{\alpha \beta}}\right|^2 \right)\left(m_k^2-m_\alpha^2 -m_\beta^2 \right)-4 m_\alpha m_\beta \text{Re}\left(a_k^{\alpha \beta} {b_k^{\alpha \beta}}^*\right) }{16 \pi m_k^3} \; ,
		\label{eq:neutralInto2fermions}
		\end{align}
		with $n_c$ being the colour factor and $\tilde{H} = (H, I)$. The couplings $a_k^{\alpha \beta}$ and $b_k^{\alpha \beta}$ are written in terms of the $\mathbf{N}_x$ matrices of Eq.~\eqref{eq:NMatrices} that control FCNCs, which in the alignment limit are given by
		\begin{align}
		a_H^{\alpha \beta} &= -\dfrac{\left(\mathbf{N}_x\right)_{\alpha \beta}}{v}  \; , \; b_H^{\alpha \beta} =  -\dfrac{\left(\mathbf{N}_x^\dagger\right)_{\alpha \beta}}{v} \; ; \nonumber \\
		a_I^{\alpha \beta} &=  n_{\pm} \dfrac{i\left(\mathbf{N}_x\right)_{\alpha \beta}}{v} \; , \; b_I^{\alpha \beta} = - n_{\pm} \dfrac{i\left(\mathbf{N}_x^\dagger\right)_{\alpha \beta}}{v} \; ,
		\label{eq:abCoupling}
		\end{align}
		where $x=d,u,e$ and $n_{\pm} = +1$ if $x=d,e$ and $n_{\pm} =-1$ if $x=u$. We also computed analytically the decay widths of a scalar into another scalar and a gauge boson following Ref.~\cite{Aguilar-Saavedra:2022rvy}, and the decay widths of $H$ and $I$ into $\gamma \gamma$ and $Z \gamma$ according to Refs.~\cite{Djouadi:2005gj,Posch:2010hx,Fontes:2014xva,Aguilar-Saavedra:2023tql}. 
		
		\item \textbf{Flavour observables:}
		All flavour constraints discussed at the beginning of this section (see Tables~\ref{tab:LFV}--\ref{tab:SemiLeptonicDecays}), were imposed at the 2$\sigma$ level. For the meson-antimeson observables, however, we allow our results to deviate by at most $10\%$ from the experimental values (see e.g. Refs.~\cite{Ferreira:2011xc,Ferreira:2019aps}). Overall, our methodology is similar to that of previous works -- see e.g. Refs.~\cite{Ferreira:2019aps,Botella:2014ska,Crivellin:2013wna}.
		
	\end{itemize}
	%
	
	\subsection{Results and discussion}
	\label{sec:resultsdiscussion}
	
	%
	The main results of our analysis are presented in Figs.~\ref{fig:phenoup} and~\ref{fig:phenodown} for the six up-decoupled and six down-decoupled models, respectively. The allowed parameter space for each model is highlighted by the black-hatched region, that satisfies all theoretical and experimental constraints discussed in Section~\ref{sec:num}. The coloured regions correspond to the parameter space that is excluded by the most restrictive constraints, i.e. the ones that shape the allowed region. The remaining flavour constraints are depicted using solid coloured contours, below which the region is excluded. To obtain these exclusion regions, we consider, besides theoretical, oblique parameters and experimental scalar constraints (see Section~\ref{sec:num}), a single flavour constraint, namely, either $Z \rightarrow b \overline{b}$ (grey), $\overline{B} \rightarrow X_s \gamma$ (bordeau), $\varepsilon_K$ (orange), $\Delta m_K$ (blue), $\Delta m_{B_d}$ (cyan), $\Delta m_{B_s}$ (green), $\Delta m_D$ (magenta) or $B_s \rightarrow \mu^- \mu^+$ (yellow). The results deserve several comments:
	\begin{itemize}
		
		\item \textbf{Scalar mass degeneracy:} Since we have three different BSM scalars, we should obtain allowed regions within the ($\tan \beta$,  $\{m_H, m_I,m_{H^\pm}\})$ parameter space. However, when $m_{H}=m_{I}=m_{H^\pm}$, certain constraints are automatically fulfilled, namely perturbative unitarity and quartic coupling perturbativity [see Eqs.~\eqref{Eq:PotParameters},~\eqref{eq:PertUnit}, and~\eqref{eq:perturbativityYuk}]. Moreover, in general, for almost degenerate BSM scalar masses the oblique parameters are in agreement with experimental data (see e.g. Ref.~\cite{Botella:2014ska}). The experimental constraints on the quark and lepton flavour observables, if not automatically satisfied in fermion decoupling scenarios, are relaxed when the scalar masses are quasi-degenerate. This is evident by looking at the theoretical expressions for the NP contributions to these processes in Eqs.~\eqref{eq:Gamma3BodyLFV_aux}, \eqref{eq:RadiativeLFV_aux}, \eqref{eq:M21NP} and~\eqref{eq:LeptonicWilson_aux}, where certain contributions are suppressed when $m_{H}\simeq m_{I} \simeq m_{H^\pm}$~\cite{Botella:2014ska,GonzalezFelipe:2014zjk,Correia:2019vbn,Camara:2020efq}. Our analysis is presented for the case where all three scalar masses are degenerate. We have numerically checked that this assumption is indeed a valid approximation and that the allowed parameter space regions for each model, in the case $m_{H}=m_{I}=m_{H^\pm}$, contains the scenarios where all three masses vary randomly (apart from extreme fine-tuning of masses~\cite{Botella:2014ska}).
		
		\item \textbf{Yukawa perturbativity bounds:} In all cases, $\tan \beta$ is bounded from above and below. This is due to the fact that, as a consequence of the flavour symmetries, each non-zero element of a mass matrix comes exclusively from one Yukawa matrix [see Eqs.~\eqref{eq:quarksmass2hdm} and~\eqref{eq:canonicalcharges}]. Thus, taking into account the Yukawa-coupling perturbativity criterium of Eq.~\eqref{eq:perturbativityYuk}, we obtain
		\begin{align}
		\tan^2 \beta \leq \dfrac{2\pi v^2}{| \left(\M_1^x\right)_{ij} |^2} - 1 ,
		\quad 
		\tan^2 \beta \geq 1/\left(\dfrac{2\pi v^2}{| \left(\M_2^x\right)_{ij} |^2} - 1 \right),
		\end{align}
		where, without loss of generality, we consider $\tan \beta$ within the first quadrant.  If a given mass matrix element originates from the first (second) Yukawa matrix, then $a=1$ ($a=2$) for $x=e,d,u$. Thus, $\tan \beta$ finds its upper and lower bounds determined by the maximum value of $| \left(\mathbf{M}_1^x\right)|$ and $| \left(\mathbf{M}_2^x\right)|$, respectively, which are identified in the figures by the left and right vertical grey-hatched regions.
		
		\item \textbf{Lepton sector constraints:} In Figs.~\ref{fig:phenoup} and~\ref{fig:phenodown}, except for $(5_1^d, \mathbf{P}_{123}{4}_{3}^u \mathbf{P}_{12})$ (top right plot in Fig.~\ref{fig:phenodown}), we select the lepton model $(5_1^e, 2_3^\nu)_{\text{NO}}$, as the conclusions are not affected by a different choice. This was numerically checked by noting  that changing the lepton model for a specific quark model leads to negligible changes in the final results. The only exception is the $(5_1^d, \mathbf{P}_{123}{4}_{3}^u \mathbf{P}_{12})$ case, where different choices for the lepton model have a significant impact. Thus, a more thorough analysis needs to be done and will be discussed in what follows.~\footnote{We also verified numerically that the BRs for the cLFV observables of Table~\ref{tab:LFV}, in the allowed parameter space regions (black-hatched) are, in general,  orders of magnitude below the sensitivities of future  experiments as MEG II~\cite{MEGII:2018kmf}, Belle II~\cite{Belle-II:2018jsg} and Mu3e~\cite{Blondel:2013ia}. Overall, the flavour constraints in the lepton sector play a negligible role in restricting the parameter space, contrary to what happens in the quark sector.}

		\begin{figure}[t!]
			\centering
			\includegraphics[scale=0.31]{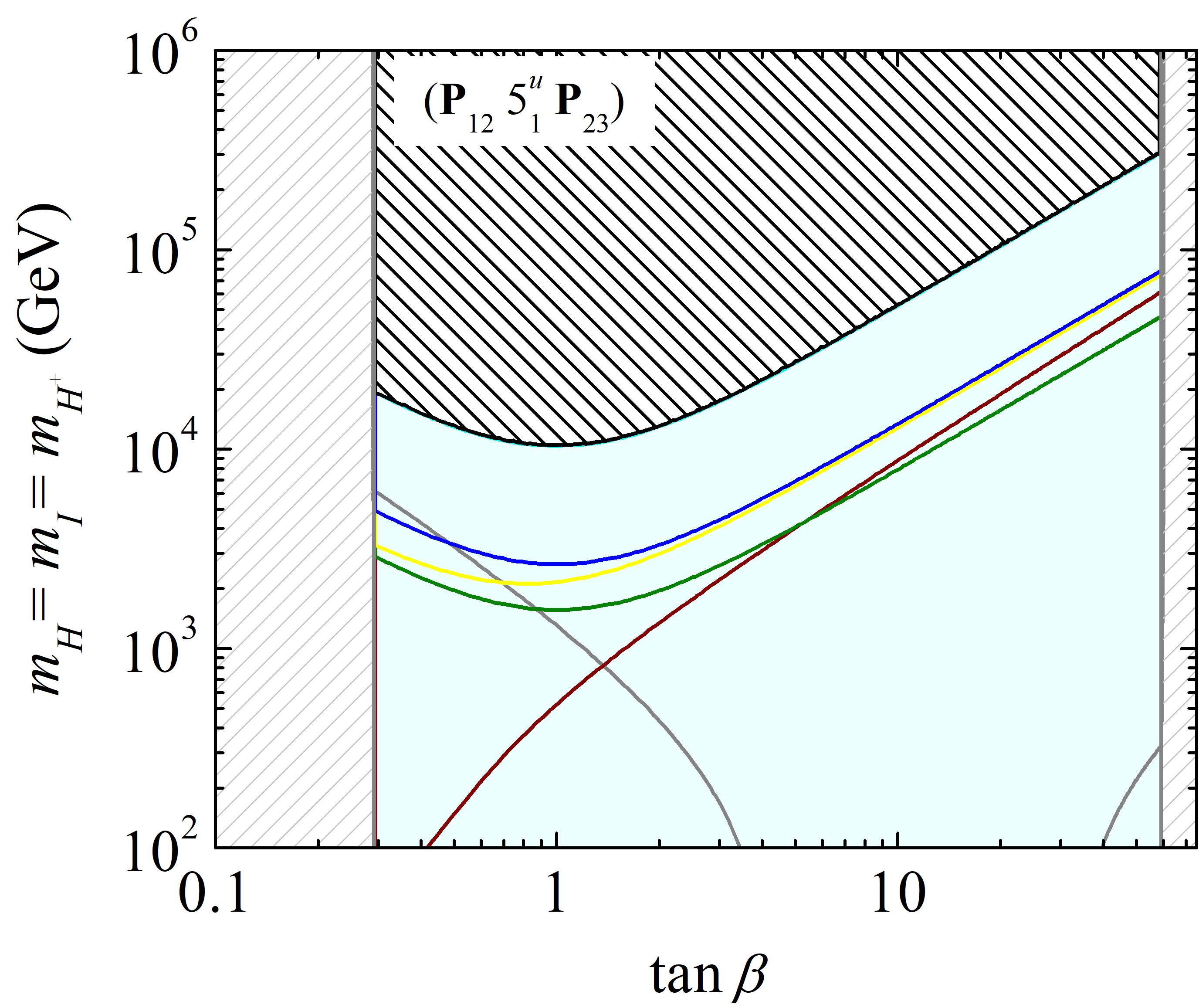} \hspace{+0.2cm} \includegraphics[scale=0.31]{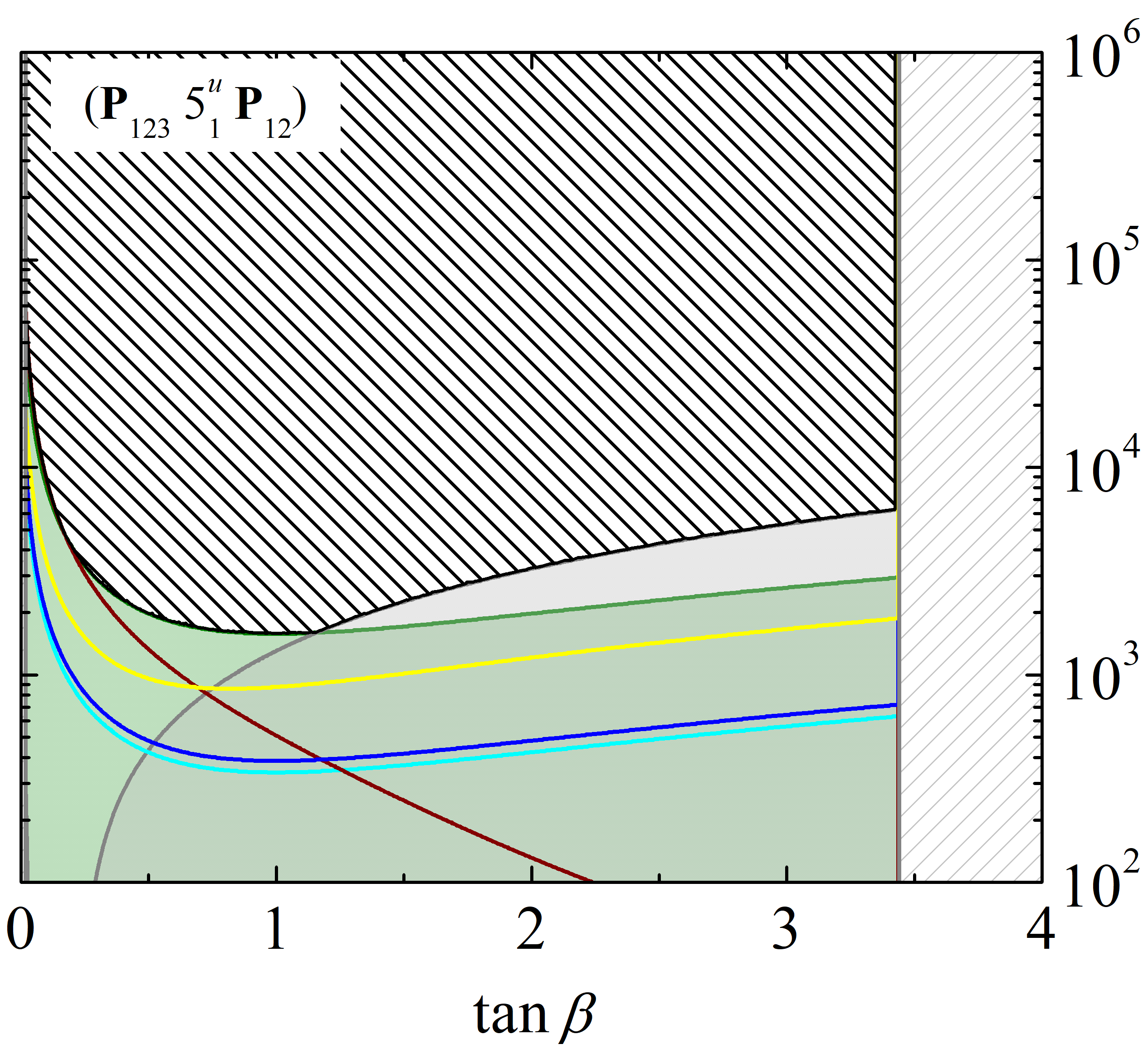}\\
			\includegraphics[scale=0.31]{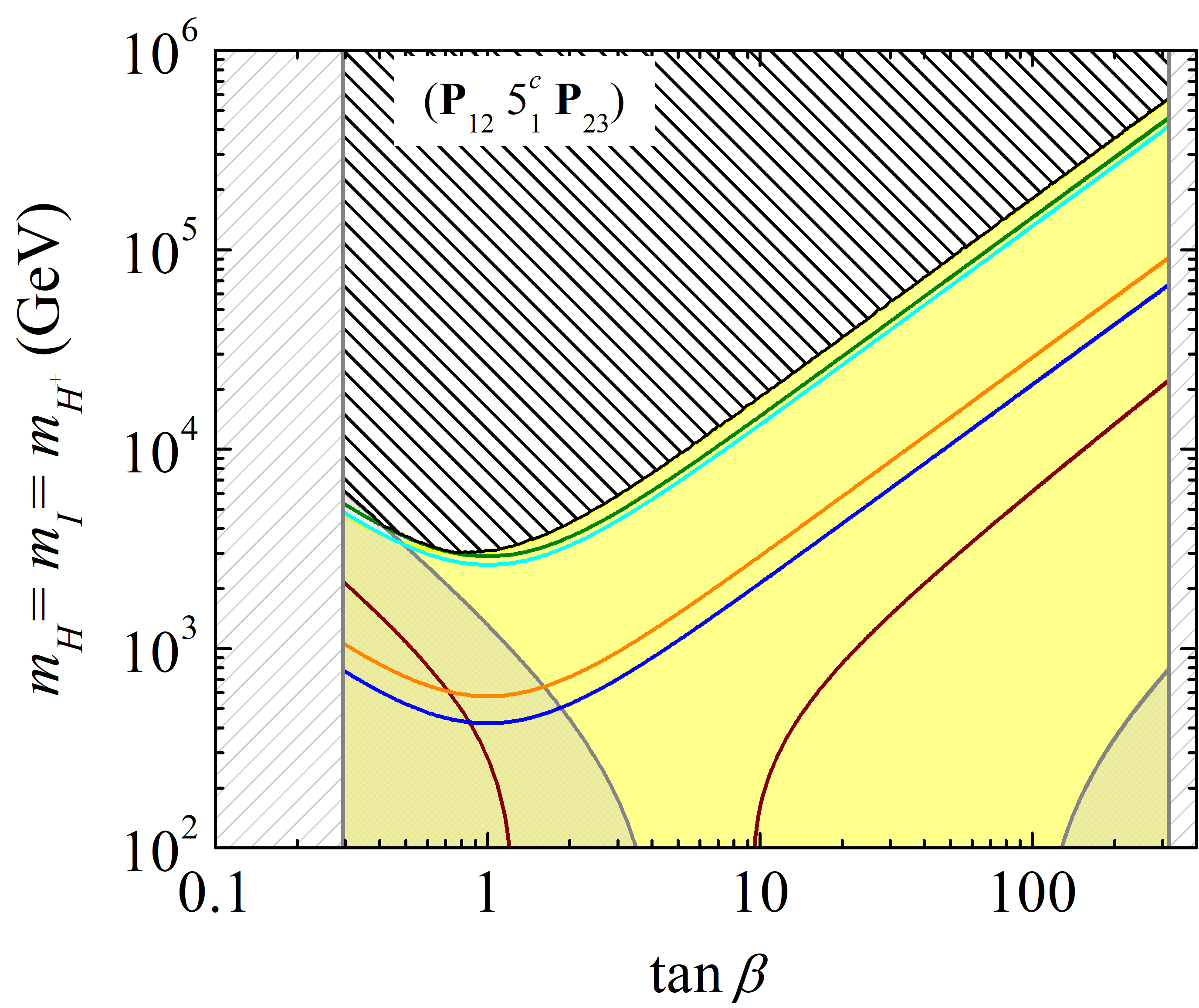} \hspace{+0.2cm} \includegraphics[scale=0.31]{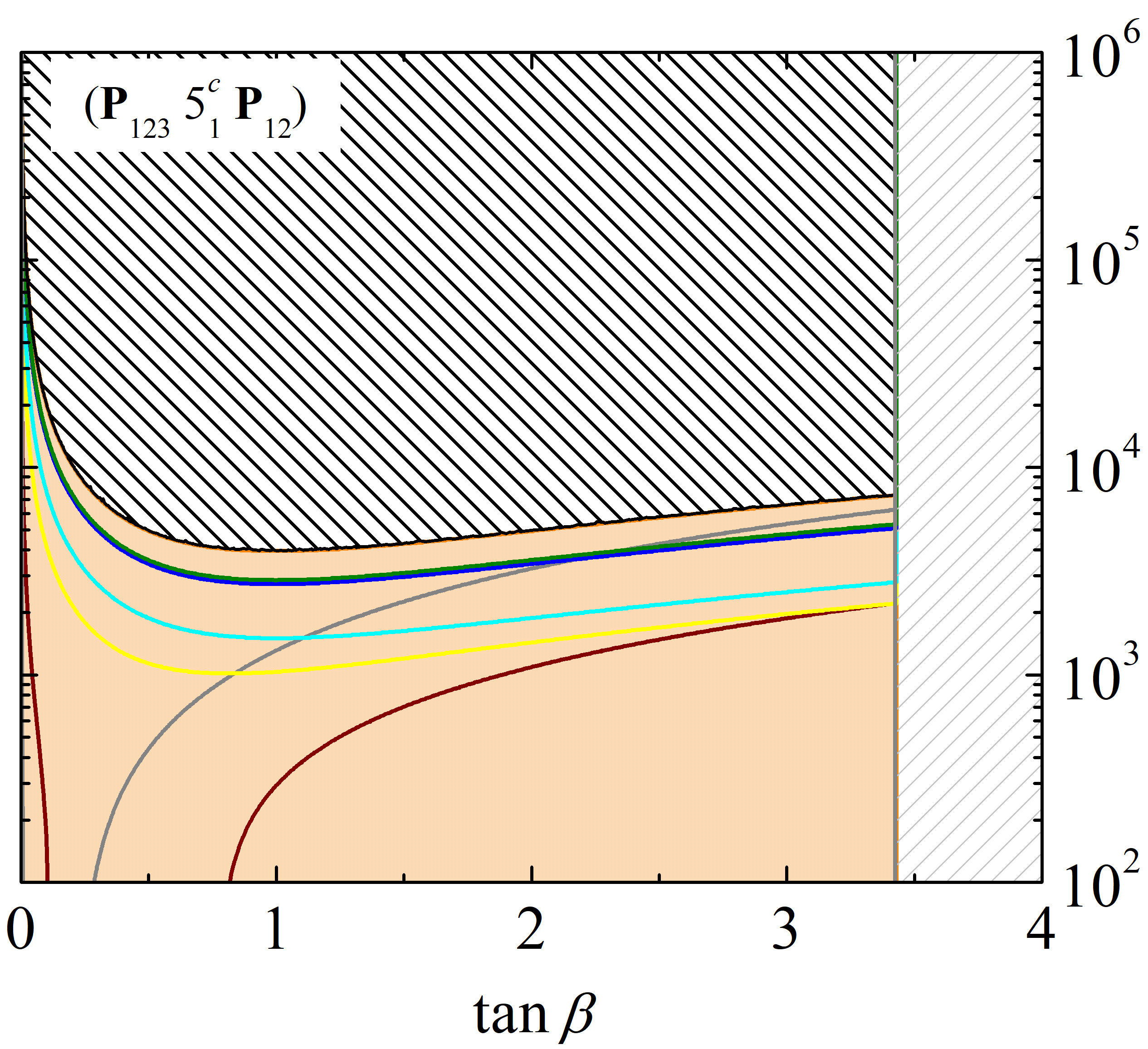}
			\\
			\includegraphics[scale=0.31]{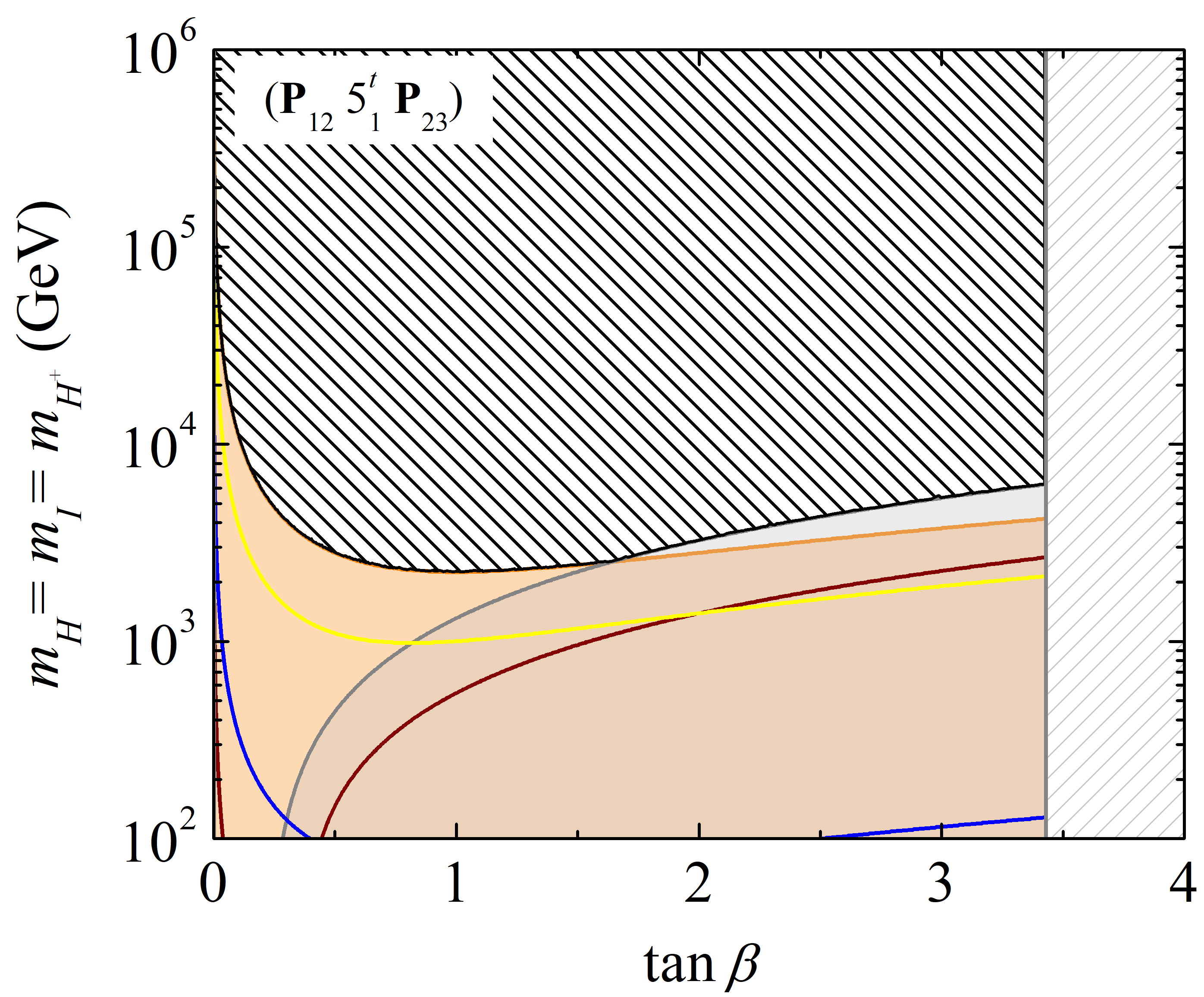} \hspace{+0.05cm} \includegraphics[scale=0.31]{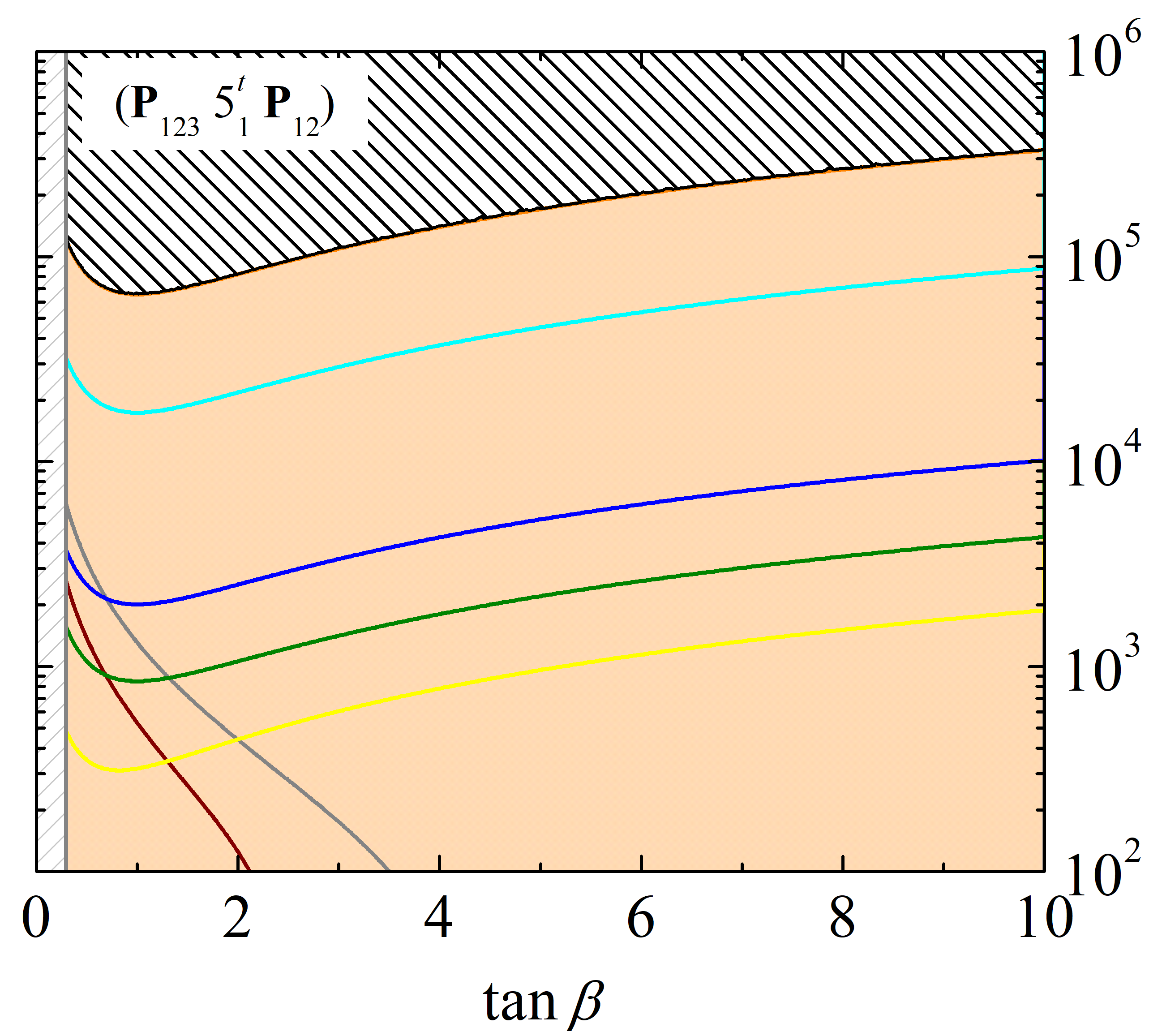} \\
			\vspace{+0.2cm}
			\includegraphics[scale=0.19]{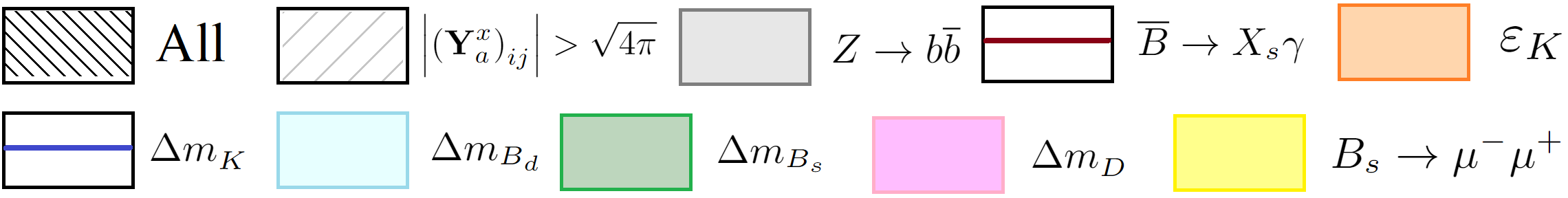}
			\caption{Allowed parameter space region satisfying all constraints of Section~\ref{sec:num} (black-hatched region) in the ($\tan \beta$, $m_H=m_I=m_{H^\pm}$) plane for cases $(4_{3}^d,\mathbf{P}_{12}{5}_{1}^u\mathbf{P}_{23})$ [$(4_{3}^d,\mathbf{P}_{123}5_{1}^u \mathbf{P}_{12})$] presented by the left [right] plots. The top (middle) [bottom] plots show cases featuring $u$($c$)[$t$]-decoupled state. The lepton sector is $(5_1^e,2_3^\nu)_{\text{NO}}$. The colour-shaded regions are excluded by the indicated flavour observable shaping the allowed black-hatched region. The exclusion regions for the other processes are below the respective solid-coloured contours (see text for details).}
			\label{fig:phenoup} 
		\end{figure} 
		\begin{figure}[t!]
			\centering
			\includegraphics[scale=0.31]{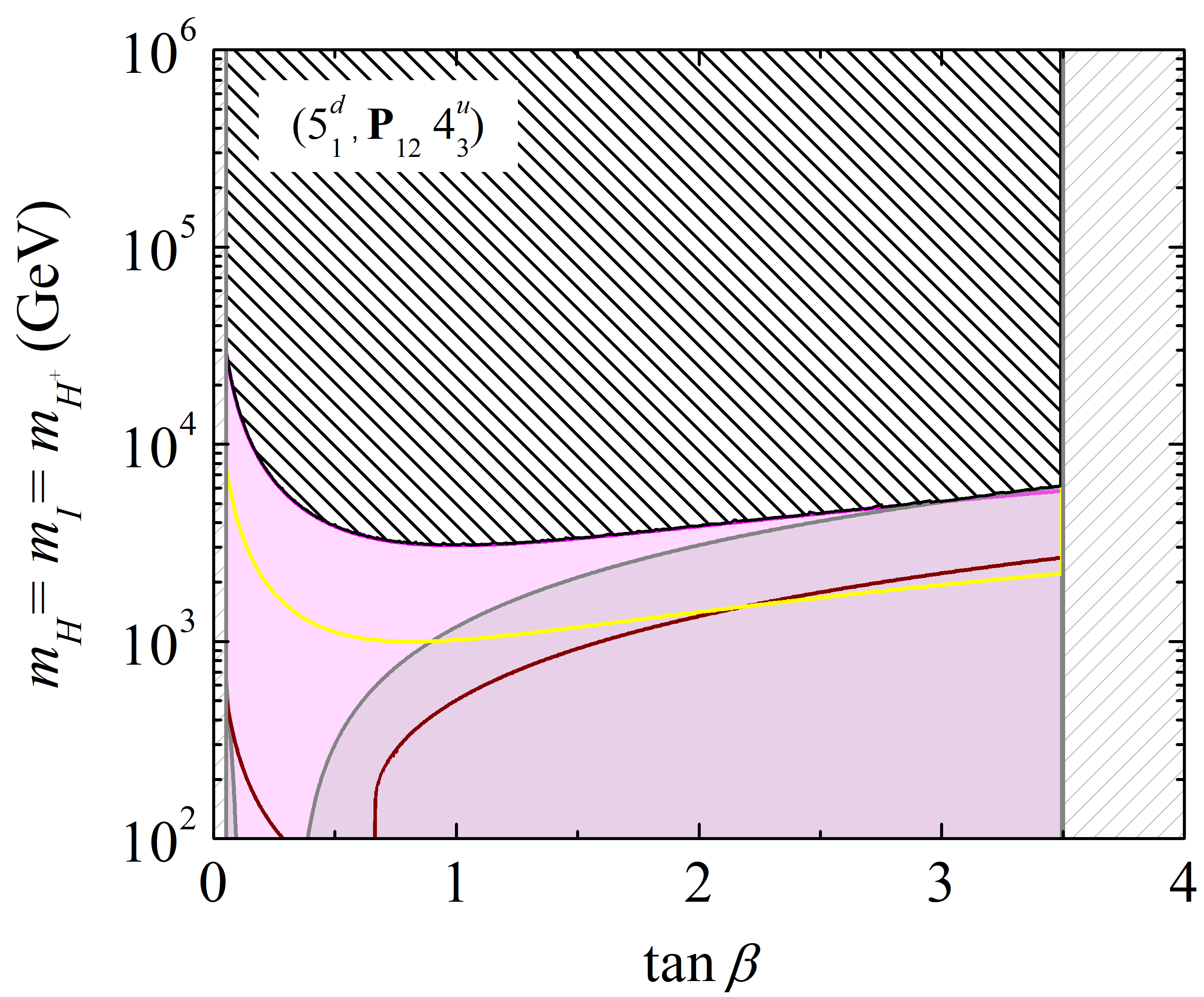} \hspace{+0.3cm} \includegraphics[scale=0.098]{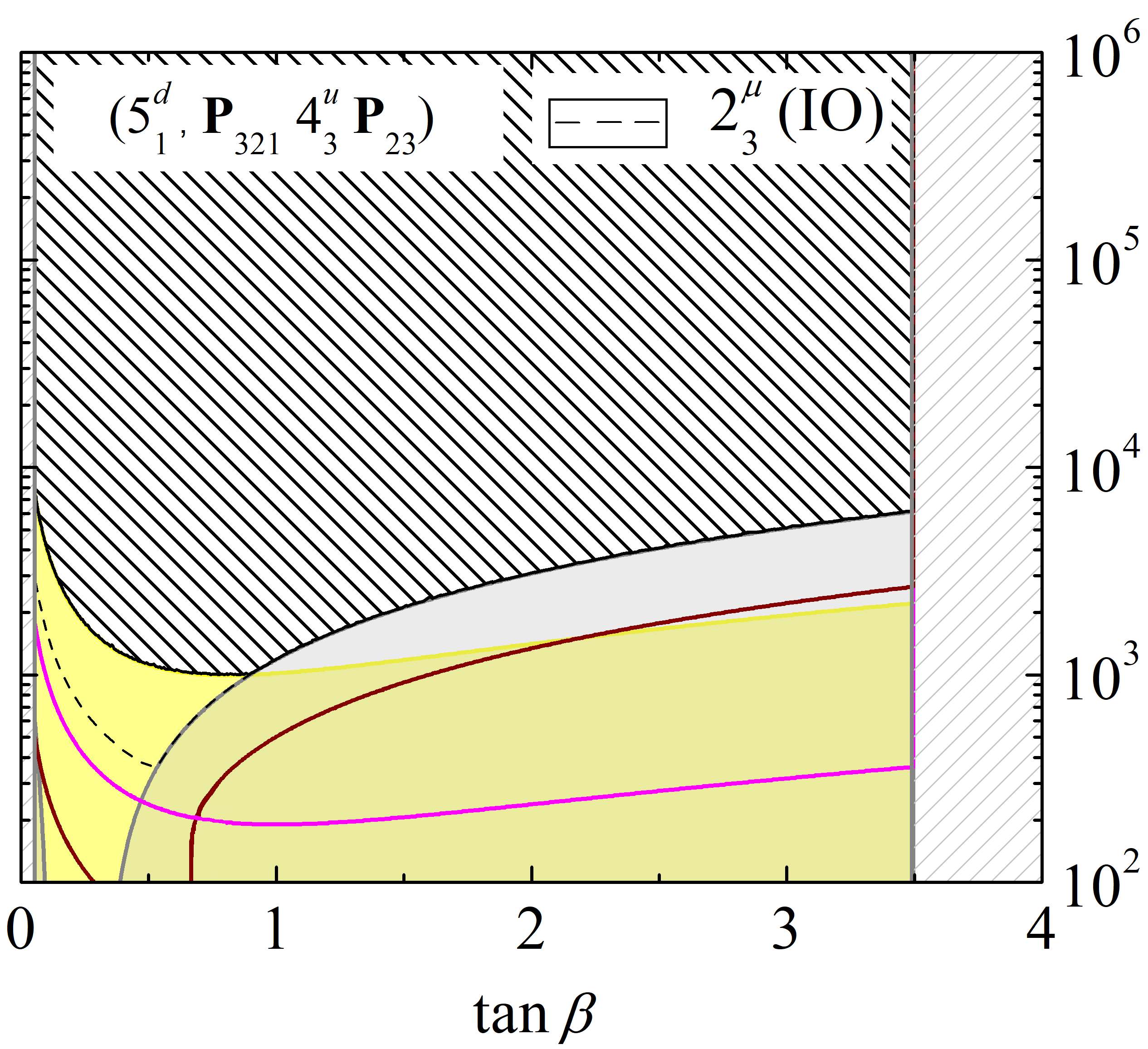} \\
			\includegraphics[scale=0.31]{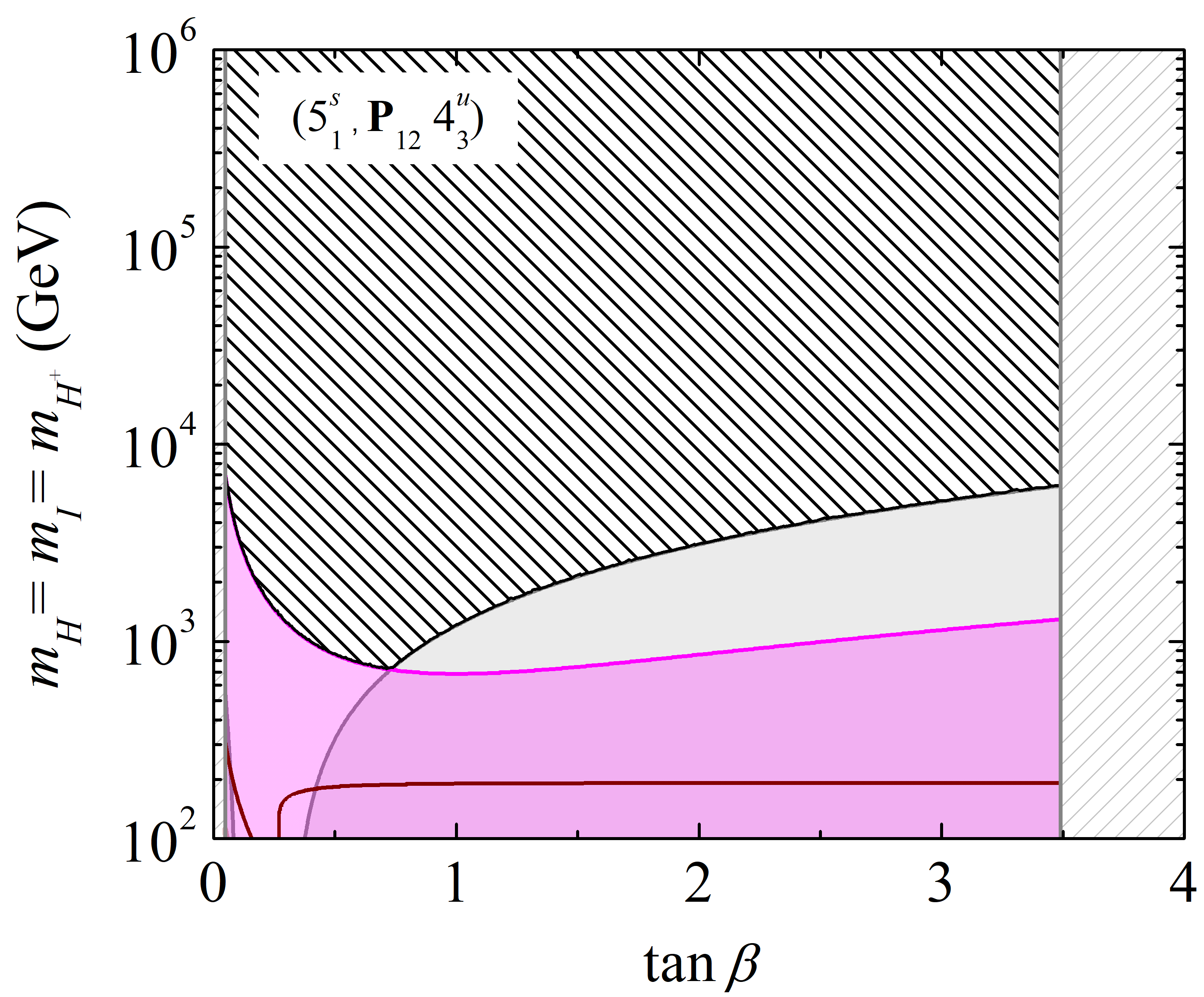} \hspace{+0.3cm} \includegraphics[scale=0.31]{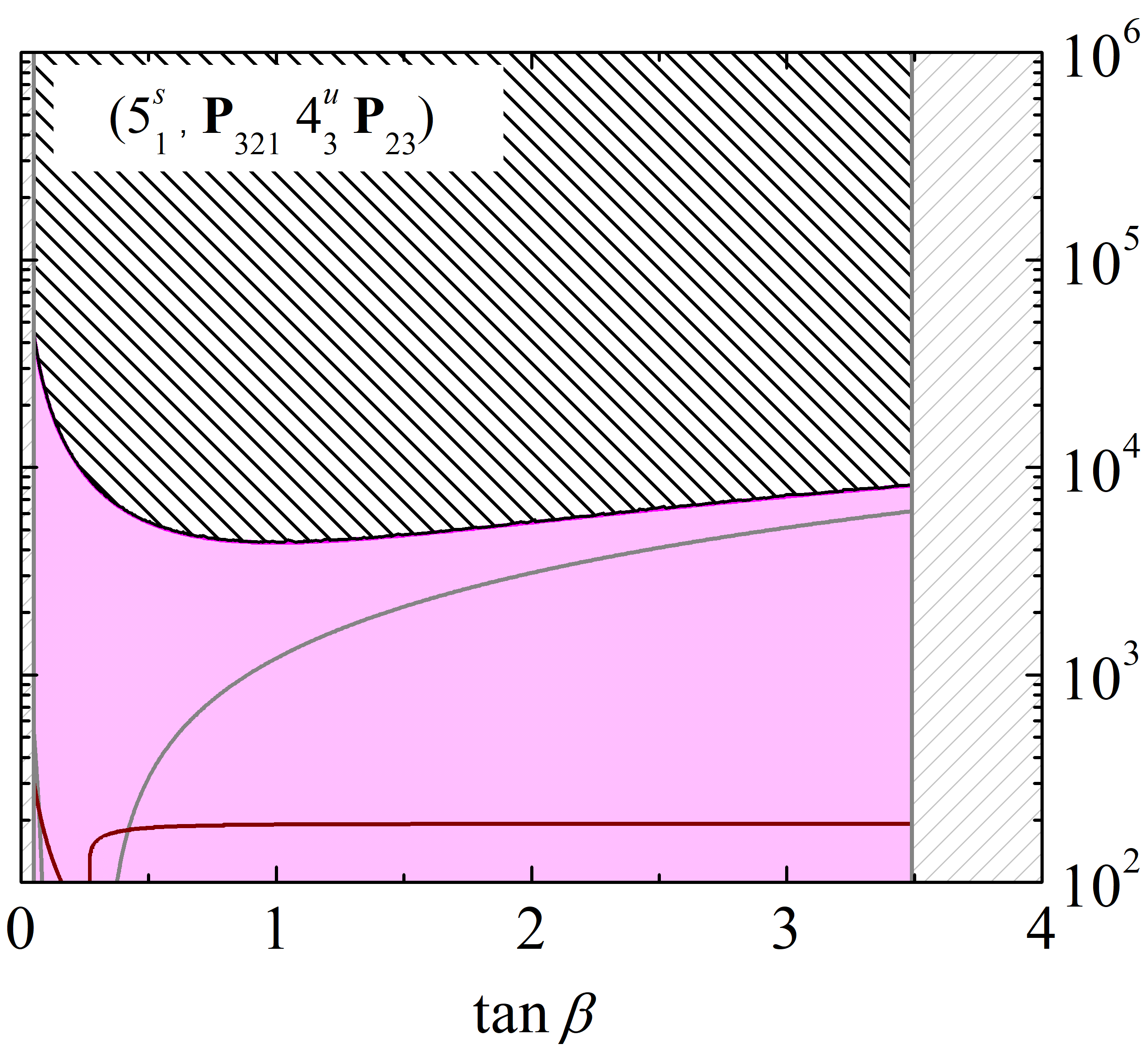}
			\\
			\includegraphics[scale=0.31]{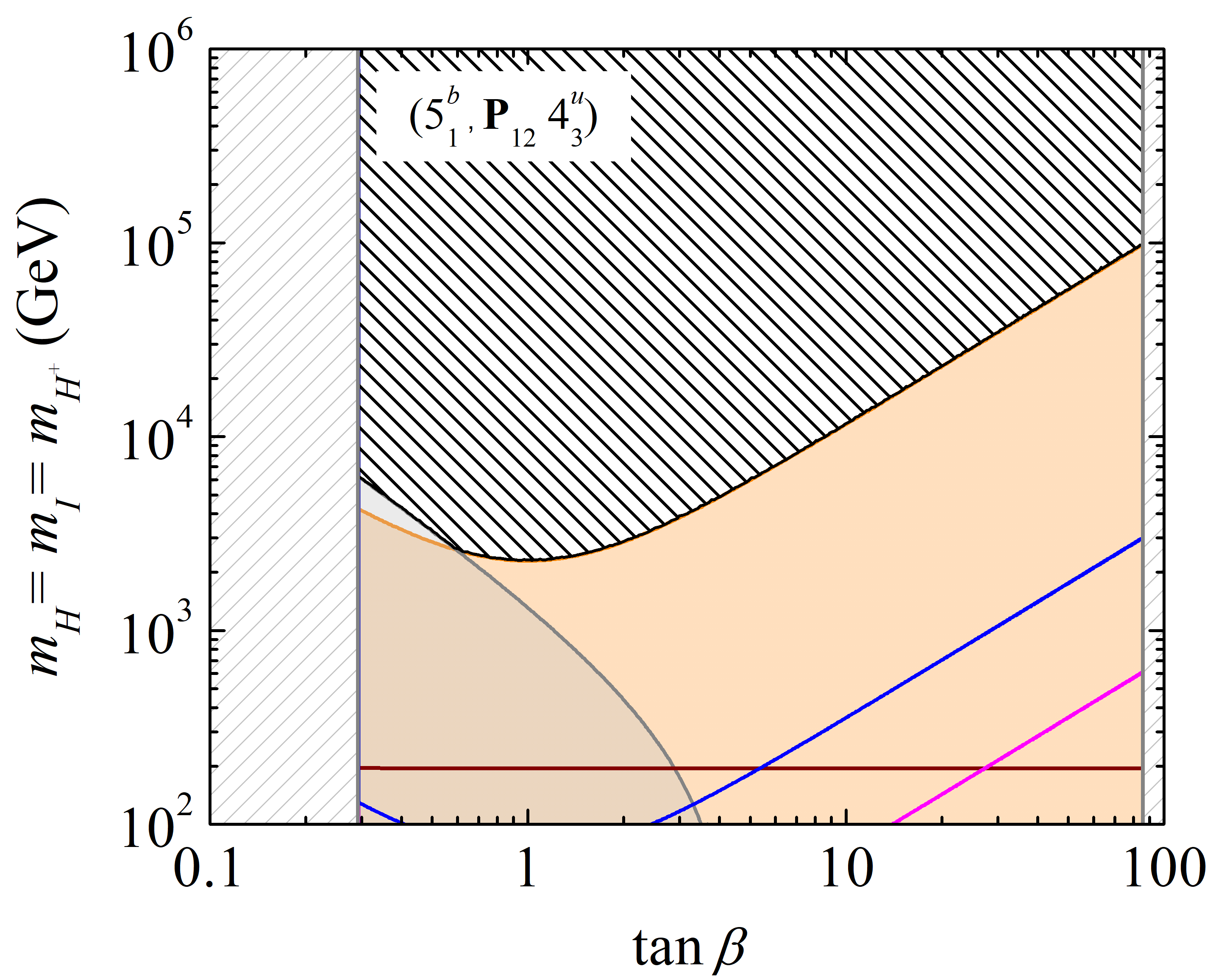} \includegraphics[scale=0.31]{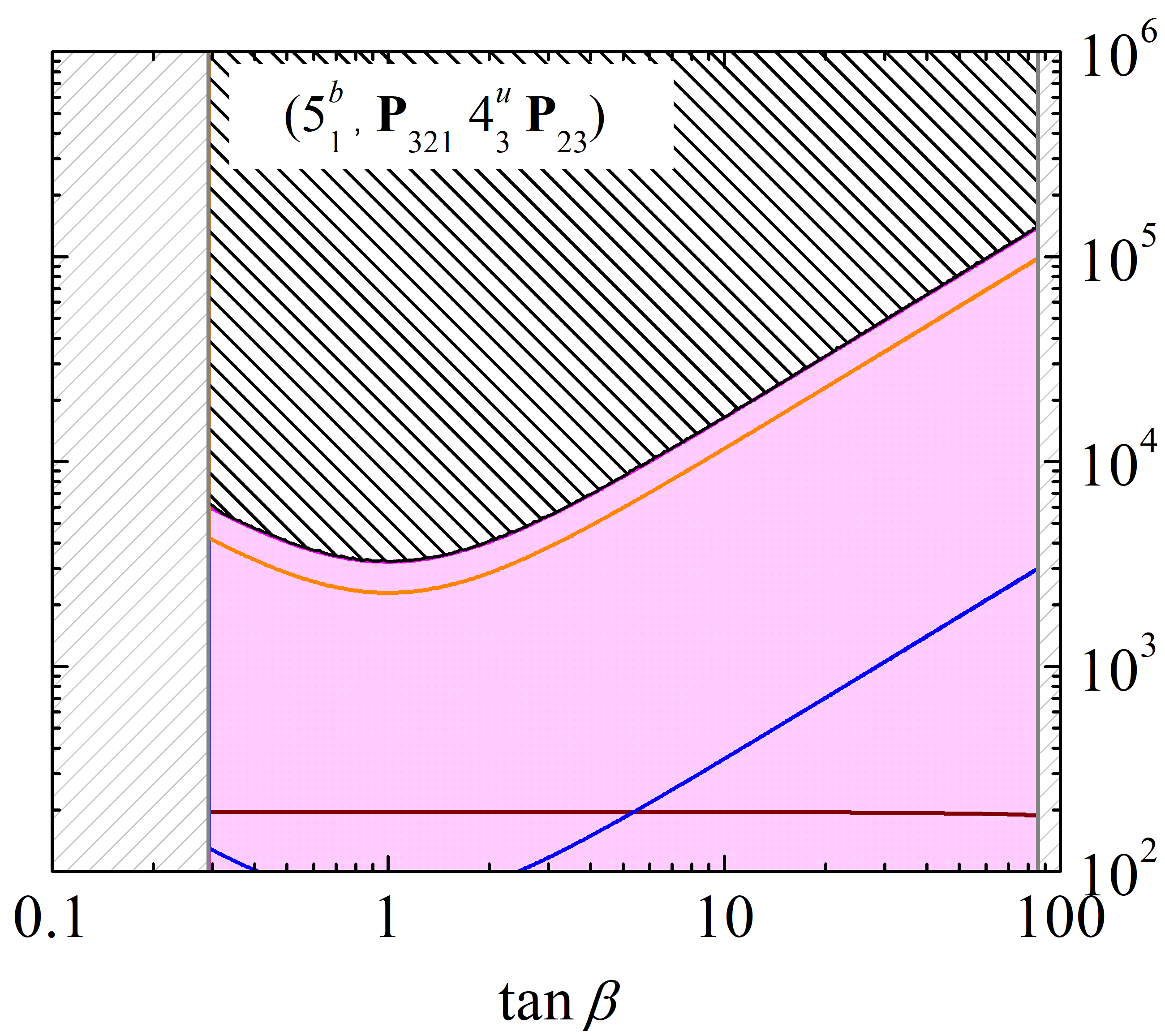} \\
			\vspace{+0.2cm}
			\includegraphics[scale=0.19]{CAPTION.png}
			\caption{Allowed parameter space region in the ($\tan \beta$, $m_H=m_I=m_{H^\pm}$) plane for cases $(5_{1}^d,\mathbf{P}_{12}{4}_{3}^u)$ [$(5_{1}^d,\mathbf{P}_{321}4_{3}^u \mathbf{P}_{23})$] presented by the left [right] plots. The top (middle) [bottom] plots show cases featuring $d$($s$)[$b$]-decoupled state. The lepton sector is $(5_1^e,2_3^\nu)_{\text{NO}}$, except for the case $(5_{1}^d,\mathbf{P}_{321}4_{3}^u \mathbf{P}_{23})$ (top right plot) where via a black-dashed contour we indicate the allowed parameter space region for $(5_1^\mu,2_3^\nu)_{\text{IO}}$ (see text for details). The presented colour scheme is the same as in Fig.~\ref{fig:phenoup}.}
			\label{fig:phenodown} 
		\end{figure}

		\item \textbf{Most restrictive constraints:} The plots in Figs.~\ref{fig:phenoup} and~\ref{fig:phenodown} show that not all constraints shape the allowed region (black-hatched). For the up-sector decoupled models (see Fig.~\ref{fig:phenoup}), the most important constraints are the $\varepsilon_K$ parameter (orange) and the mass differences of the meson-antimeson systems $B_d$ (cyan) and $B_s$ (green). These are generally not automatically fulfilled for up-sector decoupled models (see Table~\ref{tab:mesonOscillations}). One would also expect $\Delta m_K$ (blue) to be a crucial constraint for these scenarios. However, for this  observable we consider that experiments only constrain the NP contribution and not the combined SM and NP, as in the case of other meson-antimeson observables. A similar conclusion can be drawn with respect to $\Delta m_D$ (magenta), which has the additional property of being automatically satisfied for $u$ and $c$ decoupled models (see Table~\ref{tab:mesonOscillations}). On the other hand, the allowed region for $\mathbf{P}_{12} 5_1^c \mathbf{P}_{23}$ (middle left plot) is not shaped by meson-antimeson observables, but by the very restrictive constraint of $B_s \rightarrow \mu^+ \mu^-$. Note that, in addition, the parameter space of the two quark models $\mathbf{P}_{123} 5_1^u \mathbf{P}_{12}$ (top right plot) and $\mathbf{P}_{12} 5_1^t \mathbf{P}_{23}$ (bottom left plot),  is mildly constrained by $Z \rightarrow b \overline{b}$ (grey) for $\tan \beta \gtrsim 1.1$ and $\tan \beta \gtrsim 1.6$, respectively.
		
		For the down-sector decoupled models (see Fig.~\ref{fig:phenodown}), $\Delta m_D$ (magenta) is, in most cases, the defining constraint. This differs significantly from what we have found in up-sector decoupled models. Indeed, the $\Delta m_{B_{s,d}}$ and $\Delta m_K$ constraints are now automatically fulfilled according to the decoupled states shown in Table~\ref{tab:mesonOscillations}. An interesting model is $(5_1^d, \mathbf{P}_{321} 4_3^u \mathbf{P}_{23})$ (top right plot), for which the allowed region is determined by the $B_s \rightarrow \mu^- \mu^+$ and $Z \rightarrow b \overline{b}$ constraints. In this case, the lepton model can change the allowed parameter space. If we were to consider $(5_1^\mu,2_3^\nu)_{\text{IO}}$ instead of $(5_1^e,2_3^\nu)_{\text{NO}}$, the parameter space would enlarge since $B_s \rightarrow \mu^- \mu^+$ would be relaxed. It is important to mention that, in this scenario, the lower mass bound for the remaining lepton models would lie between the black-solid $[(5_1^e,2_3^\nu)_{\text{NO}}]$ and black-dashed  $[(5_1^\mu,2_3^\nu)_{\text{IO}}]$ lines. Note also that for $(5_1^b, \mathbf{P}_{12} 4_3^u)$ (bottom left plot), the allowed parameter-space region is defined by the $\varepsilon_K$ constraint which is not automatically satisfied in $b$-decoupled models (see Table~\ref{tab:mesonOscillations}). 
		
		\item \textbf{BSM scalar mass scale:} One of the most interesting features to consider is the  allowed ranges for the BSM scalar masses. Before examining any specific model, it is noteworthy that the flavour constraints coming from the up-quark sector are less restrictive than those of the down sector. Consequently, down-decoupled models, where some down sector constraints are automatically satisfied, generally allow for lower BSM scalar masses and larger parameter spaces compared to the up-decoupled scenarios. With this in mind, the most restricted case is found for the up-decoupled model $(\mathbf{P}_{123} 5_1^t \mathbf{P}_{12})$ (bottom right plot in Fig.~\ref{fig:phenoup}), for which the lowest scalar mass is around $70$~TeV. This outcome is somehow expected since,  in this model, $t$ is decoupled and none of the most restrictive constraints are automatically satisfied (see Tables~\ref{tab:mesonOscillations} and~\ref{tab:LeptonicDecays}). On the other hand, the lowest scalar mass for an up-decoupled model is approximately $2$~TeV for $\mathbf{P}_{123} 5_1^u \mathbf{P}_{12}$ (top right plot in Fig.~\ref{fig:phenoup}). In this case, the  $\varepsilon_K$ and $\Delta m_D^{\text{NP}}$ constraints are automatically fulfilled. In fact, $u$-decoupled models feature the largest number of most restrictive constraints that can be automatically satisfied.
		
		As for down-decoupled models, there are two specially interesting cases shown in Fig.~\ref{fig:phenodown}, namely $(5_1^s , \mathbf{P}_{12} 4_3^u)$ (middle left plot) and $(5_1^d , \mathbf{P}_{321} 4_3^u \mathbf{P}_{23})$ (top-right plot) that allow masses down to  $\sim 700$~GeV and $\sim 300$ GeV, respectively.  The latter is within the range of direct searches at the LHC. The lower mass bound for all remaining models lie between $3$ to $10$~TeV. In all cases, these limits are attained for $\tan \beta \sim 1$, due to the fact that, for such $\tan\beta$ values, the contributions of the Yukawa matrices $\mathbf{Y}_{1,2}^x$ ($x=d,u,e$) to FCNCs are on equal footing.
		
	\end{itemize}

	\section{Concluding remarks}
	\label{sec:concl}
	
	In this work we have tackled the flavour puzzle for the quark and lepton sectors, within the 2HDM supplemented with horizontal Abelian symmetries and neutrino mass generation via effective Weinberg operators. We performed a systematic analysis to identify the minimal 2HDMs, in which the symmetries impose the most restrictive quark and lepton flavour patterns, while being compatible with the observed fermion masses, mixing angles and CP-violating phases. We found four minimal models for quarks, with ten independent parameters matching the six quark masses, and four CKM parameters. For the lepton sector, we identified three minimal and predictive flavour models that feature a total of ten parameters compared to twelve observables, namely six lepton masses, and six parameters of the PMNS matrix, including two Majorana phases. A detailed study of the lepton sector revealed that cases $2_{3,7}^\mu$ (NO) [$2_{3,7}^\tau$ (NO)] and $2_{10}^\mu$ (IO) [$2_{10}^\tau$ (NO)] have a preference for the lower [upper] octant of the $\theta_{23}$ atmospheric mixing angle, as shown in Fig.~\ref{fig:predictionst23delta}. As far as the lightest neutrino mass predictions are concerned, a few models are in tension with cosmological Planck data, establishing lower bounds on the $m_{\text{lightest}}$ scale, namely $2_{3,7}^{\mu,\tau}$ for NO (Fig.~\ref{fig:predictionsmlightestdeltaNO}) and $2_{10}^{\mu,\tau}$ for IO (Fig.~\ref{fig:predictionsmlightestdeltaIO}). It is well known that IO neutrino masses can be fully tested by experiments that search for $0\nu\beta\beta$ decay process (see Fig.~\ref{fig:predictionsmlightestmbetabetaIO}). Remarkably, we identified four NO scenarios, namely $2_{3,7}^{\mu,\tau}$, that can be testable in such experiments as shown in Fig.~\ref{fig:predictionsmlightestmbetabetaNO}.
	
	We thoroughly investigated the phenomenology of each model with special emphasis on their flavour predictions. Our results are gathered in Figs.~\ref{fig:phenoup} and~\ref{fig:phenodown}. In our analysis we took into account all relevant theoretical, EWPO, and scalar sector constraints, and also stringent quark flavour observables such as $\overline{B} \rightarrow X_s \gamma$, $B_s \rightarrow \mu^- \mu^+$ and neutral meson oscillations. cLFV processes like $e_\alpha^{-} \rightarrow e_\beta^{-} e_\gamma^{+} e_\delta^{-}$ and $e_\alpha \rightarrow e_\beta \gamma$ were also considered. We showed that constraints in the quark sector are the most relevant and, indeed, shape the allowed parameter space for each model. In some cases, the Abelian flavour symmetries provide a natural framework to automatically suppress tree-level FCNCs. These symmetries lead to decoupled fermion states in the mass matrices, which impose restricted patterns in the $\mathbf{N}_x$ ($x=d,u,e$) matrices. In particular, zero off-diagonal $\mathbf{N}_x$ entries, that otherwise contribute to the aforementioned flavour processes (see Tables~\ref{tab:LFV}, \ref{tab:mesonOscillations} and~\ref{tab:LeptonicDecays}), are heavily constrained by experiment. 
	Thus, NP contributions are controlled or even vanishing. This feature leads to models with extended parameter spaces in the ($\tan \beta$,  $\{m_H, m_I,m_{H^\pm}\})$ planes, specifically for down-type quark decoupled cases, where BSM scalar masses can be as low as a few hundreds GeV. This is the case for the models $[(5_1^e,2_3^\nu)_{\text{NO}},(5_{1}^d,\mathbf{P}_{321}4_{3}^u \mathbf{P}_{23})]$, $[(5_1^\mu,2_3^\nu)_{\text{IO}},(5_{1}^d,\mathbf{P}_{321}4_{3}^u \mathbf{P}_{23})]$ and $[(5_1^e,2_3^\nu)_{\text{NO}}, (5_{1}^s,\mathbf{P}_{12}{4}_{3}^u)]$, as can be seen in Fig.~\ref{fig:phenodown}. 
	
	To conclude, Abelian flavour symmetries in the 2HDM stand out as one of the simplest approaches to effectively address the flavour puzzle, allowing to construct minimal quark and lepton models that are predictive with respect to low-energy masses, mixing and CP-violation parameters. Tree-level FCNCs that arise in our 2HDMs can be naturally controlled by the imposed Abelian symmetries, thus making them a simple framework for theoretical scenarios compatible with highly-constraining experimental observations. In some cases, new scalars with masses below the TeV scale are allowed, being within the reach of current experiments such as the LHC and testable at future facilities.
	
	\begin{acknowledgments}
		This work is supported by Fundação para a Ciência e a Tecnologia (FCT, Portugal) through the projects CFTP-FCT Unit UIDB/00777/2020 and UIDP/00777/2020, CERN/FIS-PAR/0019/2021, which are partially funded through POCTI (FEDER), COMPETE, QREN and EU. The work of H.B.C. is supported by the PhD FCT grant 2021.06340.BD.
	\end{acknowledgments}
	
	\appendix
	
	\section{Scalar sector of U(1) symmetric 2HDM}
	\label{sec:scalar}
	
	The maximally-restrictive textures for quarks and leptons compatible with masses, mixing and CP violation data, realisable by Abelian symmetries, are presented in Section~\ref{sec:texturesrealisable}. The minimal scalar content required to implement these textures are two Higgs doublets,
	\begin{equation}
	\Phi_a = \begin{pmatrix} \phi^+_a \\ \phi^0_a \end{pmatrix}
	= \frac{1}{\sqrt{2}}\begin{pmatrix} \sqrt{2}\phi^+_a \\ v_a e^{i \varphi_a} + \rho_a + i \eta_a\end{pmatrix}
	\; , \; 
	a = 1,2 \; ,
	\end{equation}    
	where $v_a$ are the VEVs of the $\Phi_a$ neutral component and only the phase difference $\varphi=\varphi_2-\varphi_1$ is relevant. The above fields transform under the Abelian symmetries shown in Table~\ref{tab:charges}, leading to the scalar potential~\cite{Branco:2011iw}:
	\begin{align}
	V(\Phi_1,\Phi_2) &= \mu_{11}^2 \Phi_1^\dagger \Phi_1 + \mu_{22}^2 \Phi_2^\dagger \Phi_2 +  \mu_{12}^2 \left( \Phi_1^\dagger \Phi_2 + \Phi_2^\dagger \Phi_1 \right) \nonumber \\  
	&+ \frac{\lambda_1}{2} \left(\Phi_1^\dagger \Phi_1\right)^2 + \frac{\lambda_2}{2} \left(\Phi_2^\dagger \Phi_2\right)^2 + \lambda_3 \left(\Phi_1^\dagger \Phi_1\right)\left(\Phi_2^\dagger \Phi_2\right) + \lambda_4 \left(\Phi_1^\dagger \Phi_2\right)\left(\Phi_2^\dagger \Phi_1\right) \; .
	\label{eq:Vpotential2HDM}
	\end{align}
	Note that, in order to avoid a massless Goldstone boson~(GB), which arises once the Abelian flavour symmetries are spontaneously broken by the Higgs VEVs, we add a soft-breaking parameter $\mu_{12}^2$, which can be taken real. 
	
	The minimisation of the scalar potential leads to two conditions,
	\begin{align}
	\mu_{11}^2 = -\frac{1}{2} \left[\lambda_1 v_1^2 + \left(\lambda_3 + \lambda_4\right)v_2^2 \pm 2 \mu_{12}^2 \frac{v_2}{v_1} \right] \; , \; \mu_{22}^2 = -\frac{1}{2} \left[\lambda_2 v_2^2 + \left(\lambda_3 + \lambda_4\right)v_1^2 \pm 2 \mu_{12}^2 \frac{v_1}{v_2} \right] \; ,
	\label{eq:Min_potential}
	\end{align}
	where the upper (lower) sign corresponds to $\varphi = 0\left(\pi\right)$, hence the vacuum is CP-conserving. In what follows, we set without loss of generality $\varphi = 0$. 
	
	Let us now study the physical scalar mass spectrum. The mass matrix for the charged scalars, in the $( \phi_1^\pm, \phi_2^\pm)$ basis, is given by
	\begin{equation}
	\mathbf{M}^2_{\pm} = -\frac{v_1 v_2 \lambda_4 + 2 \mu_{12}^2}{2}
	\begin{pmatrix}
	\dfrac{v_2}{v_1} & - 1 \\
	- 1 & \dfrac{v_1}{v_2}
	\end{pmatrix} \; ,
	\end{equation}
	which is diagonalised by the rotation matrix
	\begin{equation}
	\textbf{R} = 
	\begin{pmatrix}
	c_\beta &  s_\beta \\
	-s_\beta & c_\beta
	\end{pmatrix}  \rightarrow \begin{pmatrix} H_1 \\ H_2\end{pmatrix} = \textbf{R} \begin{pmatrix} \Phi_1 \\ \Phi_2\end{pmatrix} \; , \; c_\beta \equiv \cos \beta = \frac{v_1}{v}  \; , \; s_\beta \equiv \sin \beta = \frac{v_2}{v}  \; ,
	\label{eq:Higgsbasis}
	\end{equation}
	where $v = \sqrt{v_1^2 + v_2^2} \simeq 246 \; \text{GeV}$. Besides the~$G^\pm$ GB, we obtain a physical charged Higgs state~$H^\pm$ with mass
	\begin{equation}
	m^2_{H^\pm} = -\frac{v^2 \lambda_4}{2} - \frac{2\mu_{12}^2}{\sin\left(2\beta\right)} \; .
	\label{eq:chargedhiggsmass}
	\end{equation}
	The above transformation brings the weak doublets $\Phi_{1,2}$ into the Higgs basis with $H_{1,2}$ given by~\cite{Branco:2011iw}
	\begin{equation}
	H_1 = \begin{pmatrix} G^+ \\ H_1^0 = \dfrac{v + H^0 + i G^0}{\sqrt{2}} \end{pmatrix}\; , \;
	H_2 = \begin{pmatrix} H^+ \\ H_2^0 = \dfrac{R + i I}{\sqrt{2}} \end{pmatrix} \; ,
	\end{equation}
	where $\left<H_1^0\right>=v/\sqrt{2}$ and $\left<H_2^0\right> = 0$. Note that, the $\rho_a$ and $\eta_a$ ($a=1,2$) components of the doublets do not mix. Thus, the CP-even neutral scalar mass matrix in the $(\rho_1, \rho_2)$ basis is
	\begin{equation}
	\mathbf{M}^2_\text{CP-even} = 
	\begin{pmatrix}
	\mathbf{M}_{11}^2 & \mathbf{M}_{12}^2\\
	\mathbf{M}_{12}^2 & \mathbf{M}_{22}^2
	\end{pmatrix}  = 
	\begin{pmatrix}
	v_1^2 \lambda_1 - \dfrac{v_2}{v_1} \mu_{12}^2 &  v_1 v_2 \left(\lambda_3 + \lambda_4\right) + \mu_{12}^2 \\
	v_1 v_2 \left(\lambda_3 + \lambda_4\right) + \mu_{12}^2 & v_2^2 \lambda_2 - \dfrac{v_1}{v_2} \mu_{12}^2
	\end{pmatrix}  \; ,
	\end{equation}
	being diagonalised through the rotation
	\begin{equation}
	\begin{pmatrix} H \\ h \end{pmatrix} = 
	\begin{pmatrix}
	c_\alpha & s_\alpha \\
	-s_\alpha & c_\alpha
	\end{pmatrix}
	\begin{pmatrix} \rho_1 \\ \rho_2 \end{pmatrix} =  
	\begin{pmatrix}
	c_{\beta-\alpha} & - s_{\beta-\alpha} \\
	s_{\beta-\alpha} & c_{\beta-\alpha}
	\end{pmatrix}
	\begin{pmatrix} H^0 \\ R \end{pmatrix} \; , \; \tan\left(2 \alpha \right) = \frac{v^2 \left(\lambda_3 + \lambda_4\right)\sin\left(2\beta\right) + 2 \mu_{12}^2}{v^2\left(\lambda_1 c_\beta^2 - \lambda_2 s_\beta^2\right) + 2 \mu_{12}^2\cot\left(2\beta\right)} \; .
	\label{eq:tan_alpha}
	\end{equation}
	This procedure results in the physical CP-even scalars masses
	\begin{equation}    
	m^2_{H, h} = \frac{1}{2}\left(\mathbf{M}_{11}^2 + \mathbf{M}_{22}^2 \pm \sqrt{\left(\mathbf{M}_{11}^2 - \mathbf{M}_{22}^2\right)^2 + 4 \left(\mathbf{M}_{12}^2\right)^2} \right) \; .
	\label{eq:evenscalarmass}
	\end{equation}
	Note that the setting $\beta-\alpha = \pi/2$ is known as the alignment limit~\cite{Branco:2011iw}, where $H^0$ coincides with $h$ -- the SM Higgs boson with mass $m_{h}= 125.25$ GeV~\cite{ParticleDataGroup:2022pth}. For the CP-odd scalars the mass matrix is
	\begin{equation}
	\mathbf{M}^2_\text{CP-odd} = - \mu_{12}^2
	\begin{pmatrix}
	\dfrac{v_2}{v_1} & - 1 \\
	- 1      & \dfrac{v_1}{v_2}
	\end{pmatrix} \; ,
	\end{equation}
	being diagonalised by the rotation matrix $\mathbf{R}$ of eq.~\eqref{eq:Higgsbasis}. Besides the $G^0$ GB, the above diagonalisation leads to a physical CP-odd scalar $I$ with mass
	\begin{equation}
	m^2_{I} = m_{H^\pm}^2 + v^2 \frac{\lambda_4}{2} = - \frac{2\mu_{12}^2}{\sin\left(2\beta\right)} \; ,
	\label{eq:pseudoscalarmass}
	\end{equation}
	proportional to the soft-breaking parameter $\mu_{12}^2$.
	
	\section{Scalar-fermion interactions}
	\label{sec:interactions}
	
	In this appendix we present the scalar-fermion interactions in the mass-eigenstate basis of fermions and scalars, within the 2HDM framework with the Yukawa Lagrangian given in Eq.~\eqref{eq:Lyuk2hdm}. Namely, the interactions between fermions and charged Higgs $H^{\pm}$, as well as, neutral scalars~$h$, $H$ and $I$, are given by
	\begin{align}
	-\mathcal{L}_{H^\pm} &= \frac{\sqrt{2} \; H^+}{v} \; \left[\overline{u} \left(\textbf{N}_u^\dagger\textbf{V} P_L - \textbf{V}\textbf{N}_d P_R\right) d - \overline{\nu} \; \textbf{U}^\dagger \textbf{N}_e P_R \; e \right] + \mathrm{H.c.} \; , \\
	-\mathcal{L}_{h} &= \frac{h}{v} \Bigg\{\overline{u} \left[\left(s_{\beta-\alpha} \; \textbf{D}_u - c_{\beta-\alpha} \; \textbf{N}_u \right) P_R + \left(s_{\beta-\alpha} \; \textbf{D}_u - c_{\beta-\alpha} \; \textbf{N}_u^\dagger\right) P_L\right] u \nonumber \\
	& + \overline{d} \left[\left(s_{\beta-\alpha} \; \textbf{D}_d - c_{\beta-\alpha} \; \textbf{N}_d \right) P_R + \left(s_{\beta-\alpha} \; \textbf{D}_d - c_{\beta-\alpha} \; \textbf{N}_d^\dagger\right) P_L\right] d  \nonumber \\ 
	&+ \overline{e} \left[\left(s_{\beta-\alpha} \; \textbf{D}_e - c_{\beta-\alpha} \; \textbf{N}_e \right) P_R + \left(s_{\beta-\alpha} \; \textbf{D}_e - c_{\beta-\alpha} \; \textbf{N}_e^\dagger\right) P_L\right] e \Bigg\} \; , \\
	-\mathcal{L}_H &= \frac{H}{v} \Bigg\{\overline{u} \left[\left(c_{\beta-\alpha} \; \textbf{D}_u + s_{\beta-\alpha} \; \textbf{N}_u \right) P_R + \left(c_{\beta-\alpha} \; \textbf{D}_u + s_{\beta-\alpha} \; \textbf{N}_u^\dagger\right) P_L\right] u \nonumber \\
	& + \overline{d} \left[\left(c_{\beta-\alpha} \; \textbf{D}_d + s_{\beta-\alpha} \; \textbf{N}_d \right) P_R + \left(c_{\beta-\alpha} \; \textbf{D}_d + s_{\beta-\alpha} \; \textbf{N}_d^\dagger\right) P_L\right] d \nonumber \\
	& + \overline{e} \left[\left(c_{\beta-\alpha} \; \textbf{D}_e + s_{\beta-\alpha} \; \textbf{N}_e \right) P_R + \left(c_{\beta-\alpha} \; \textbf{D}_e + s_{\beta-\alpha} \; \textbf{N}_e^\dagger\right) P_L\right] e \Bigg\} \; , \\
	-\mathcal{L}_I &= \frac{i \; I}{v} \; \left[ \overline{u} \left(\textbf{N}_u P_R - \textbf{N}_u^\dagger P_L\right) u
	- \overline{d} \left(\textbf{N}_d P_R - \textbf{N}_d^\dagger P_L\right) d
	- \overline{e} \left(\textbf{N}_e P_R - \textbf{N}_e^\dagger P_L\right) e
	\right] \; ,
	\end{align}
	with the $\textbf{N}_{d,u,e}$ matrices defined in Eq.~\eqref{eq:NMatrices}, while the angles $\beta$ and $\alpha$ are defined in Eqs.~\eqref{eq:Higgsbasis} and~\eqref{eq:tan_alpha}, respectively.
	
	
	
\end{document}